\documentstyle[10pt,epsf,epsfig,hangcaption,xspace,amssymb,amsfonts,amsmath,amsthm,cite,
dp_delphititle,lineno]{dp_delphi}
\makeindex
\def\DpPaperGroup{EP}
\def\DpPaperRef{2003-008}
\def\DpDate{11 February 2003}
\def\DpAuthors{DELPHI Collaboration}
\def\DpSubmit{(Eur. Phys. J. C32 (2004) 145-183)}
\def\DpTitle{{\bf Final results from DELPHI on the searches for
SM and MSSM Neutral Higgs bosons}}
\def\DpComment{  }
\def\DpEMail{ }

\newcommand{\Zz} {\mbox{Z}}

\newcommand{\W} {\mbox{$ {\mathrm W}^{\pm} \,$}}
\newcommand{\Hz} {\mbox{H}}
\newcommand{\hz} {\mbox{h}}

\newcommand{\MA} {\mbox{$ m_{\mathrm A} $}}

\newcommand{\MH} {\mbox{$ m_{\mathrm H} \, $}}

\newcommand{\mh} {\mbox{$ m_{\mathrm h} $}}

\newcommand{\mtop} {\mbox{$ m_{\mathrm top} $}}
\newcommand{\ee}{\mbox{${\mathrm e}^+{\mathrm e}^-$}}

\newcommand{\qqbar}{\mbox{${\mathrm q}\bar{\mathrm q}$}}
\newcommand{\bbbar}{\mbox{${\mathrm b}\bar{\mathrm b}$}}
\newcommand{\ccbar}{\mbox{${\mathrm c}\bar{\mathrm c}$}}

\newcommand{\toto}{\mbox{$\tau^+ \tau^-$}}

\newcommand{\btagpep}{\mbox{$P^{+}_{\mathrm E}$} }

\newcommand{\xb}{\mbox{$x_{\mathrm b}$}}
\newcommand{\xbi}{\mbox{$x_{\mathrm b}^{i}$}}
\newcommand{\hmm}{\mbox{\Hz$ \mu^+ \mu^-$}}
\newcommand{\hee}{\mbox{\Hz${\mathrm {e^+ e^-}}$}}
\newcommand{\hnn}{\mbox{\Hz$ \nu \bar{\nu}$}}
\newcommand{\hqq}{$\Hz{\mathrm {q \bar{q}}}$}
\newcommand{\lhqq}{$\hz{\mathrm {q \bar{q}}}$}
\newcommand{\ttqq}{$\tau^+\tau^- {\mathrm q \bar{q}}$ }
\newcommand{\nnqq}{$\nu\bar{\nu}{\mathrm q \bar{q}}$ }
\newcommand{\hAtt}{${\mathrm {hA}} \rightarrow \tau^+\tau^- {\mathrm b \bar{b}}$ }
\newcommand{\hAbb}{hA$\rightarrow $\bbbar \bbbar }
\newcommand{\Abb}{A$\rightarrow $\bbbar}
\newcommand{\Acc}{A$\rightarrow $\ccbar}
\newcommand{\hAA} {${\mathrm{h}}\rightarrow {\mathrm{AA}}$}
\newcommand{\Zqq} {${\mathrm{Z}}\rightarrow$ \qqbar}

\newcommand{\qqg} {\mbox{$ {\mathrm q}\bar{\mathrm q}(\gamma) $}}
\newcommand{\qqgg} {\mbox{$ {\mathrm q}\bar{\mathrm q}\gamma(\gamma) $}}
\newcommand{\gaga}{\mbox{$\gamma \gamma$}}

\newcommand{\WW} {\mbox{${\mathrm W}^+{\mathrm W}^-$}}

\newcommand{\ZZ} {${\mathrm {ZZ}}$}
\newcommand{\ZH} {\mbox{${\mathrm {HZ}}$}}
\newcommand{\hA} {\mbox{$ {\mathrm h} {\mathrm A}$}}
\newcommand{\hZ} {\mbox{$ {\mathrm h} {\mathrm Z}$}}

\newcommand{\eeqq} {\mbox{\ee \qqbar }}
\newcommand{\mmqq} {\mbox{$\mu^+ \mu^- $\qqbar }}
\newcommand{\tautauqq}{\mbox{$\tau^+ \tau^- $\qqbar }}

\newcommand{\llqq} {\mbox{${\ell^+\ell^- }$\qqbar }}

\newcommand{\tauvqq} {\mbox{$\tau\nu$${\mathrm q'}\bar{\mathrm q}$ }}

\newcommand{\tbeta} {\mbox{$\tan \beta$}}
\newcommand{\MeV} {\mbox{${\mathrm{MeV}} $}}
\newcommand{\MeVc} {\mbox{${\mathrm{MeV}}/c $}}

\newcommand{\GeV} {\mbox{${\mathrm{GeV}} $}}
\newcommand{\GeVc} {\mbox{${\mathrm{GeV}}/c $}}
\newcommand{\GeVcc} {\mbox{${\mathrm{GeV}}/c^2 $}}

\newcommand{\dgree} {\mbox{$^{\circ}$}}
\newcommand{\mydeg} {$^{\circ}$}

\newcommand{\pbinv} {\mbox{pb$^{-1}$}}

\newcommand{\sqrts }{\mbox{$ \sqrt{s} \,$}}

\newcommand{\likear}{\mbox{$\cal Q$}}

\newcommand{\CLb}{\mbox{$\rm CL_{\rm b}$}}
\newcommand{\CLs}{\mbox{$\rm CL_{\rm s}$}}
\newcommand{\CLsb}{\mbox{$\rm CL_{\rm s + b}$}}

\newcommand{\HZ} {\mbox{$ {\mathrm H} {\mathrm Z} $}}

\newcommand{\zee} {\mbox{$ {{\mathrm Ze}}^+{{\mathrm e}}^- $}}

\def\EPJ#1#2#3{{ Eur. Phys. J.} {\bf{C#1}} (#2) #3}
\def\CPC#1#2#3{{ Comp. Phys. Comm.} {\bf{#1}} (#2) #3}
\def\NPB#1#2#3{{ Nucl.~Phys.} {\bf{B#1}} (#2) #3}
\def\PLB#1#2#3{{ Phys.~Lett.} {\bf{B#1}} (#2) #3}
\def\PRD#1#2#3{{ Phys.~Rev.} {\bf{D#1}} (#2) #3}

\def\ZPC#1#2#3{{ Z.~Phys.} {\bf C#1} (#2) #3}

\def\NIMA#1#2#3{{ Nucl.~Instr.~and~Meth.} {\bf A#1} (#2) #3}
\newcommand{\rs}{\mbox{$\sqrt{s}$}}

\begin{document}
\makeatletter
\newcount\@tempcntc
\def\@citex[#1]#2{\if@filesw\immediate\write\@auxout{\string\citation{#2}}\fi
  \@tempcnta\z@\@tempcntb\m@ne\def\@citea{}\@cite{\@for\@citeb:=#2\do
    {\@ifundefined
       {b@\@citeb}{\@citeo\@tempcntb\m@ne\@citea\def\@citea{,}{\bf ?}\@warning
       {Citation `\@citeb' on page \thepage \space undefined}}%
    {\setbox\z@\hbox{\global\@tempcntc0\csname b@\@citeb\endcsname\relax}%
     \ifnum\@tempcntc=\z@ \@citeo\@tempcntb\m@ne
       \@citea\def\@citea{,}\hbox{\csname b@\@citeb\endcsname}%
     \else
      \advance\@tempcntb\@ne
      \ifnum\@tempcntb=\@tempcntc
      \else\advance\@tempcntb\m@ne\@citeo
      \@tempcnta\@tempcntc\@tempcntb\@tempcntc\fi\fi}}\@citeo}{#1}}
\def\@citeo{\ifnum\@tempcnta>\@tempcntb\else\@citea\def\@citea{,}%
  \ifnum\@tempcnta=\@tempcntb\the\@tempcnta\else
   {\advance\@tempcnta\@ne\ifnum\@tempcnta=\@tempcntb \else \def\@citea{--}\fi
    \advance\@tempcnta\m@ne\the\@tempcnta\@citea\the\@tempcntb}\fi\fi}
 
\makeatother
\begin{titlepage}
\pagenumbering{roman}
\CERNpreprint{\DpPaperGroup}{\DpPaperRef} 
\date{{\small\DpDate}} 
\title{\DpTitle} 
\address{\DpAuthors} 
\begin{shortabs} 
\noindent
\noindent
%
\noindent

These final results from  DELPHI searches
for the Standard Model ({\sc SM}) Higgs boson, together with
benchmark
scans  of the Minimal Supersymmetric Standard Model ({\sc MSSM}) neutral Higgs
bosons, used data taken at centre-of-mass energies between 
 200 and 209~\GeV\ with a total integrated luminosity of 
 224~\pbinv.  
The data from 192 to 202~\GeV\ are reanalysed with improved b-tagging
for MSSM final states decaying to four b-quarks. The 95\%
confidence level lower mass bound on the 
 Standard Model Higgs boson is 114.1~\GeVcc.
 Limits are also given on the lightest scalar and pseudo-scalar Higgs
bosons of the MSSM. 
\end{shortabs}
\vfill
\begin{center}
\DpSubmit \ \\ 
\DpComment \ \\
\DpEMail \ \\
\end{center}
\vfill
\clearpage
\headsep 10.0pt
\addtolength{\textheight}{10mm}
\addtolength{\footskip}{-5mm}
\begingroup
%
\newcommand{\DpName}[2]{\hbox{#1$^{\ref{#2}}$},\hfill}
\newcommand{\DpNameTwo}[3]{\hbox{#1$^{\ref{#2},\ref{#3}}$},\hfill}
\newcommand{\DpNameThree}[4]{\hbox{#1$^{\ref{#2},\ref{#3},\ref{#4}}$},\hfill}
\newskip\Bigfill \Bigfill = 0pt plus 1000fill
\newcommand{\DpNameLast}[2]{\hbox{#1$^{\ref{#2}}$}\hspace{\Bigfill}}
%
\footnotesize
\noindent
\DpName{J.Abdallah}{LPNHE}
\DpName{P.Abreu}{LIP}
\DpName{W.Adam}{VIENNA}
\DpName{P.Adzic}{DEMOKRITOS}
\DpName{T.Albrecht}{KARLSRUHE}
\DpName{T.Alderweireld}{AIM}
\DpName{R.Alemany-Fernandez}{CERN}
\DpName{T.Allmendinger}{KARLSRUHE}
\DpName{P.P.Allport}{LIVERPOOL}
\DpName{U.Amaldi}{MILANO2}
% ...
\normalsize
\endgroup
\titlefoot{Department of Physics and Astronomy, Iowa State
     University, Ames IA 50011-3160, USA
    \label{AMES}}
\titlefoot{Physics Department, Universiteit Antwerpen,
     Universiteitsplein 1, B-2610 Antwerpen, Belgium \\
     \indent~~and IIHE, ULB-VUB,
     Pleinlaan 2, B-1050 Brussels, Belgium \\
     \indent~~and Facult\'e des Sciences,
     Univ. de l'Etat Mons, Av. Maistriau 19, B-7000 Mons, Belgium
    \label{AIM}}
\titlefoot{Physics Laboratory, University of Athens, Solonos Str.
     104, GR-10680 Athens, Greece
    \label{ATHENS}}
\titlefoot{Department of Physics, University of Bergen,
     All\'egaten 55, NO-5007 Bergen, Norway
    \label{BERGEN}}
\titlefoot{Dipartimento di Fisica, Universit\`a di Bologna and INFN,
     Via Irnerio 46, IT-40126 Bologna, Italy
    \label{BOLOGNA}}
\titlefoot{Centro Brasileiro de Pesquisas F\'{\i}sicas, rua Xavier Sigaud 150,
     BR-22290 Rio de Janeiro, Brazil \\
     \indent~~and Depto. de F\'{\i}sica, Pont. Univ. Cat\'olica,
     C.P. 38071 BR-22453 Rio de Janeiro, Brazil \\
     \indent~~and Inst. de F\'{\i}sica, Univ. Estadual do Rio de Janeiro,
     rua S\~{a}o Francisco Xavier 524, Rio de Janeiro, Brazil
    \label{BRASIL}}
\titlefoot{Coll\`ege de France, Lab. de Physique Corpusculaire, IN2P3-CNRS,
     FR-75231 Paris Cedex 05, France
    \label{CDF}}
\titlefoot{CERN, CH-1211 Geneva 23, Switzerland
    \label{CERN}}
\titlefoot{Institut de Recherches Subatomiques, IN2P3 - CNRS/ULP - BP20,
     FR-67037 Strasbourg Cedex, France
    \label{CRN}}
\titlefoot{Now at DESY-Zeuthen, Platanenallee 6, D-15735 Zeuthen, Germany
    \label{DESY}}
\titlefoot{Institute of Nuclear Physics, N.C.S.R. Demokritos,
     P.O. Box 60228, GR-15310 Athens, Greece
    \label{DEMOKRITOS}}
\titlefoot{FZU, Inst. of Phys. of the C.A.S. High Energy Physics Division,
     Na Slovance 2, CZ-180 40, Praha 8, Czech Republic
    \label{FZU}}
\titlefoot{Dipartimento di Fisica, Universit\`a di Genova and INFN,
     Via Dodecaneso 33, IT-16146 Genova, Italy
    \label{GENOVA}}
\titlefoot{Institut des Sciences Nucl\'eaires, IN2P3-CNRS, Universit\'e
     de Grenoble 1, FR-38026 Grenoble Cedex, France
    \label{GRENOBLE}}
\titlefoot{Helsinki Institute of Physics, P.O. Box 64,
     FIN-00014 University of Helsinki, Finland
    \label{HELSINKI}}
\titlefoot{Joint Institute for Nuclear Research, Dubna, Head Post
     Office, P.O. Box 79, RU-101 000 Moscow, Russian Federation
    \label{JINR}}
\titlefoot{Institut f\"ur Experimentelle Kernphysik,
     Universit\"at Karlsruhe, Postfach 6980, DE-76128 Karlsruhe,
     Germany
    \label{KARLSRUHE}}
\titlefoot{Institute of Nuclear Physics,Ul. Kawiory 26a,
     PL-30055 Krakow, Poland
    \label{KRAKOW1}}
\titlefoot{Faculty of Physics and Nuclear Techniques, University of Mining
     and Metallurgy, PL-30055 Krakow, Poland
    \label{KRAKOW2}}
\titlefoot{Universit\'e de Paris-Sud, Lab. de l'Acc\'el\'erateur
     Lin\'eaire, IN2P3-CNRS, B\^{a}t. 200, FR-91405 Orsay Cedex, France
    \label{LAL}}
\titlefoot{School of Physics and Chemistry, University of Lancaster,
     Lancaster LA1 4YB, UK
    \label{LANCASTER}}
\titlefoot{LIP, IST, FCUL - Av. Elias Garcia, 14-$1^{o}$,
     PT-1000 Lisboa Codex, Portugal
    \label{LIP}}
\titlefoot{Department of Physics, University of Liverpool, P.O.
     Box 147, Liverpool L69 3BX, UK
    \label{LIVERPOOL}}
\titlefoot{Dept. of Physics and Astronomy, Kelvin Building,
     University of Glasgow, Glasgow G12 8QQ
    \label{GLASGOW}}
\titlefoot{LPNHE, IN2P3-CNRS, Univ.~Paris VI et VII, Tour 33 (RdC),
     4 place Jussieu, FR-75252 Paris Cedex 05, France
    \label{LPNHE}}
\titlefoot{Department of Physics, University of Lund,
     S\"olvegatan 14, SE-223 63 Lund, Sweden
    \label{LUND}}
\titlefoot{Universit\'e Claude Bernard de Lyon, IPNL, IN2P3-CNRS,
     FR-69622 Villeurbanne Cedex, France
    \label{LYON}}
\titlefoot{Dipartimento di Fisica, Universit\`a di Milano and INFN-MILANO,
     Via Celoria 16, IT-20133 Milan, Italy
    \label{MILANO}}
\titlefoot{Dipartimento di Fisica, Univ. di Milano-Bicocca and
     INFN-MILANO, Piazza della Scienza 2, IT-20126 Milan, Italy
    \label{MILANO2}}
\titlefoot{IPNP of MFF, Charles Univ., Areal MFF,
     V Holesovickach 2, CZ-180 00, Praha 8, Czech Republic
    \label{NC}}
\titlefoot{NIKHEF, Postbus 41882, NL-1009 DB
     Amsterdam, The Netherlands
    \label{NIKHEF}}
\titlefoot{National Technical University, Physics Department,
     Zografou Campus, GR-15773 Athens, Greece
    \label{NTU-ATHENS}}
\titlefoot{Physics Department, University of Oslo, Blindern,
     NO-0316 Oslo, Norway
    \label{OSLO}}
\titlefoot{Dpto. Fisica, Univ. Oviedo, Avda. Calvo Sotelo
     s/n, ES-33007 Oviedo, Spain
    \label{OVIEDO}}
\titlefoot{Department of Physics, University of Oxford,
     Keble Road, Oxford OX1 3RH, UK
    \label{OXFORD}}
\titlefoot{Dipartimento di Fisica, Universit\`a di Padova and
     INFN, Via Marzolo 8, IT-35131 Padua, Italy
    \label{PADOVA}}
\titlefoot{Rutherford Appleton Laboratory, Chilton, Didcot
     OX11 OQX, UK
    \label{RAL}}
\titlefoot{Dipartimento di Fisica, Universit\`a di Roma II and
     INFN, Tor Vergata, IT-00173 Rome, Italy
    \label{ROMA2}}
\titlefoot{Dipartimento di Fisica, Universit\`a di Roma III and
     INFN, Via della Vasca Navale 84, IT-00146 Rome, Italy
    \label{ROMA3}}
\titlefoot{DAPNIA/Service de Physique des Particules,
     CEA-Saclay, FR-91191 Gif-sur-Yvette Cedex, France
    \label{SACLAY}}
\titlefoot{Instituto de Fisica de Cantabria (CSIC-UC), Avda.
     los Castros s/n, ES-39006 Santander, Spain
    \label{SANTANDER}}
\titlefoot{Inst. for High Energy Physics, Serpukov
     P.O. Box 35, Protvino, (Moscow Region), Russian Federation
    \label{SERPUKHOV}}
\titlefoot{J. Stefan Institute, Jamova 39, SI-1000 Ljubljana, Slovenia
     and Laboratory for Astroparticle Physics,\\
     \indent~~Nova Gorica Polytechnic, Kostanjeviska 16a, SI-5000 Nova Gorica, Slovenia, \\
     \indent~~and Department of Physics, University of Ljubljana,
     SI-1000 Ljubljana, Slovenia
    \label{SLOVENIJA}}
\titlefoot{Fysikum, Stockholm University,
     Box 6730, SE-113 85 Stockholm, Sweden
    \label{STOCKHOLM}}
\titlefoot{Dipartimento di Fisica Sperimentale, Universit\`a di
     Torino and INFN, Via P. Giuria 1, IT-10125 Turin, Italy
    \label{TORINO}}
\titlefoot{INFN,Sezione di Torino, and Dipartimento di Fisica Teorica,
     Universit\`a di Torino, Via P. Giuria 1,\\
     \indent~~IT-10125 Turin, Italy
    \label{TORINOTH}}
\titlefoot{Dipartimento di Fisica, Universit\`a di Trieste and
     INFN, Via A. Valerio 2, IT-34127 Trieste, Italy \\
     \indent~~and Istituto di Fisica, Universit\`a di Udine,
     IT-33100 Udine, Italy
    \label{TU}}
\titlefoot{Univ. Federal do Rio de Janeiro, C.P. 68528
     Cidade Univ., Ilha do Fund\~ao
     BR-21945-970 Rio de Janeiro, Brazil
    \label{UFRJ}}
\titlefoot{Department of Radiation Sciences, University of
     Uppsala, P.O. Box 535, SE-751 21 Uppsala, Sweden
    \label{UPPSALA}}
\titlefoot{IFIC, Valencia-CSIC, and D.F.A.M.N., U. de Valencia,
     Avda. Dr. Moliner 50, ES-46100 Burjassot (Valencia), Spain
    \label{VALENCIA}}
\titlefoot{Institut f\"ur Hochenergiephysik, \"Osterr. Akad.
     d. Wissensch., Nikolsdorfergasse 18, AT-1050 Vienna, Austria
    \label{VIENNA}}
\titlefoot{Inst. Nuclear Studies and University of Warsaw, Ul.
     Hoza 69, PL-00681 Warsaw, Poland
    \label{WARSZAWA}}
\titlefoot{Fachbereich Physik, University of Wuppertal, Postfach
     100 127, DE-42097 Wuppertal, Germany \\
\noindent
{$^\dagger$~deceased}
    \label{WUPPERTAL}}
\addtolength{\textheight}{-10mm}
\addtolength{\footskip}{5mm}
\clearpage
\headsep 30.0pt
\end{titlepage}
%
\pagenumbering{arabic} 
\setcounter{footnote}{0} %
\large
\section{Introduction}

  This paper presents the final results of the DELPHI collaboration on
the search for the Standard Model ({\sc SM}) Higgs boson, together with
benchmark
scans  of the Minimal Supersymmetric Standard Model ({\sc MSSM}) neutral Higgs
bosons.  
Results are presented for the SM Higgs particle in the mass range
from 12 to 120~\GeVcc,
and for the A and h bosons of the {\sc MSSM} in a similar range.
With the data taken up to $\sqrt{s} = 201.7$~\GeV, 
DELPHI excluded a  {\sc SM} Higgs boson with mass from zero to 
107.3~\GeVcc~\cite{ref:pap99} at the 95\% confidence level.
The results obtained for a high mass {\sc SM} signal with the data taken by 
DELPHI in the last year of LEP operation, 2000,  and analyzed with preliminary
calibration constants can be found in Ref.~\cite{ref:D2000}. 
In that year there was considerable interest caused by the observation
of an excess of events when the combined results of all the LEP
collaborations were considered ~\cite{ref:lep2000}.
Results on MSSM Higgs bosons have not previously been published using
the DELPHI from the year 2000.

The present work contains a more thorough analysis of the 2000 data, and
is combined with the results already published from previous
years~\cite{ref:pap99}.
It might be compared with the final results on Neutral Higgs
bosons from the other LEP collaborations~\cite{ref:lep-final}.
It benefits from many improvements when compared to
the originally published results, including  a revised data processing
with improved calibrations and significant
improvements in the simulation of signal and especially background
processes. 
These analyses concentrate on masses between 105 and 120~\GeVcc, but
they are also applied to lower masses, 
down to the \bbbar\ threshold, in order
to derive a constraint on the production cross-section of a {\sc SM}-like
Higgs boson as a function of its mass.
The revised data processing and the extension towards low masses 
implied changes to the analysis selection criteria that were tuned
on simulated samples. The high mass optimizations remain the same
as in  Ref.~\cite{ref:D2000}.

The dominant production mechanism at LEP for a scalar Higgs boson, such as the
{\sc SM} predicts, is  the  s-channel process
\ee $\rightarrow \mbox{Z}^* \rightarrow $\ZH, but there are additional
t-channel diagrams in the \hnn\ and \hee\ final states, which proceed
through \WW\ and \ZZ\ fusions, respectively. 
In the {\sc MSSM}, the
production of the lightest scalar Higgs boson, h, proceeds 
through the same processes as in the {\sc SM}.
The data from the search for the {\sc SM} Higgs boson also provide
information on  the h boson. However, in the {\sc MSSM}  
the production cross-section is  smaller than the {\sc SM} one 
and can even vanish in certain  regions of the {\sc MSSM} parameter space.
There is  also a CP-odd
pseudo-scalar, A, which would be produced mostly in the
\ee $\rightarrow \mbox{Z}^* \rightarrow {\mathrm h} {\mathrm A}$
process at LEP2. 
This channel is therefore also considered in this paper.
For {\sc MSSM} parameter values for which
single h production is suppressed, the associated \hA\ production
is enhanced (if kinematically permitted). 
Previous 95\% CL limits from DELPHI on the masses of h and
A
were 85.9~\GeVcc\ and 
86.5~\GeVcc\ respectively~\cite{ref:pap99}. 
 The present analysis in the \hA\ channel covers  masses between
40 and 100~\GeVcc.
The {\sc MSSM} interpretations rely on theoretical calculations with
limited 
second-order radiative corrections. They will be updated in a separate
paper using more complete corrections.

In the \ZH\ channel, all known decays of the Z boson
 (hadrons, charged leptons and neutrinos) have been taken into account,
while the analyses have been optimized for decays of the Higgs 
particle into \mbox{${\mathrm b}\bar{\mathrm b}$},
making use of the expected high branching fraction of this mode,
and for Higgs boson decays into a pair of \mbox{$\tau$} particles, which is
the second main decay channel in the  {\sc SM} and in most of the  
{\sc MSSM} parameter space.
The sensitivity of the  four-jet search to the  decay \hAA\ has
been measured and included.
The \hA\ production has been searched for in the two main decay channels, 
namely the \bbbar\bbbar\  and \mbox{${\mathrm b}\bar{\mathrm b}\tau^+ \tau^-$} 
final states. 
An  extended MSSM search, including more signal channels, will be 
reported separately. 

The detector description and the data samples  are discussed in
section~\ref{sec:data}, and the simulations with which they are compared are
described
in section~\ref{sec:mc}. Techniques common to more than one analysis are
presented in 
section~\ref{sec:common}, while the analyses themselves are described in
sections \ref{sec:hee} to \ref{sec:4jet}. The systematic errors are
discussed in section~\ref{sec:systematics}, and the results and
conclusions are in sections~\ref{sec:results} and \ref{sec:conclusions}
respectively.

\section{Data samples and detector overview}
\label{sec:data}

DELPHI recorded a total of 224~\pbinv\ of data in the year 2000. 

A short description of the detector can be found in Ref.~\cite{ref:pap97},
while more details can be found in Ref.~\cite{ref:delsim,ref:perfo}
for the original setup and in Ref.~\cite{ref:upgrade} for the LEP2 upgrade
of the silicon tracking detector.

The whole detector was unchanged from the previous operational period, 
except that one of the twelve sectors of the Time Projection Chamber (TPC)
suffered a failure in September 2000.
The reconstruction software for charged particle tracks in data
collected after this time  was adjusted to
make best use of the Silicon Tracker and Inner Detector both
placed closer to the beam than the TPC and the Outer Detector
and Barrel Rich placed outside the outer radius of the TPC.
As a result, the impact of the malfunctioning of that TPC sector
on the determination of  jet momenta was not large 
but the ${\mathrm b}$-tagging in that
twelfth of the detector remained significantly degraded.

\begin{figure}[htbp]
\begin{center}
\epsfig{figure=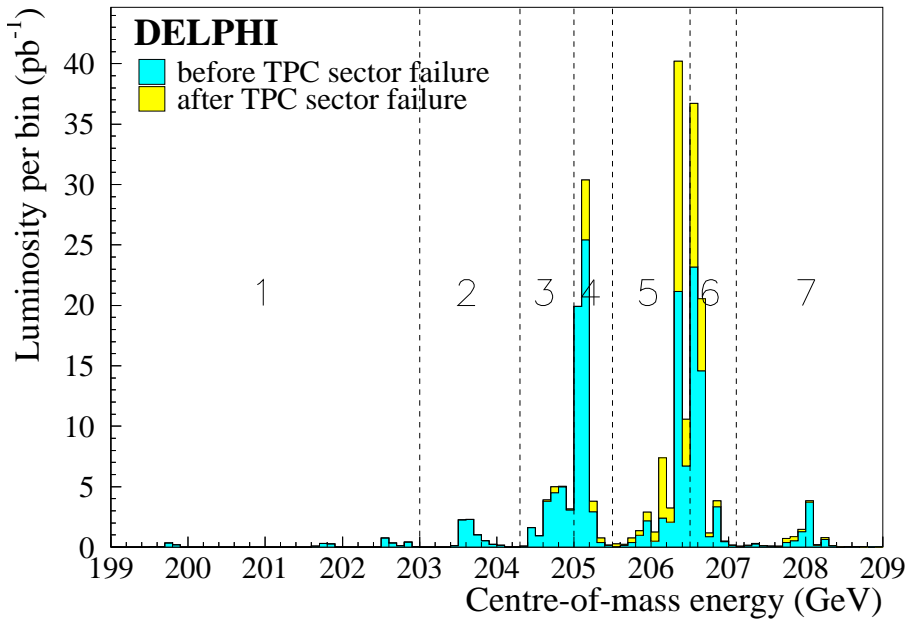,width=13cm} \\
\caption[]{The LEP energy distribution in 2000. The data analysis 
has been divided into two periods. The darker grey
shows the data taken in the first period, and the data taken in the
second period is in a lighter grey.
The vertical lines and numbers  show the energy bins into which the
data were grouped.
}
\label{fig:ebeam}
\end{center}
\end{figure}

LEP was run with a beam energy which was optimized to maximize
the sensitivity to the SM Higgs boson. The resulting spectrum
is shown in Fig.~\ref{fig:ebeam}, which also shows 
seven windows into which the analysis was divided. Data with a beam energy
falling into a particular window  was treated as if it had the mean energy
for that window, giving the values listed in Table~\ref{ta:ewindow}.
These windows were selected to give accurate results without complicating the
statistical analysis.

DELPHI recorded  164.1~\pbinv\ with a fully operational
detector, and 60.1~\pbinv after the TPC problem occurred, as 
shown in Fig.~\ref{fig:ebeam}.
The analyses described here make a distinction between data collected 
before and after this event, which are referred to as 
 {\em the first operational period} 
and {\em the second operational period}.
In the second operational period there were no data in the first two energy
windows, resulting in twelve data sets in total.
The requirement of adequate detector performance reduces the luminosities
in the \hee\ and \hnn\ samples by 0.5\% and 3.8\% respectively in the first
period, and 1.7\% and 4.3\% respectively when the TPC sector was off.

The data have been reprocessed since our previous publication~\cite{ref:D2000}.
This reprocessing was primarily motivated by an improved calibration of the
TPC.

\begin{table}[htbp]
{\small
\begin{center}
\begin{tabular}{cccccccc}     \hline
                  & \multicolumn{7}{c}{ Energy windows} \\
\hline \hline 
Number            & 1    &2   & 3 & 4 & 5 & 6 & 7 \\
Low edge,   (\GeV)  & --    &203.0  & 204.3 & 205.0 & 205.5 & 206.5 & 207.1 \\
Mean energy, (\GeV) &~201.80~&~203.64~&~204.73~&~205.10~&~206.28~&~206.59~&~207.93 \\
Luminosity,~(\pbinv)& 2.92  & 6.64  & 19.72 & 54.97 & 68.10 & 62.92 & 8.91 \\
\hline
\end{tabular}
\caption[]{
The energy windows into which the recorded data were grouped.
}
\label{ta:ewindow}
\end{center}
}
\end{table}

\section{Simulation software}
\label{sec:mc}

The DELPHI simulation software has been significantly upgraded
with respect to the version described in Ref.~\cite{ref:D2000}. 
New Monte Carlo generator software has been used for both two-fermion
and four-fermion
background processes, and the signal simulations have also been updated. 
The generated events were passed through the DELPHI detector 
simulation program~\cite{ref:delsim}. 
These samples typically correspond to more than 100 times the
luminosity of the collected data, with $10^6$ hadronic two-fermion and
four-fermion background events at each of the following centre of mass
energies: 
203.7, 205.0, 206.5 and 208.0~\GeV.
Simulated samples allowing estimation of the effect of the TPC
problem were also produced at 206.5~\GeV.
Two-fermion background events were generated with 
{\tt KK2f}~\cite{ref:kk2f} for hadronic events and muon pairs
and with {\tt KORALZ}~\cite{ref:koralz}  for $\tau^+\tau^-$ 
final states.
The four-fermion events, 
which originate from  a coherent sum of many processes
whose main components are referred to as $\Zz\gamma^*$, 
 \WW\ and \ZZ\ in the following,
 were generated with
{\tt WPHACT}~\cite{ref:wphact}, which includes low-mass hadronic resonances
and
use of the full CKM matrix.
For  all of these,  the hadronisation was handled
by {\tt PYTHIA}~\cite{ref:pythia}, version 6.156.

{\tt PYTHIA} and {\tt BDK}~\cite{ref:bdk} with {\tt PYTHIA} 6.143 fragmentation
were used for two-photon processes (hereafter denoted as $\gamma \gamma$)
and {\tt BHWIDE}~\cite{ref:bhwide} for
Bhabha events in the main acceptance region.

The  \Zz\Zz\ production process, especially if at least one of the \Zz\
particles decays
to b-quarks, is an essentially  irreducible background process
in all signal channels since it has many features in common with the signal.
It is therefore a relevant check on the DELPHI detector 
that this process can be
accurately modelled. This has been demonstrated in Ref.~\cite{ref:delphi_zz}.

Signal events were produced using the {\tt HZHA}~\cite{ref:hzha} generator,
 which includes the \WW\ and \ZZ\ fusion processes in the
\hnn\ and \hee\ channels respectively, and the interference with \Hz\Zz.
Fragmentation using {\tt PYTHIA 6.156} was used to allow for the
scalar nature
of the Higgs particle, which increases the gluon radiation by some 10\%
compared with that for a vector boson\cite{ref:gluon10}.
For the \ZH\ process, the H mass was varied from 12 to 120~\GeVcc,
with steps of 5~\GeVcc\ above 80~\GeVcc, and wider steps at lower
masses. Extra points were inserted at 114 and 116~\GeVcc. 
For the \hA\ process,
samples were generated over a
 grid of more than 60 points 
in (\mh, \MA). Equal mass points were generated from 12 to
100~\GeVcc\ with a 5~\GeVcc\ step above 80~\GeVcc, and wider steps
below. For non-equal (\mh, \MA) points, the lower mass was varied 
in the same mass range with a step double that of the equal mass
points and, 
for each of the values of the lower mass, the higher mass was varied
up to the kinematic
limit with a 20~\GeVcc\ step. Extra points were generated with a 10 or
5~\GeVcc\ granularity around 80~\GeVcc.
In all samples, the Higgs boson widths were 
set below 1~\GeVcc\, 
which is 
consistent with the expectations of the {\sc MSSM} in most of the
parameter space, that is for \tbeta\ (the ratio of the vacuum 
expectation values of the two Higgs field doublets of the 
{\sc MSSM}) below 20. However, for \tbeta\ above 20, the h and A widths 
increase rapidly to reach several~\GeVcc\ at \tbeta~=~50, 
thus exceeding the experimental mass resolution which is
typically around 5~\GeVcc\ on the sum of the masses in the \hA\ channels. 
Because of this,
 a second set of simulations was performed at \tbeta~=~50
with \MA\ varied according to the same pattern as for
the equal mass point simulations. This fixes the h mass, 
which is almost equal to \MA\ at such a large value of \tbeta.


The \ZH\ simulated samples were classified
according to the Higgs and Z boson decay modes. For \hee, \hmm\ and 
\hnn\ the natural  {\sc SM} mix of \Hz\ decay modes into fermions was 
generated. 
As final states with hadrons and two $\tau$ particles benefit from a dedicated
analysis, the $\tau\tau$ decay mode was removed in the 
\hqq\ channel simulations, and
 the two \ZH\ channels involving $\tau$ leptons, for which one 
of the
bosons is forced to decay to a $\tau$ pair and the other hadronically,
were generated separately.
 Finally, three sets of 
 \hA\ simulations were generated, covering final states involving either
four b-quarks  or two b-quarks and two $\tau$ particles, with either the
h or the A
decaying into two leptons. These were then combined 
giving equal weight to each channel.
Efficiencies were defined relative
to these states. The size of these samples was normally 5000 
events and they were produced at the same centre-of-mass energies as the
background samples.

Although the signal simulations described above cover most of the expected
final states in the {\sc SM} and  {\sc MSSM}, they were complemented by
two additional sets 
at~206.5~\GeV, one with a fully 
 operational detector
 and the other one with one TPC sector missing.
 These samples were of  \hZ\  production with \hAA, as
 expected in restricted regions of the  {\sc MSSM} parameter space. 
 The A (h) mass was varied from 12~\GeVcc\ (50~\GeVcc)
 up to the kinematic limit.
The final states simulated were 
 hadronic decays of the Z boson and
 either four b or four c quarks from the A pair.
 The results obtained from these samples were assumed also to be 
 valid  at the other centre-of-mass energies.



\section{Features common to all analyses}
\label{sec:common}

\subsection{Particle selection}

In all analyses, charged particles were selected if their momentum was greater
than 100~\MeVc\ and if they originated from the interaction region 
(within  4 cm in the transverse plane and within
4 cm / $\sin\theta$ along the beam direction, where $\theta$ is the 
particle polar angle). Neutral
particles were defined either as energy clusters 
in the calorimeters not associated to charged particle tracks,
or as reconstructed vertices of photon conversions, interactions of 
neutral hadrons
or decays of neutral particles in the tracking volume.
All neutral clusters of energy greater than 200 or 300~\MeV\ (depending on
the calorimeter) were used, except in the searches with missing energy,
where 300 or 400~\MeV\ was required. 
The $\pi^{\pm}$ mass was used for all charged particles except identified 
leptons,
while  zero mass was used for electromagnetic clusters and the K$^0$ mass was
assigned to neutral hadronic clusters. 

\subsection{Jets and Constrained fits}

The DURHAM\cite{ref:durham} algorithm was used to reconstruct jets,
which were taken as estimators of the quark momenta.
A constrained fit~\cite{ref:pufitc} was performed to reconstruct 
the Higgs boson mass. 
The constraints of energy and momentum conservation were applied,
and the \Zz\ mass was fixed to its central value, except in the \hee\
and \hmm\ channels where a Breit-Wigner width was allowed.
An algorithm has been developed~\cite{ref:sprime} in order to estimate the
effective energy of the \ee\ collision. This algorithm makes use of 
a three-constraint kinematic fit in order to test the presence of an 
initial state photon 
along one of the beam directions and hence lost in the beam pipe. 
This effective 
centre-of-mass energy is called \mbox{$ \sqrt{s'}$} throughout this paper,
and is used to remove most of the events 
radiatively returning to the \Zz.

\subsection{b-quark identification }
\label{sec:btag}

The method of separation of  b-quarks from other flavours is described 
in detail in Ref.~\cite{ref:btag_new}, where the various differences between
B hadrons and other particles are accumulated into a single variable,
hereafter denoted 
\xb\ for an event and \xbi\ for the $i^{th}$ jet of particles.
 An important contribution to this combined 
variable is the probability $P_{i}^+$ that all tracks with a positive 
lifetime-signed impact parameter in the jet led to a product of
track significances as large as that observed, if these tracks
originated from the interaction point; (\btagpep\ is the same, but for all
tracks in an event). 
A low value of this probability is a signature for a B hadron. 
The likelihood ratio technique was then used to construct
\xbi\ by combining $P_{i}^+$ with 
 the transverse momentum 
(with respect to the jet axis) of any lepton belonging to the jet
and with the following information from any
secondary vertex found in the jet:
the mass computed from the particles assigned to the secondary vertex,
the momentum transverse to the line joining the secondary vertex to the 
primary,
the rapidity of the secondary  particles, and the fraction of the jet momentum 
carried by them.
The event variable, \xb, is $x_{\mathrm b}^{1} + x_{\mathrm b}^{2}$ for a 
two jet event, or the sum of the two largest \xbi\ in the case of a multi-jet
configuration.
Increasing values of \xb\ (or \xbi) correspond to increasingly 
`b-like' events (or jets).

Specifically for the four-jet channels, a further improvement of 
the  ${\mathrm b}$-tagging procedure was made.
The purity of the sample defined by a given b-tagging value had a 
dependence on various properties of the jet.
The ${\mathrm b}$-tagging was equalised (see Ref.~\cite{ref:btag_new})
to remove this effect explicitly for
the following variables: 
   the polar angle of the jet direction,
   the jet energy,
   the charged multiplicity of the jet,
   the angle between the jet direction and the nearest other jet,
   the average transverse momentum of charged particles with respect
   to the jet direction,
   the number of particles with negative impact parameter, and
  the invariant mass of the jet.
Including this dependence in the tagging algorithm significantly improved
the rejection of the light quark background events.
This technique required specifying the signal hypothesis.
For the \hZ\ search this was defined using \mh=110~\GeVcc\ at
 \sqrts=206.7~\GeV,
while in the case of  \hA\ a mixture of A masses (80 to 95~\GeVcc)
and beam energies (205 to 208~\GeV) was used.

The impact parameter resolutions were measured   using tracks with negative
lifetime signed impact parameters taken from \Zz\ calibration events.
The  overall calibrations were tuned~\cite{ref:btag_tune} 
using tracks with negative lifetime signed
impact parameters taken from high energy four-jet events. 
Tuning the Monte Carlo to match the data in this way introduced very
little bias as  such tracks
were only used in the final b-tagging for the equalization corrections
described above.

\begin{figure}[htbp]
\begin{center}
\epsfig{figure=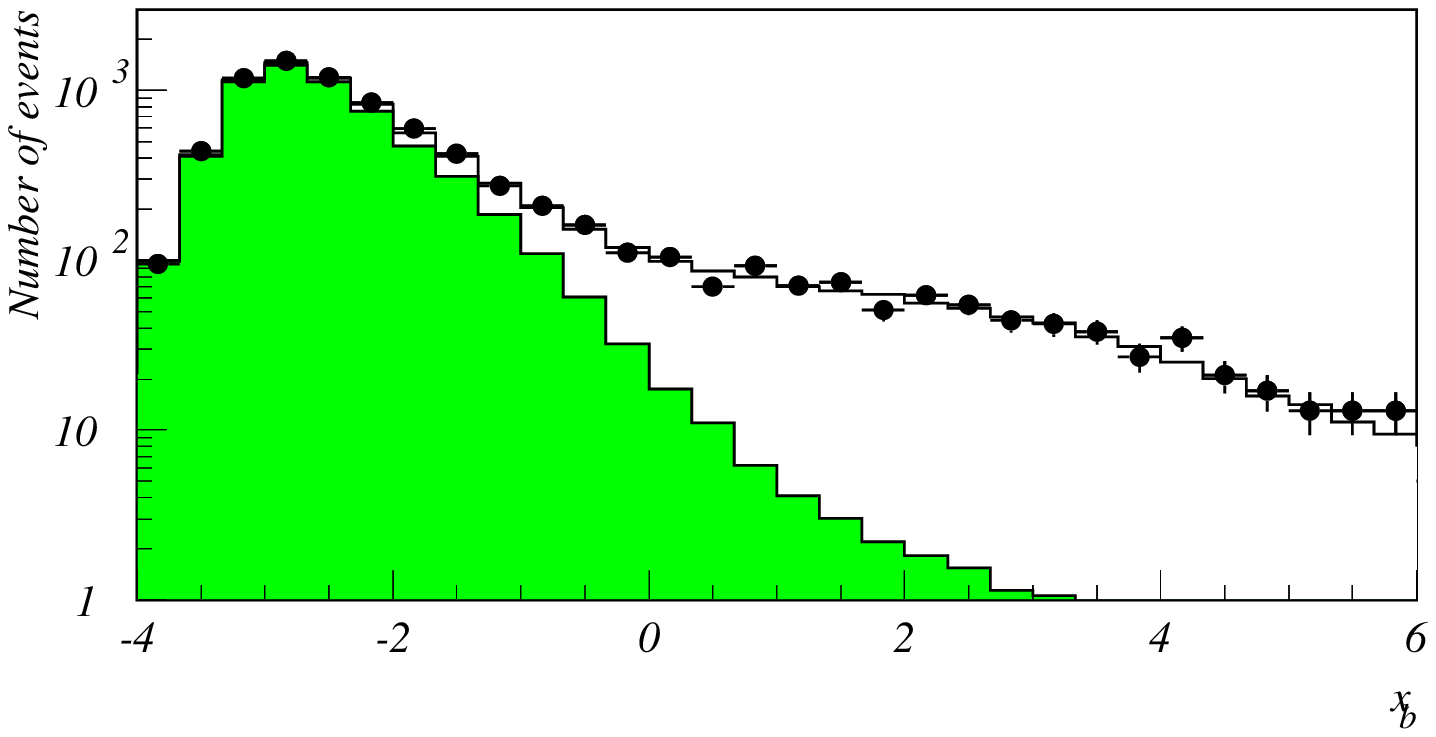,width=13cm} \\
\vspace{-1.9cm}
\epsfig{figure=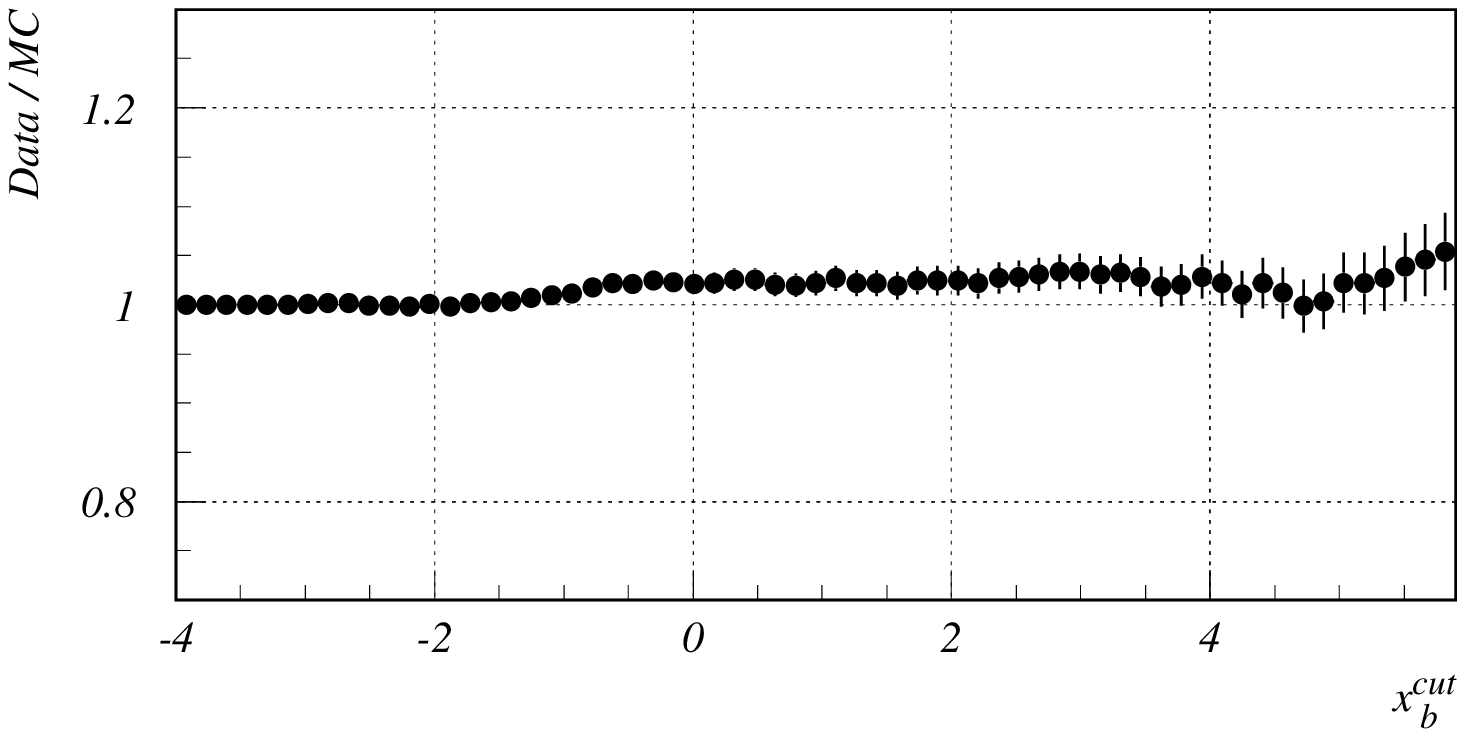,width=13cm} \\
\vspace{-1.9cm}
\epsfig{figure=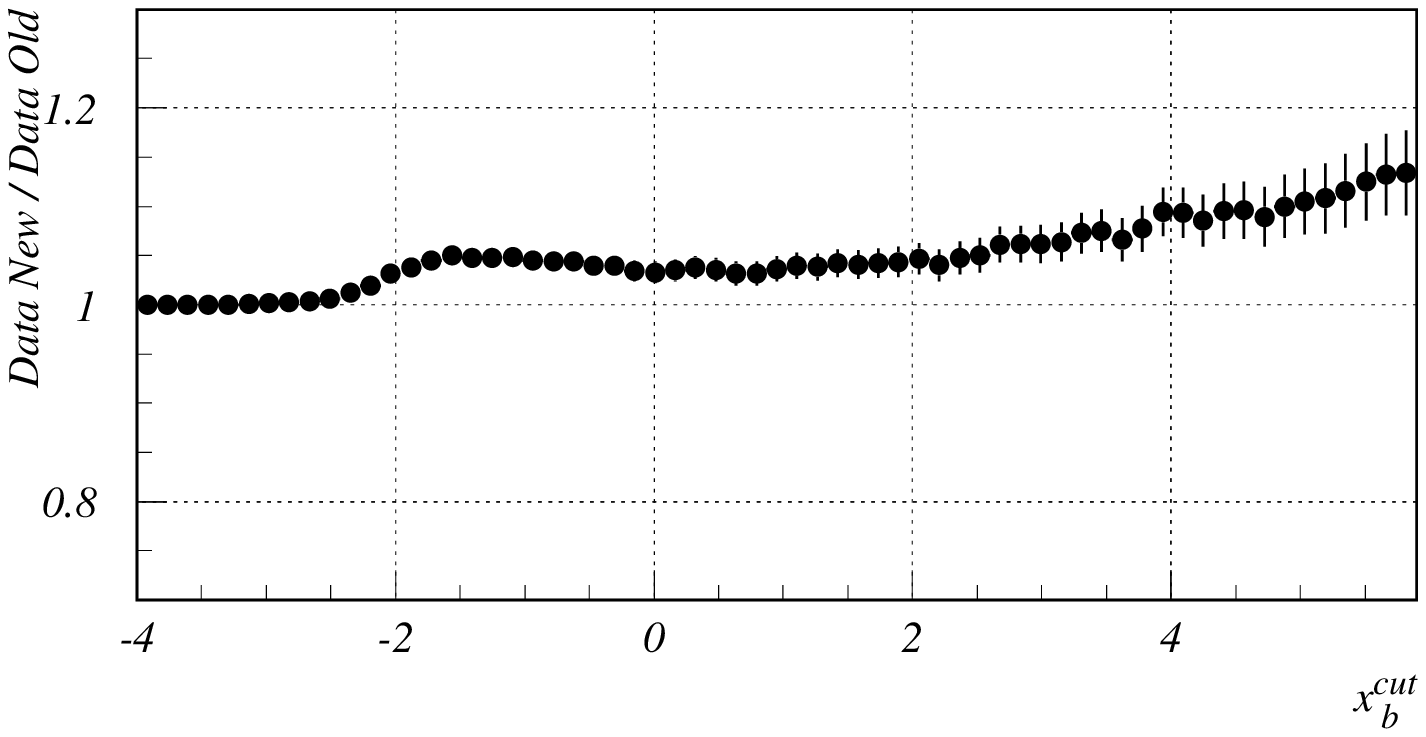,width=13cm} \\
\caption[]{
Top: distributions of the combined ${\mathrm b}$-tagging variable \xb, in 
events including radiative return to the \Zz\ from 2000 data
(dots) and simulation (histogram). 
The contribution of udsc-quarks is shown as the dark histogram. 
Middle: ratio of integrated tagging rates in 91~\GeV\ \Zz\ data
and simulation on application of the selection criterion
$\xb > \xb^{\rm cut}$, 
as a function of the $\xb^{\rm cut}$.
Bottom:
the ratio of the current rate of tagging of \Zz\ events to
that used in our previous publication, as a function of the cut value
 $\xb^{\rm cut}$.
Comparison with the previous processing shows that the efficiency for
${\mathrm b}$-tagging has increased by about 5\%.
}
\label{fig:btag-dataMC}
\end{center}
\end{figure}

\begin{figure}[htbp]
\begin{center}
\epsfig{figure=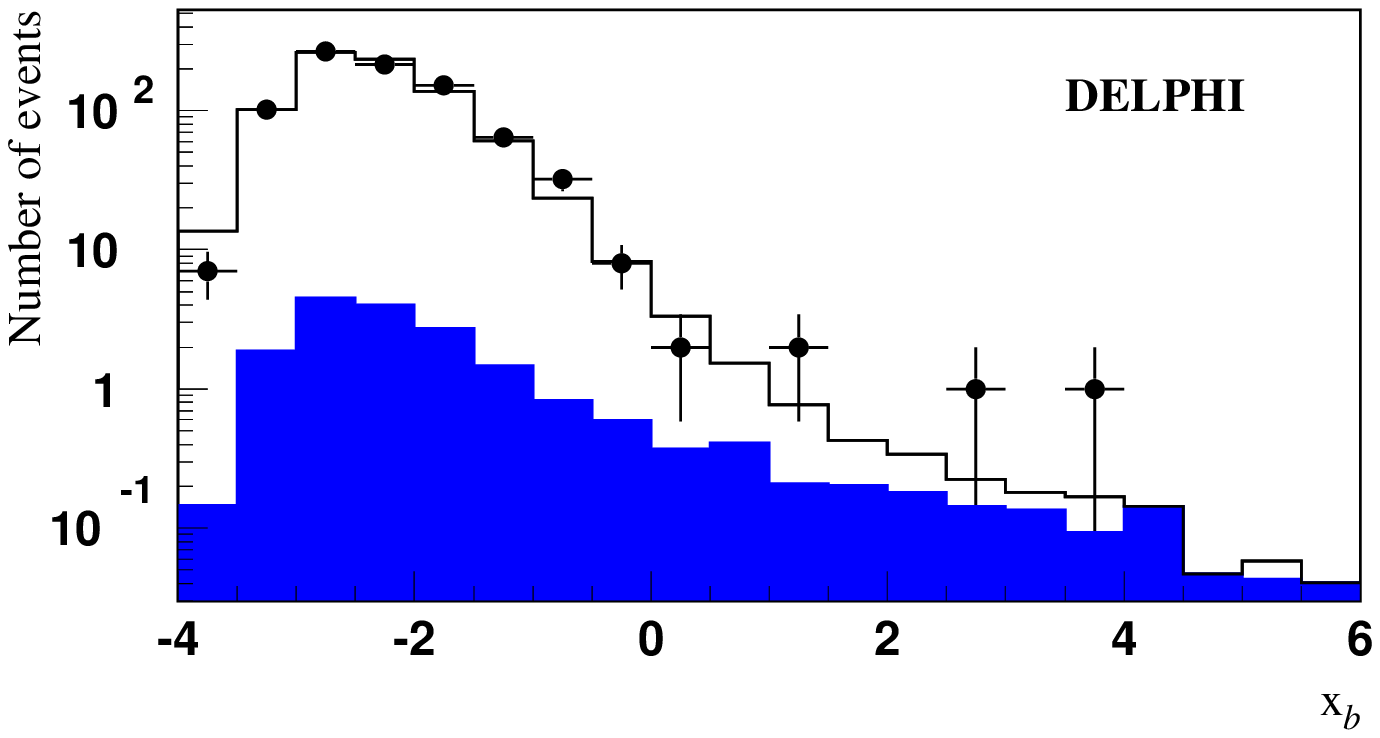,width=13cm} \\
\caption[]{
The b-tag obtained for semileptonic \WW\ decays, where the low level 
of b-quarks expected makes them a good sample for checking the
mistagging. The solid points are the data, the open histogram the total
of the simulation and the  black the contribution from processes other than
semileptonic \WW\ decays.
}
\label{fig:btag-ww}
\end{center}
\end{figure}

The agreement between data and simulation found in a  
sample of events returning radiatively  to the \Zz, and in data taken on
the \Zz\ peak, 
is shown in Fig.~\ref{fig:btag-dataMC}, and for semileptonic WW events
in Fig.~\ref{fig:btag-ww}.
The overall agreement in the ${\mathrm b}$-tagging between data and
simulation is
 better than 5\% in the whole range of cut values.
Figure~\ref{fig:btag-dataMC} also illustrates the increase in the
fraction of jets tagged as b-jets
for  \Zz\ peak data taken in the year 2000 from this  processing compared
to our previous publication.

\begin{figure}[htbp]
\begin{center}
\epsfig{figure=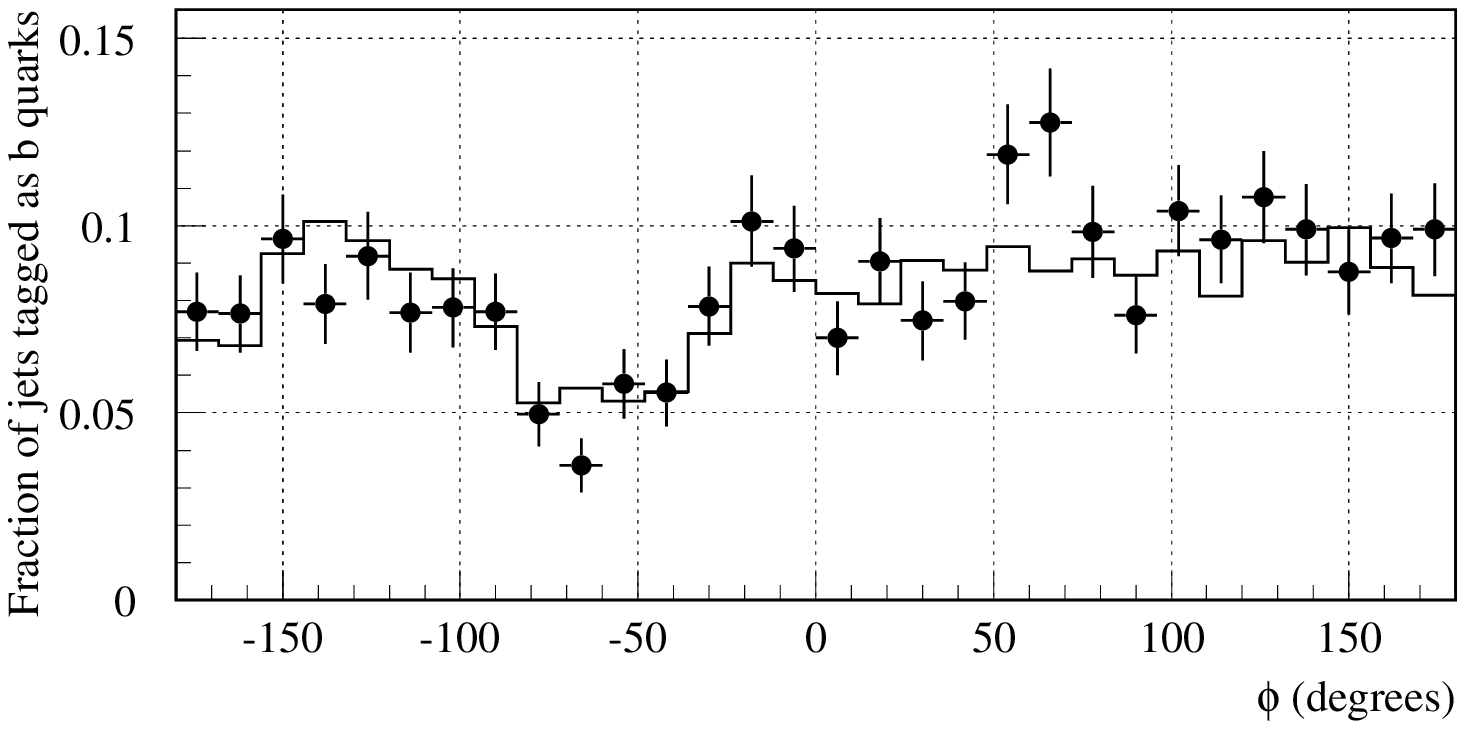,width=13cm}
\caption[]{
The fraction of jets tagged as b-jets
 as a function of azimuthal angle   in the 
hemisphere centred on the positron beam direction,  
after the TPC sector located from 
-90$^\circ$ to -30$^\circ$ failed.
The jets are from on peak \Zz\ events, and have
\xbi\ greater than -0.5.
The points are the data, and the line shows the simulation.
The opposite hemisphere is not affected.
}
\label{fig:btag-sector}
\end{center}
\end{figure}

Also shown in Fig.~\ref{fig:btag-sector} is the fraction of jets tagged as
b-jets as a function
of azimuthal angle for jets from \Zz\ particles which are in the
hemisphere centred on the positron beam direction,  
 for data taken when the TPC sector was off.
A significant degradation is seen in this small region, well matched by
simulation.

\subsection{Structure of the analysis}

The analysis for each channel takes the same basic pattern. A fairly loose
selection is applied which results in many candidates, 
 which are used to calculate the overall likelihood of the signal hypothesis.
The densities of signal and background 
processes for any measured combination of discriminant variable
(channel dependent) and candidate mass are estimated using Monte Carlo
simulation for the  centre-of-mass energies and Higgs masses which have
been discussed in sections~\ref{sec:data} and \ref{sec:mc}. 
These are interpolated to give
signal and background densities corresponding to the required beam
energy and Higgs mass hypothesis under consideration.
To simplify the analysis, the data events are treated as having the
 mean energy of the energy bin into which they fell, so all events
in that bin are treated together.

The estimated signal and background densities at the event are used to
find how much more probable the event is if the signal existed.
The use of a likelihood fit to extract the results means that regions
with low signal purities can be included in the selected data, and
each improves the separation. 
Loose cuts were made
on the  discriminant variables, with the result that
from the total search over one hundred events were expected from
background processes while retaining maximal sensitivity to
a {\sc SM}  Higgs signal. 
This procedure has the additional advantage that
the analysis is less dependent on biases from selection cuts.

%

\subsection{Confidence level definitions and calculations}
\label{sec:limits}
The confidence levels are calculated using a modified frequentist
technique based on the extended maximum likelihood ratio~\cite{ref:alex}
which has also been adopted by the LEP Higgs working group.

The basis of the calculation is the likelihood ratio test-statistic, \likear:
\[ \ln\likear = -S + \sum_i \ln \frac{s_i+b_i}{b_i} \]
where the $S$ is the total signal expected and $s_i$ and $b_i$ are the
signal and background densities for event $i$.
These densities are constructed using 
two-dimensional discriminant information  in all channels, 
as in our previous publication~\cite{ref:pap99}.
The first variable is the reconstructed Higgs boson mass (or the sum of
the reconstructed h and A masses in the \hA\ channels),
the second one is channel-dependent, as specified in the 
following sections.


The observed value of \likear\ is 
calculated using the two-dimensional
Probability Density Functions (PDFs) of the variables chosen for each channel.
The PDFs for \likear, which is naturally one-dimensional, are  built using
Monte Carlo sampling making the assumption that background processes
only or that both signal and
background are present, and the  confidence levels \CLb\ and \CLsb\ are
their integrals from $-\infty$ to the observed value of \likear.
Systematic uncertainties in the rates of signal or background events are taken 
into account in the calculation of the PDFs for \likear\ by 
randomly varying the expected rates while generating the
distribution~\cite{ref:CousinsAndHighland},
which has the effect of broadening the
expected \likear\ distribution and therefore 
making any extreme total observed likelihood  value seem more probable.

\CLb\ is the p-value for the  hypothesis that only background processes
are 
present, i.e. the probability
of obtaining a result as background-like
or more so than the one observed if the background hypothesis is correct.
It will tend toward one if there is a signal present, and is typically 0.5
if only background is there.
Similarly,
the confidence level for the hypothesis that both signal and background 
are present,
\CLsb, is the probability, in this hypothesis,
to obtain more background-like results than those observed. 
It is the  p-value for the hypothesis that both signal and background 
events were present.
The quantity \CLs\ is defined 
as the ratio of these
two probabilities, \CLsb/\CLb. It is not a true
confidence level, but a conservative pseudo-confidence level
for the signal hypothesis.
The division by \CLb\ means that \CLs\ is similar to \CLsb\ when 
\CLb\ is close to one, but increases the signal confidence when 
a result in the background-like region is obtained.
It is always larger than \CLsb, such that it reflects how many
times less likely the result is if the signal is present, and so gives a
more conservative limit which is designed to avoid the possibility of
excluding the Higgs in an experiment which has no sensitivity to it.
That is to say, the use of \CLs\ increases the signal confidence interval
in the background-like region  compared to \CLsb.
1-\CLs\ measures the confidence with which the
signal hypothesis can be rejected and, 
because it is conservative, 
 will be  at least 95\% for an
exclusion confidence of 95\%. 

\subsubsection{Estimation of distributions of mass and tag variable.}

The Probability Density Functions (PDFs) of the mass and the 
channel-dependent Higgs tagging
variable are required to check the consistency of the data with the
 background
and signal processes. They were treated as having two components:
the overall normalization and the shape of the
distribution.

In the case of a background process PDF, the normalization was
calculated from
the number of simulated events of each background class passing the cuts.
For the
signal the measured efficiencies had also to be interpolated
to estimate efficiencies at Higgs boson masses which were
not simulated. In most cases this was done using one polynomial
to describe the slow rise, and a second to handle the kinematic cut-off,
which can be much more abrupt.
For the cases where two signal masses must
be allowed,  a two-dimensional  parameterization was used.

 The shapes of the PDFs were derived  using two-dimensional histograms
which are taken from the simulated events. The two dimensions were the
Higgs boson mass estimator 
and a channel-dependent Higgs tag.
 These distributions were
 smoothed using a two-dimensional kernel, which consists of  a 
Gaussian distribution with a small component of a longer tail.
 The global covariance of
the distribution was used to determine the relative scale factors 
of the two axes. The width of the kernel
varied from point to
point, such that the statistical error on the estimated background
processes
was constant at 20\%. 
Finally multiplicative correction factors (each a one-dimensional
distribution for one of the two
dimensions of the PDF) were derived
such that when projected onto
either axis the PDF has the same distribution 
as would have been observed if it had been projected onto the axis 
first and then smoothed.
This makes better use of the simulation statistics if there are features 
which are essentially one-dimensional, such as mass peaks.

The error parameter fixed to 20\% was an important choice. It was set by
dividing
the background simulation into two subsamples, generating a PDF with one and
using the
other to test for over-training by calculating the \CLb\ obtained
from simulation of background events. This should be 0.5 if the results are not
to be biased, and a value of 20\% for
the error gave the closest approximation to 0.5. 
An
accurate description of the background is very important in a search for a 
new particle.

The simulations were made at fixed beam energies and
Higgs boson masses, but in order  to test a continuous range of masses and
 beam energies,  interpolation software~\cite{ref:alex-interp} was used to
create  signal PDFs at arbitrary masses and at the correct
centre-of-mass energies as well as background process PDFs at the correct
centre-of-mass energies.
 This
was done by linearly interpolating the cumulative distributions.
The procedure was essentially the same whether it is the beam energy or the 
signal mass which is being interpolated, and has been tested over ranges 
up to 40~\GeVcc\ in mass. The actual shifts were up to 0.3~\GeV\ in
  $\sqrt{s}$,
 and 5~\GeVcc\ in mass for the Standard Model Higgs overall, but less
than 0.5~\GeVcc\ for Higgs boson masses between 113.5 and 116.5~\GeVcc.
Comparisons of simulated and interpolated distributions
for a given mass show good agreement.


\section{Higgs boson searches in events with jets and electrons}
\label{sec:hee}

The analysis used a cut based method to separate signal from 
background.
It was very similar to that used in~\cite{ref:pap99}, but it has
been modified to increase the sensitivity to  low mass signals.
The event b-tag variable was used as the second variable in the
CL calculations.

The preselection required at least 8 charged particles,
a total energy above 0.12\rs\ and at least one pair 
of charged particles with energies above
 10~\GeV\ (where the energy was determined from
the tracking information and, when available, the calorimeter measurement) 
and track impact parameters
below 2~mm (1~cm) in the transverse plane (along the beam direction).
These tracks were required to have either an associated shower in
the electromagnetic
calorimeter
(tight electron candidates) or point to an insensitive calorimeter
region (loose electron candidates). The tight candidates had to have a
total associated energy in the last three  layers of the hadron calorimeter
of less 
than 1.6~\GeV\ and an E/p ratio above 0.3. The loose candidates had to have 
a normalised measured ionization energy loss in the TPC above 1.4.
The total energy of other particles within 5$^\circ$ of each
 candidate electron had to be less than 8~\GeV.
The sum of the calorimetric energies of the two candidates was required to
exceed 10~\GeV.
After removing the electron candidates, the remaining particles were forced
into two jets, and it was required that each 
of them contained at least 3 charged particles.

Bhabha events  showering in the detector material
were vetoed by rejecting cases where the  charged
multiplicity was less than or equal to 12 if a candidate electron had
an energy above 70~\GeV\ 
and an angle with respect to either of the beams below 25$^\circ$ or if the 
acoplanarity\footnote{The acoplanarity is defined
as the supplement of the angle between the transverse momenta 
(with respect to the beam axis) of the two electrons.}
was below 3$^\circ$ and both electron candidates had an energy
above 40~\GeV.

To reduce the contributions from the $\Zz\gamma^*$ and \qqg\ backgrounds,
the sum of the di-electron and hadronic system unfitted masses
had to  be above 50~\GeVcc, while the missing
momentum was required to be below 50~\GeVc\ if its direction was
within 10$^\circ$ of the beam axis. 

After this preselection, each pair of electron candidates with opposite
charges was subjected to further cuts. The electron identification was
first tightened, allowing at most one electron candidate in the 
insensitive  regions of the calorimeters.
The two electrons were required to have energies above 20~\GeV\ and 
15~\GeV, respectively. 
Electron isolation angles with respect to the closest jet were 
required to be more than 20$^\circ$ for the more isolated electron 
and more than 8$^\circ$ for the other one. 

There were two different mass estimators  used in this analysis: 
a four-constraint
kinematic fit imposing energy and momentum conservation,
and a five-constraint kinematic fit 
taking into account the 
Breit-Wigner shape of the \Zz\ resonance~\cite{ref:pap98}.
The latter  was used to  test the compatibility of
the \ee\ invariant mass with the \Zz\ mass and provided
a better resolution in case of signal events.
Events with a 5C fit probability below 10$^{-8}$ were rejected. 
If the 5C fitted Higgs boson mass was greater than 60~\GeVcc,  the event was
accepted as a
candidate for a high mass signal. To reduce the background it was 
required that the sum of the 4C
fitted masses of the electron pair and of the hadronic system 
was above 150~\GeVcc.
If the fitted mass was less than 60~\GeVcc,  the requirement
on the sum of the masses was relaxed to 100~\GeVcc\ 
to improve efficiency for
low mass signals.
The difference between the hadronic and the di-electron mass was required 
to be below 100~\GeVcc.
The 5C fitted hadronic mass and the ${\mathrm b}$-tagging variable
\xb\ were used in the two-dimensional calculation of the confidence 
levels.

\begin{table}[htbp]
\begin{center}
\begin{tabular}{cccccc}     \hline
Selection & Data & Total  & \qqg  & 4 fermion & Efficiency (\%)\\
          &      & background        &       &           &    \\ 
\hline \hline
\multicolumn{6}{c} {\hee\  channel 163.3~\pbinv } \\ \hline
Preselection        &  936  &  942 $\pm$ 2 &   604 &  333 & 79.5 \\
Electron identification &   69  &   67.8 $\pm$ 0.4 &    17.9 &  49.3  & 67.0 \\
Candidate selection &   11  &   10.5 $\pm$ 0.1 &    0.7  &   9.7  & 59.0 \\ 
\hline
\multicolumn{6}{c} {\hmm\ channel 164.1~\pbinv } \\ \hline
Preselection        &  2678 & 2688 $\pm$ 6  & 1833 & 801 & 80.6 \\
Muon identification &   14  & 12.8 $\pm$ 0.2 &  0.2  & 12.6 & 71.5 \\ 
Candidate selection &    6  &  8.39 $\pm$ 0.14 &  0.04  & 8.35  & 67.0 \\ 
\hline
\multicolumn{6}{c} {Tau channel 163.7~\pbinv } \\ \hline
Preselection        &  6862 &  6534 $\pm$ 4&   3894&   2639&    96.1 \\
\llqq\              &    14 &  15.1 $\pm$ 0.12&  0.5 &  14.6&    18.4 \\
Candidate selection &    6  &   5.1  $\pm$ 0.07& 0.1 &  5.0&   16.3 \\
\hline
\multicolumn{6}{c} {\hnn\ channel 157.8~\pbinv (Low mass analysis)} \\ \hline
Anti \gaga          & 13038 & 12890 $\pm$ 10 & 9669 &2929 & 85.6 \\
Preselection        &   787 & 786  $\pm$ 4 & 463  & 290  & 70.7 \\
Candidate selection &   68  & 67.0 $\pm$ 0.8 &  31.5 & 35.5 & 55.3  \\ 
\hline
\multicolumn{6}{c} {\hnn\ channel 157.8~\pbinv (High mass analysis)} \\ \hline
Anti \gaga          & 13546 & 13361 $\pm$ 11 & 10023 & 2964 & 86.2 \\
Preselection        & 672   & 621 $\pm$ 3 & 328  & 280  & 66.3 \\
Candidate selection &  71   & 72.6$\pm$ 0.9 &  32.1 & 40.5 & 59.0  \\ 
\hline
\hline
\multicolumn{6}{c} {\hqq\ channel 163.7~\pbinv } \\ \hline
Preselection        & 1701  & 1686  $\pm$ 2&   473&  1213&    85.0 \\
Candidate selection &   31  &   35.5 $\pm$ 0.3&    12.1&    23.6&    56.5 \\
\hline
\hline
\end{tabular}
\caption[]{
 Effect of the selection cuts  for the {\sc SM} channels on data,
 simulated  background processes and simulated signal events during the first
operational period.

 Efficiencies are given for a signal with  \MH~=~115~\GeVcc\ at 206.5 \GeV.
 The quoted errors are statistical only. For each channel, the
 first line shows the integrated luminosity used; the line labelled `candidate
selection' shows the data selection used for calculating the confidence levels.
The total background can include small contributions from sources, such as
the two photon interaction process, not listed explicitly.

The excess in the tau channel is discussed in sectrion~\ref{sec:htau}.
}
\label{ta:hzsum2000e}
\end{center}
\end{table}

\begin{table}[htbp]
\begin{center}
\begin{tabular}{cccccc}     \hline
Selection & Data & Total  & \qqg  & 4 fermion & Efficiency (\%)\\
          &      & background        &       &           &    \\ 
\hline \hline
\multicolumn{6}{c} {\hee channel 59.1~\pbinv } \\ \hline
Preselection         &  348 &  352   $\pm$ 1.3 &     226 &    124 &  78.1 \\
Electron identification &   17 &   23.9 $\pm$ 0.2 &    6.4  &  17.5  &  62.4 \\
Candidate selection &    4 &    3.7 $\pm$ 0.1 &    0.3  &   3.4  &  55.0 \\ 
\hline
\multicolumn{6}{c} {\hmm channel 60.1~\pbinv } \\ \hline
Preselection         &  1142  & 1156 $\pm$ 6  &  788   &   317 & 81.7 \\
Muon identification  &    4   &  4.92  $\pm$ 0.08 &  0.11  & 4.81  & 72.0 \\ 
Candidate selection &    2   &  3.15  $\pm$ 0.06 &  0.02  & 3.12  & 67.1 \\ 
\hline
\multicolumn{6}{c} {Tau channel 60.1~\pbinv } \\ \hline
Preselection        &  2636& 2395 $\pm$ 4&  1398&  997&    95.9 \\
\llqq\              &    3 &  4.9 $\pm$ 0.2&   0.3&  4.6&    16.2 \\
Candidate selection &   1 &  2.08 $\pm$ 0.12&  0.1&  1.9&    15.1 \\
\hline
\multicolumn{6}{c} {\hnn\ channel 57.5~\pbinv (Low mass analysis)} \\ \hline
Anti \gaga       & 4475 & 4539 $\pm$ 6 & 3388 & 1060 & 85.1 \\
Preselection     &  303 & 288 $\pm$ 2.7 & 168  & 107  & 70.4 \\
Candidate selection &   22 & 25.0 $\pm$ 0.4  &11.3  &13.7  &53.6  \\ 
\hline
\multicolumn{6}{c} {\hnn\ channel 57.5~\pbinv (High mass analysis)} \\ \hline
Anti \gaga          & 4571 &4828 $\pm$ 7 & 3617  & 1080   & 86.5 \\
Preselection        &  234 & 236 $\pm$ 2.2 & 125 & 104  & 66.4 \\
Candidate selection &   28 &  28.0 $\pm$ 0.4 &  11.7 & 14.8   & 58.1  \\ 
\hline
\hline
\multicolumn{6}{c} {\hqq\ channel  60.1~\pbinv } \\ \hline
Preselection        &  577 &  619 $\pm$ 1.2&   174&   446&    85.2 \\
Candidate selection &    9 &   12.2 $\pm$ 0.2&     4.2&     8.0&    55.0 \\
\hline
\end{tabular}
\caption[]{
 As in Table~\ref{ta:hzsum2000e}, during the second operational period.}
\label{ta:hzsum2000u}
\end{center}
\end{table}

\begin{figure}[htbp]
\begin{center}
\epsfig{figure=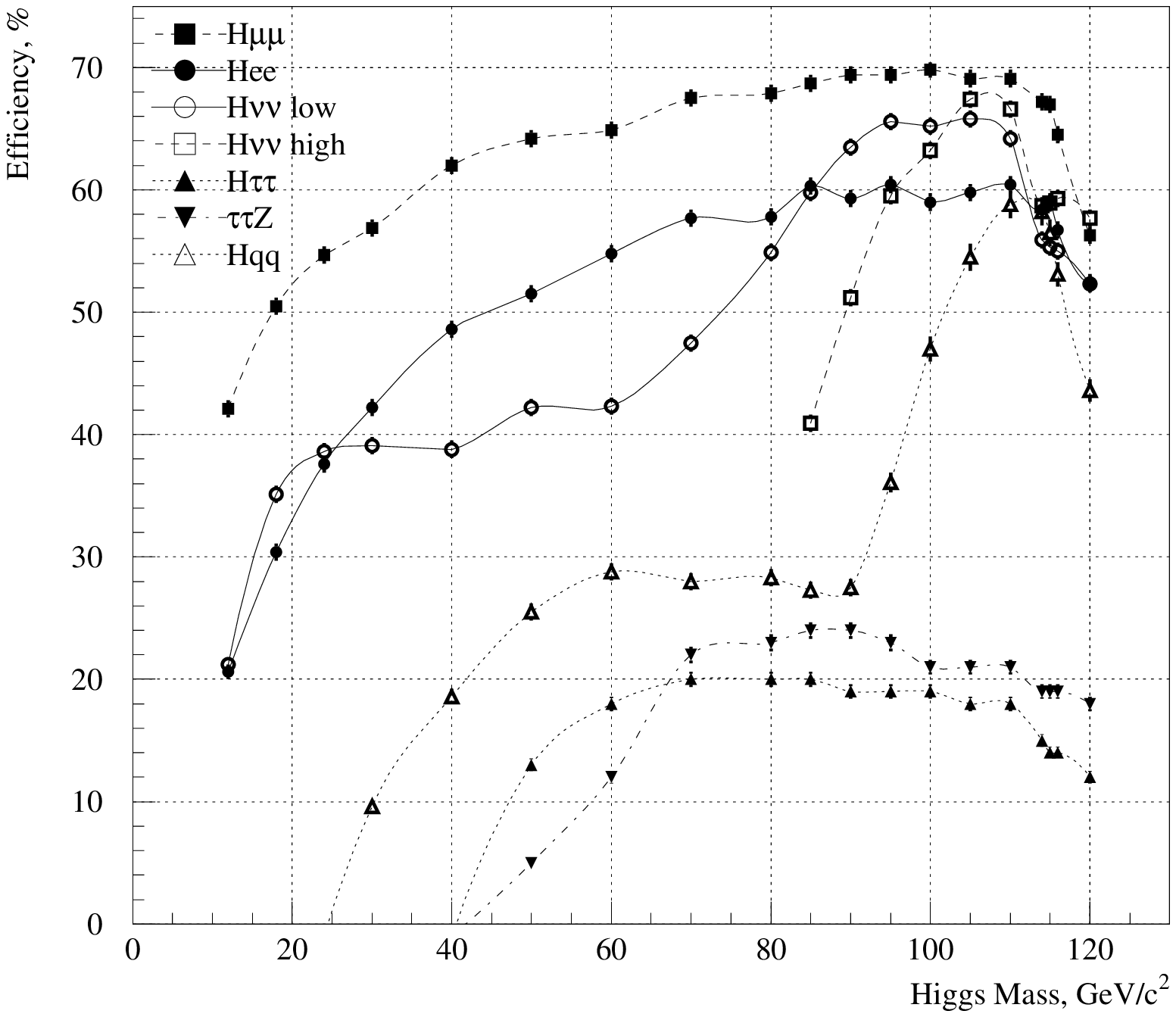,width=15.9cm}\\
\caption[]{
 Efficiencies
of the candidate level selection in the first data
taking period, as a function of the mass of  the Higgs boson and at \sqrts=206.5~\GeV.
The errors are statistical only,
and the curves drawn to guide the eye.
 Only efficiencies higher than 5\% are shown.
 The efficiency of the high-mass \hnn\ analysis is only shown above
a  mass value of 80~\GeVcc.
}
\label{fig:hzeff}
\end{center}
\end{figure}

   The effect of the selections on data and simulated samples  
are detailed in Tables~\ref{ta:hzsum2000e} and \ref{ta:hzsum2000u},
while the efficiencies at the end of the analysis in the first period are
shown as a function of the Higgs boson mass 
in Fig.~\ref{fig:hzeff} and for both periods in Table~\ref{ta:hzeff}. 
The efficiency in the later period is typically within 2\% absolute of that in
the  earlier.


\begin{figure}[htbp]
\hspace{-1.cm}
\epsfig{figure=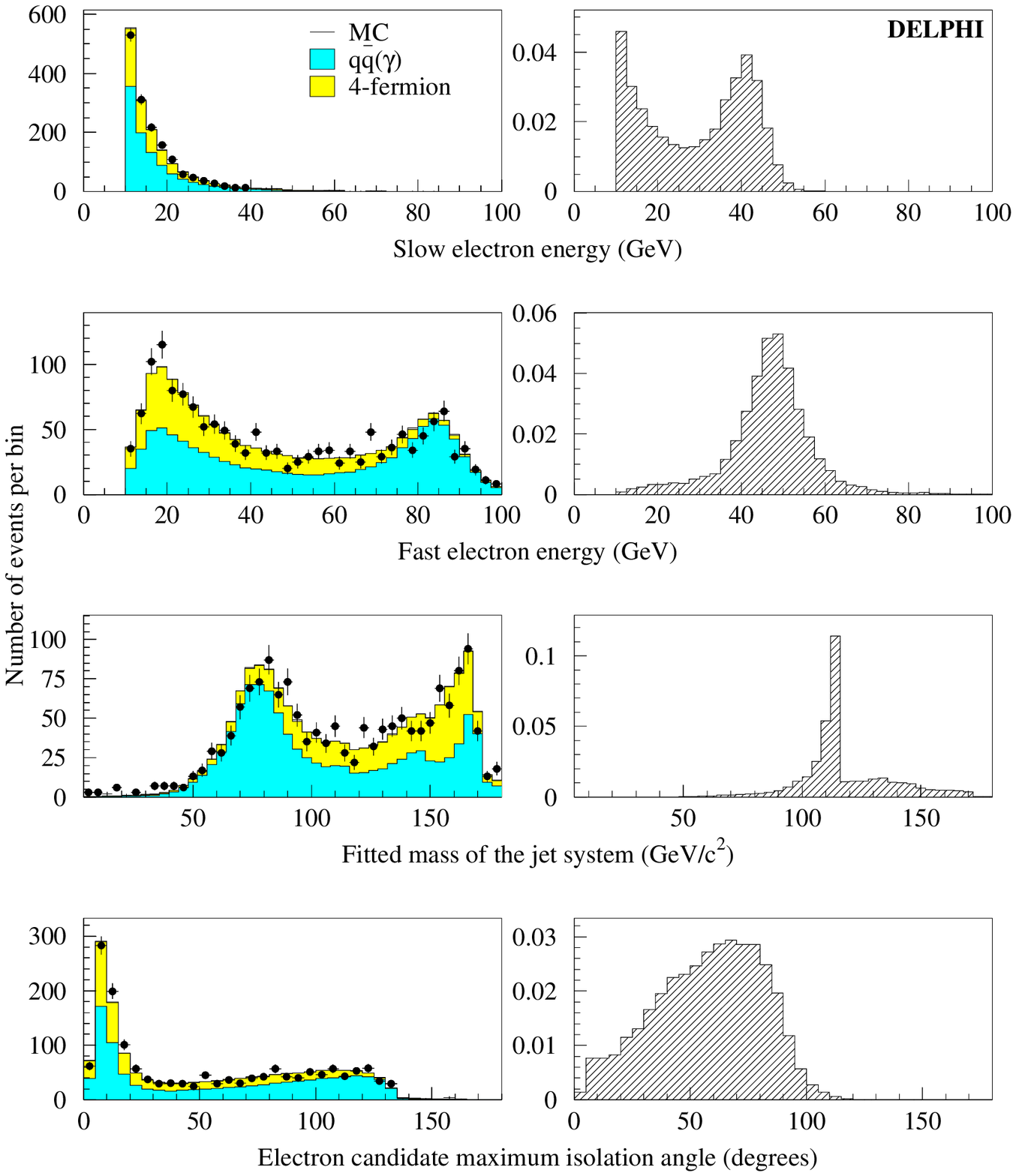,width=16cm}
\caption[]{\hee\ channel: 
  distributions of four analysis variables,
  as described in the text, at preselection level. Data from
the year 2000 (dots) are compared with  {\sc SM} background
   process 
  expectations (left-hand side histograms). The expected distributions
  for a 115~\GeVcc\ signal are shown in the right-hand side plots.}
\label{fig:hee}
\end{figure}

The agreement between data and simulation at the preselection level
is illustrated in Fig.~\ref{fig:hee} which shows
the distributions of the electron energies, the 5C fitted mass of the jet 
system and the isolation angle of the more isolated electron candidate.
At the end of the analysis, 15 events were selected in the data for 
a total expected background rate of 
\mbox{$14.2 \pm 0.12 (stat.)$} events coming mainly from the \eeqq\ 
process. 

 

\section{Higgs boson searches in events with jets and muons}

The analysis used a primarily cut based method to separate signal from 
background.
It followed the analysis published in~\cite{ref:pap97,ref:pap98,ref:pap99}, 
with
slight modifications to improve the sensitivity for low Higgs boson masses.
The event b-tag variable was used as the second variable in the
CL calculations.

The preselection in the first (second) operational period required
at least two (one) high quality track(s) of particle(s)
with a transverse momentum greater than 5~\GeVc.
High quality tracks have impact
parameters less than 100~$\mu$m in the transverse plane and 
less than 500~$\mu$m along the beam direction.
Furthermore, there had to be at least 9 
charged particles with two of them in the central part of the detector, 
$40^\circ < \theta <140^\circ $.
The final requirement of the preselection was that there be
at least two particles of opposite charges 
with momenta greater than 15~\GeVc.

The rest of the analysis was based upon the same muon identification
algorithm and discriminant variables as in~\cite{ref:pap97}, but 
the selection criteria were re-optimised.
At least two charged particles were required with 
opposite charges and an opening angle larger than 10\dgree.
The muon identification algorithm~\cite{ref:pap97}, which yields five
different levels of identification, was then applied to both particles of
such pairs.
The minimum level of muon identification required here corresponds to an 
efficiency of 88\% per pair of muon candidates, with 8.8\% of the pairs 
containing at least one pion.
A jet reconstruction  algorithm was then applied to the hadronic system 
recoiling from the muon pair, as explained in~\cite{ref:pap97}. 
In contrast with  previous analyses, no selection was  applied on
the number of jets in the recoiling system, nor on the number of
particles in these jets, in order to increase the sensitivity to low
Higgs boson masses. This leads to no significant increase of the
background.

The muons were  required to have momenta greater than 
28~\GeVc\ and 21~\GeVc, and their angles with respect to the closest jet axis
had to be greater than 12\dgree\ for the more isolated muon and greater than
9\dgree\ 
for the other one. 

A five-constraint kinematic (5C) fit taking into account 
energy and momentum conservation and the Breit-Wigner shape of
the Z resonance was  performed to test 
the compatibility of the di-muon mass with the Z mass 
in a window of $\pm$ 30~\GeVcc\ around the \Zz\ pole.
Events were kept only if 
the fit converged in this mass window.
A second similar four-constraint fit (4C) was performed afterwards
to take into account the possible loss of an ISR photon produced in the
beam direction. The results of the 4C procedure superseded that of 
the 5C one 
if the momentum of the fitted ISR photon was greater than 
10~\GeVc\ and if the 4C fit probability was greater than that of the 5C fit.
As in the \hee\ channel, 
the fitted mass of the hadronic system and the ${\mathrm b}$-tagging
variable \xb\
were chosen as the discriminant variables 
for the two-dimensional calculation of the confidence levels.

\begin{figure}[htbp]
\begin{center}
\epsfig{figure=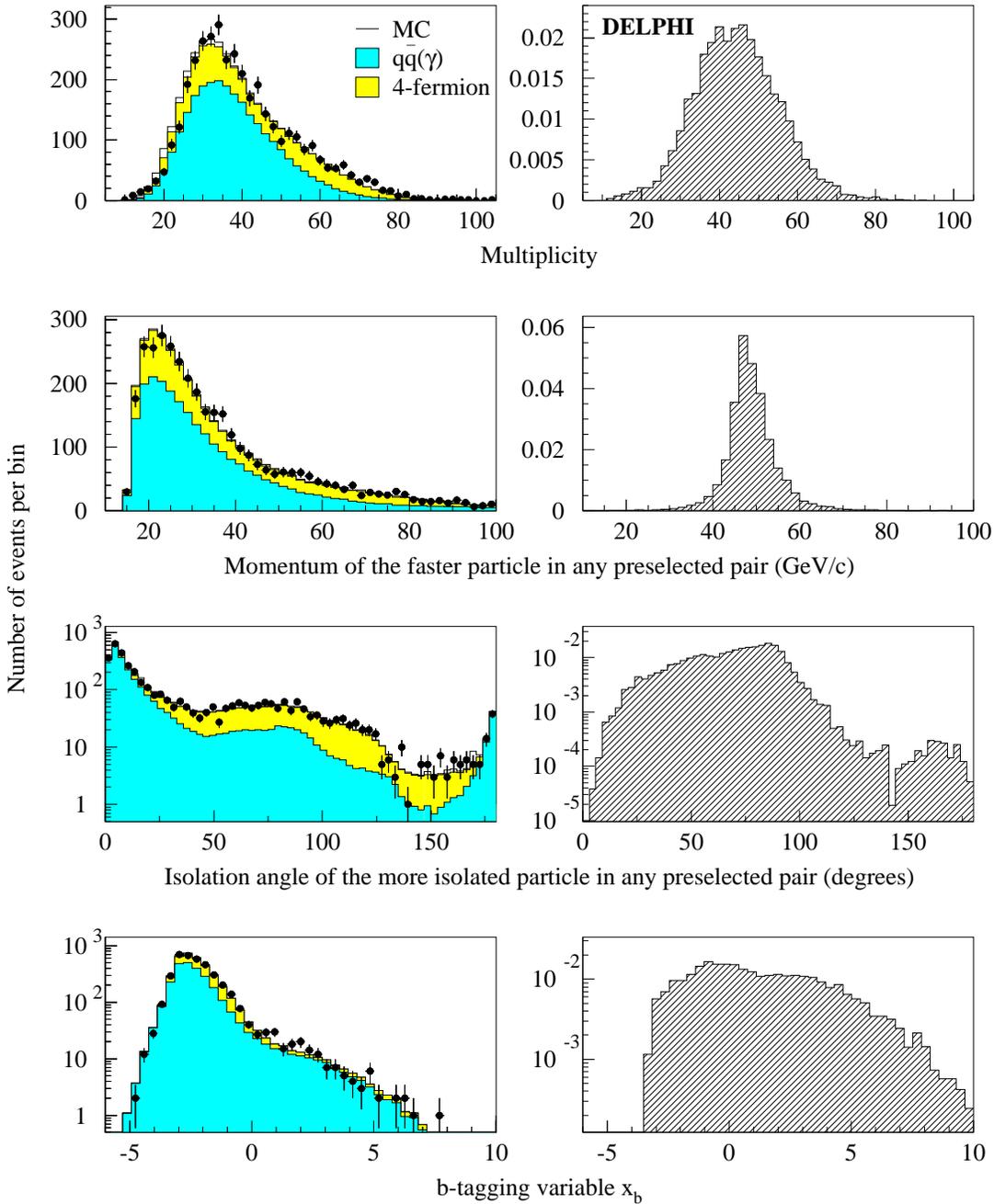,width=15.9cm} 
\caption[]{\hmm\ channel: 
  distributions of four analysis variables,
  as described in the text, at preselection level. Data from the year 2000
  (dots) are compared with  {\sc SM} background process
  expectations (left-hand side histograms). The expected distributions
  for a 115~\GeVcc\ Higgs signal are shown in the right-hand side plots.
}
\label{fig:hmumu}
\end{center}
\end{figure}

The effect of the selections on  data and simulated samples  
for the two periods of data taking are detailed in Tables~\ref{ta:hzsum2000e} 
and \ref{ta:hzsum2000u}. The signal  efficiencies for the first period
are shown as a function of Higgs boson mass
in  Fig.~\ref{fig:hzeff} and for both periods in 
Table~\ref{ta:hzeff}.
The rise of efficiency in the second
period is due to the relaxation of the track quality
cuts as described above.
 The agreement of simulation with data is quite good, as illustrated 
at preselection level in Fig.~\ref{fig:hmumu},
which shows the multiplicity of the charged particles, 
the momentum of the higher-momentum particle in any preselected pair,
the isolation angle of the more isolated particle in any preselected pair
and the ${\mathrm b}$-tagging variable \xb.

\section{Higgs boson searches in events with jets and taus} 
\label{sec:htau}

The analysis used a cut based method to identify tau pairs and 
jet pairs, and then a likelihood variable based on kinematics
and ${\mathrm b}$-tagging  as the second variable in the
CL calculations. 
Four channels are covered by these searches, two for the \ZH\ channel,
depending on which boson decays into \toto, and similarly two for the
\hA\ channel. One data set is selected, containing events from all
decay channels.
The analysis, almost identical to that described in~\cite{ref:pap98}, 
selected hadronic events by requiring at least ten charged particles, 
a total reconstructed energy greater than $0.4\sqrts$, a reconstructed 
charged energy above $0.2\sqrts$ and 
\mbox{$ \sqrt{s'}$} greater than 120~\GeV.

\begin{figure}[htbp]
\begin{center}
\begin{picture}(453,450)
\put(3,-20){\mbox{
\epsfig{figure=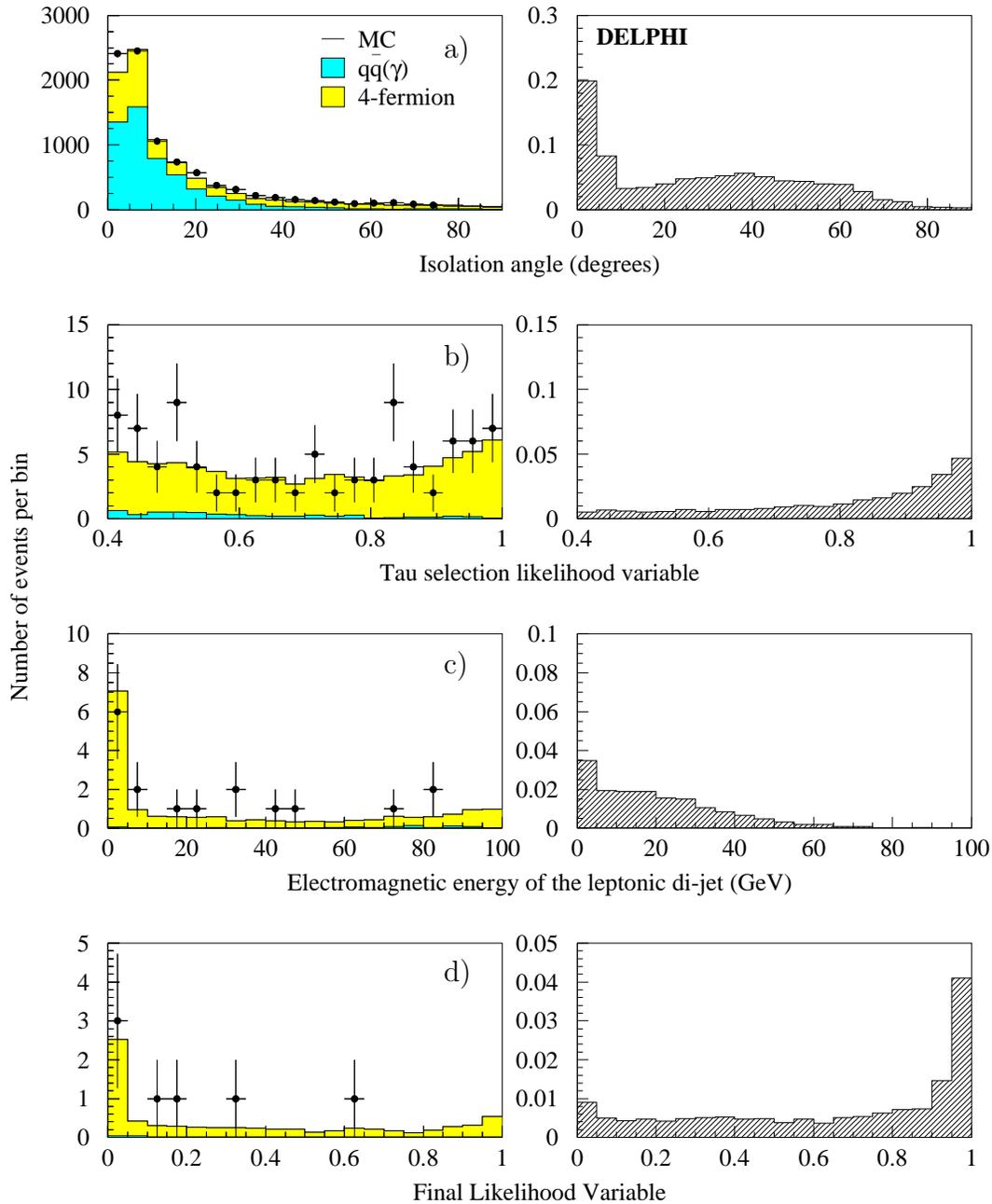,width=15.6cm}
}}
\put(190,465){\mbox{a)}}
\put(190,340){\mbox{b)}}
\put(190,215){\mbox{c)}}
\put(190,090){\mbox{d)}}
\end{picture}
\caption{\tautauqq\ channel: 
  distributions of four analysis variables at different levels of the
  selection, as described in the text. Data from
  the year 2000 (dots) are compared with  {\sc SM} background 
   process
  expectations (left-hand side histograms). The expected distributions
  for a 115~\GeVcc\ Higgs signal in the 
 (H$\rightarrow$\toto)(Z$\rightarrow$\qqbar) channel
  are shown in the right-hand side plots.
}
\label{fig:ttqq}
\end{center}
\end{figure}

A search for $\tau$ lepton candidates was then performed using a 
likelihood ratio technique. 
Single charged particles were preselected 
if they were isolated from all other charged particles by more than 
10\mydeg, 
if their momentum was above 2~\GeVc\ and if all neutral particles in a 
10\mydeg\ 
cone around their direction made an invariant mass below 2~\GeVcc.
The likelihood variable was calculated for the preselected particles using
distributions of the particle momentum,  isolation angle and 
the probability that it came from the primary vertex. 
Fig.~\ref{fig:ttqq}a shows the distribution of
the isolation angle of the preselected charged particle with the highest
$\tau$ likelihood variable in the event
for data and simulation.
There is an excess of data seen at very low isolation angles. The
simulation is known to underestimate the contributions from Bhabha 
events and the two photon interaction process in this region,
which is therefore cut away.

Pairs of $\tau$ candidates were then selected requiring opposite charges, 
an opening angle greater than 90\mydeg\ and a product of the
$\tau$ likelihood variables above 0.45. If more than one pair was
selected, only the pair with the highest product was kept. 
The distribution of the highest product of two $\tau$ likelihood 
variables 
in the event is given in Fig.~\ref{fig:ttqq}b. The discrimination
between the Higgs signal and the {\sc SM} background processes
is clearly visible.
The percentage of $\tau$ pairs correctly identified was over
90\% in simulated Higgs events.

Two slim jets were reconstructed with all 
neutral particles  
inside a 10\mydeg\ cone around the directions of the $\tau$ candidates.
The rest of the event was forced into two jets.
The slim jets were required to be in the
\mbox{20\mydeg$\le~\theta_{\tau}~\le$ 160\mydeg} polar angle region to 
reduce the \zee\ background, while the hadronic jet pair
invariant mass was required to be between 20 and 110~\GeVcc\ in order
to reduce the \qqg\ and \mbox{${\mathrm Z}\gamma^*$} backgrounds.
The jet energies and masses were then rescaled, imposing energy and 
momentum conservation, to give a better estimate of the masses 
of both jet pairs (\toto\ and \qqbar). The 
 rescaled masses were required to be above 20~\GeVcc, and
below \sqrts\ to discard
unphysical solutions of the rescaling procedure. 
Both  hadronic jets had to have rescaling factors in the range 0.4 to 1.5.

\begin{figure}[htbp]
\begin{center}
\epsfig{figure=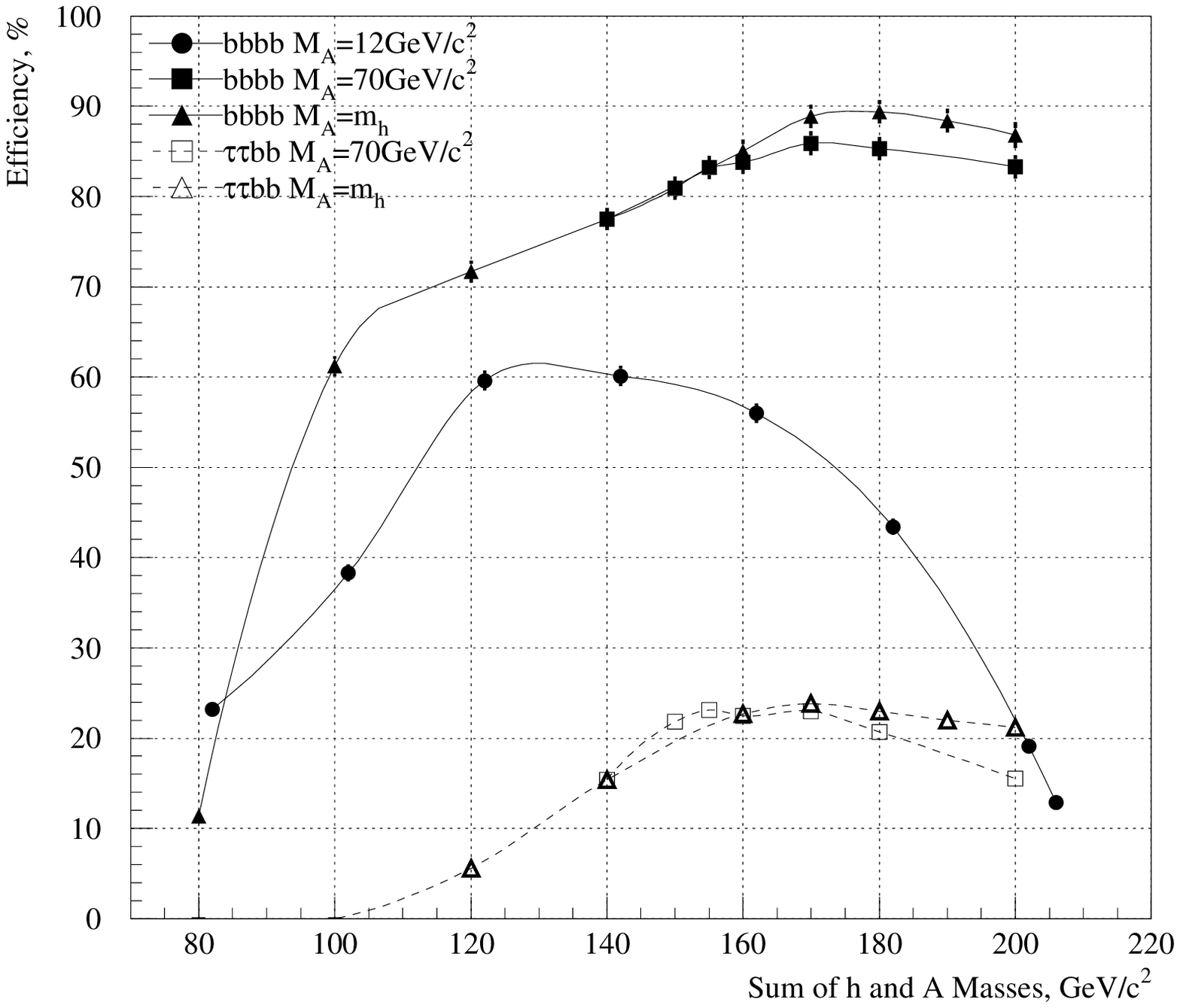,width=14cm} \\
\caption{hA channels: 
efficiencies at the candidate selection level at \sqrts=206.5~\GeV\ in
the first data taking period.
 The open symbols are for the \hAtt\ analysis and the solid ones
for the \hAbb. 
  The different shapes  correspond to three different mass difference criteria
for the h and A:
triangles are for equal masses, squares are for the A mass fixed
to 70~\GeVcc\ and
circles for $m_A = 12$~\GeVcc. The errors shown are statistical only.
Only efficiencies above 5\% are shown;
the \hAtt\ channels have no points for \MA~=~12~\GeVcc.
The curves are drawn to guide the eye.
}
\label{fig:4b_eff}
\end{center}
\end{figure}

The remaining background processes were mostly genuine \llqq\ events. 
In order  to reject leptonic Z decays producing \eeqq\ and \mmqq,
 the measured mass of 
the leptonic system was required to be between 10 and 80~\GeVcc\ and its 
electromagnetic energy to be below 60~\GeV\ (see Fig.~\ref{fig:ttqq}c). 
This concluded the selection procedure. 
The effect of the selections on data and simulated samples  
is detailed in Tables~\ref{ta:hzsum2000e} and \ref{ta:hzsum2000u}.
Efficiencies for the SM process in the first period can be found
as a function of Higgs boson mass in
Fig.~\ref{fig:hzeff} and for both periods in Table~\ref{ta:hzeff}. 
The efficiencies for the MSSM channels for some selected points are given
in  Table~\ref{ta:haeff} and in Fig.~\ref{fig:4b_eff}.

At the end of the analysis, 7 events were selected in data for a 
total expected background  of 
\mbox{$7.2 \pm 0.1 (stat.)$} events, coming mainly from the 
\tautauqq\ (predominantly ZZ) and \tauvqq\ (predominantly WW) processes. 


The two-dimensional calculation of the confidence levels uses the
reconstructed mass given by the sum of the \toto\ and 
\qqbar\ jet pair masses after rescaling and 
a likelihood variable built from the distributions of the 
rescaling factors of the $\tau$ jets, the $\tau$ momenta and the 
${\mathrm b}$-tagging variable, \xb. 
The distribution of this likelihood variable at the end of the analysis
is shown in Fig.~\ref{fig:ttqq}d.
Since all the  \tautauqq\ signals are covered by the same 
analysis, the corresponding channels cannot be considered as independent in the
confidence level computation. Therefore 
they were combined into one global \tautauqq\ channel:
at each test point, the signal expectations (rate, two-dimensional 
distribution) in this channel were obtained by summing the contributions 
from all the  original signals weighted by their expected production rates.


\section{Higgs boson searches in events with missing energy and jets } 
\label{sec:hnn}

  The signal topology in this channel is characterised by two acollinear and
acoplanar jets 
and a measured energy significantly less than 
 the centre-of-mass energy, due to neutrinos coming either from the
decay of a Z boson or from the \WW\ fusion process. In addition to the
irreducible \nnqq\ four-fermion events, several other background
processes can
lead to similar topologies; for example events due to particles from one
beam only,  or the
\qqgg\ process with initial state radiation photons emitted along the beam
axis. The signal  topology and hence the dominant background processes
are somewhat different when the Higgs boson mass 
is very close to the kinematic threshold for \Hz\Zz\ production 
compared with lower masses. DELPHI chose to use two separate analyses,
one optimised for masses close to the HZ kinematic threshold and the
other covering masses down to the \bbbar\ threshold. These two analyses are
hereafter referred to as the high mass and low mass analyses.
The results are never used simultaneously:
instead the sensitivity of the two searches is compared for any given signal
and the more powerful analysis is selected.
This comparison is done independently for each data set, 
with the result that for \MH\ below 99~\GeVcc\ only the low mass analysis
is used, while above 116~\GeVcc\ all results are taken from  the high mass
analysis. In the region 
where the limit is set, two of the twelve data sets use the low mass
analysis.

Both analyses followed the same procedure. First, a set of preselection 
criteria was applied to reject most of the single-beam, Bhabha and 
$\gamma \gamma$ events, and to reduce the contamination of \qqg\
and \WW\ events. For the final step of the analysis,
the separation between the signal and the  background channels
was achieved through a multidimensional variable built with the likelihood 
ratio method. After a loose cut on the likelihood variable,
the two-dimensional calculation of the confidence levels used
the final multi-variable likelihood 
and the reconstructed Higgs boson mass, 
defined as the visible mass given by a one-constraint fit, where the 
recoil system is an on-shell Z boson.

A third analysis provided a cross-check of the high mass analysis.
 This analysis uses preselection criteria similar to the two others,
but the multidimensional variable was derived using a two step discriminant
analysis.

The three analyses are presented in the next sections, but they all
use the following algorithms.
 Events were forced into two jets 
 (the so called ``two-jet configuration'') but for each
event jets were also reconstructed 
using a distance of $y_{cut}=0.005$ (the so-called ``free-jet configuration'') 
and general variables of each jet 
(such as  multiplicities and momenta) were calculated in both configurations.
In order to tag isolated particles from semi-leptonic decays 
of \WW\ pairs, an energy fraction was defined which was  the total
energy emitted at angles between 5\mydeg\ and 25\mydeg\ of the 
direction of the particle under study
divided by the energy of that particle.
This was calculated for the most isolated and the most energetic particles,
and the smaller of these two normalised energies defined the 
anti-\WW\ isolation variable, 
which was used in the three analyses in the
determination of the final discriminant variable.

\subsection{Low mass Higgs bosons with missing energy}

The low mass analysis is essentially the same as 
that described in~\cite{ref:pap99}.
The preselection criteria remain unchanged. 

The discriminant variable is a likelihood, constructed by multiplying
the likelihoods (and neglecting correlations) from eleven discriminant
variables : 
the angle between the missing momentum and the closest jet in the free-jet 
configuration, 
the polar angle of the more forward jet in the two-jet configuration, 
the polar angle of the missing momentum, 
the acoplanarity
in the two-jet
configuration,\footnote{The acoplanarity is defined
as the supplement of the angle between the transverse momenta 
(with respect to the beam axis) of the two jets.}
the ratio between \mbox{$\sqrt{s'}$} and the centre-of-mass energy, 
the missing mass of the event, the anti-\WW\ isolation variable, 
the largest transverse momentum with respect to its 
jet axis of any charged particle in the two-jet configuration, 
the minimum jet charged multiplicity in the free-jet configuration,
the  ${\mathrm b}$-tagging variable \xb\ and
the event lifetime probability \btagpep. 
The first five variables discriminate the signal from the \qqg\
channel and the other variables  discriminate against
\WW\ pairs. Compared to the analysis described in~\cite{ref:pap99},
 one variable (the DURHAM distance for the transition between 
the two-jet and three-jet configurations) was removed because
it was found to reduce 
sensitivity to a low mass signal.

The likelihood functions for the eleven variables were calculated for
the two operational periods separately. In each case,
PDFs were obtained from simulated events, using half of
the available statistics in all backgrounds and signals
of masses 95, 100, 105 and 110~\GeVcc. 

\begin{figure}[htbp]
\begin{center}
\epsfig{figure=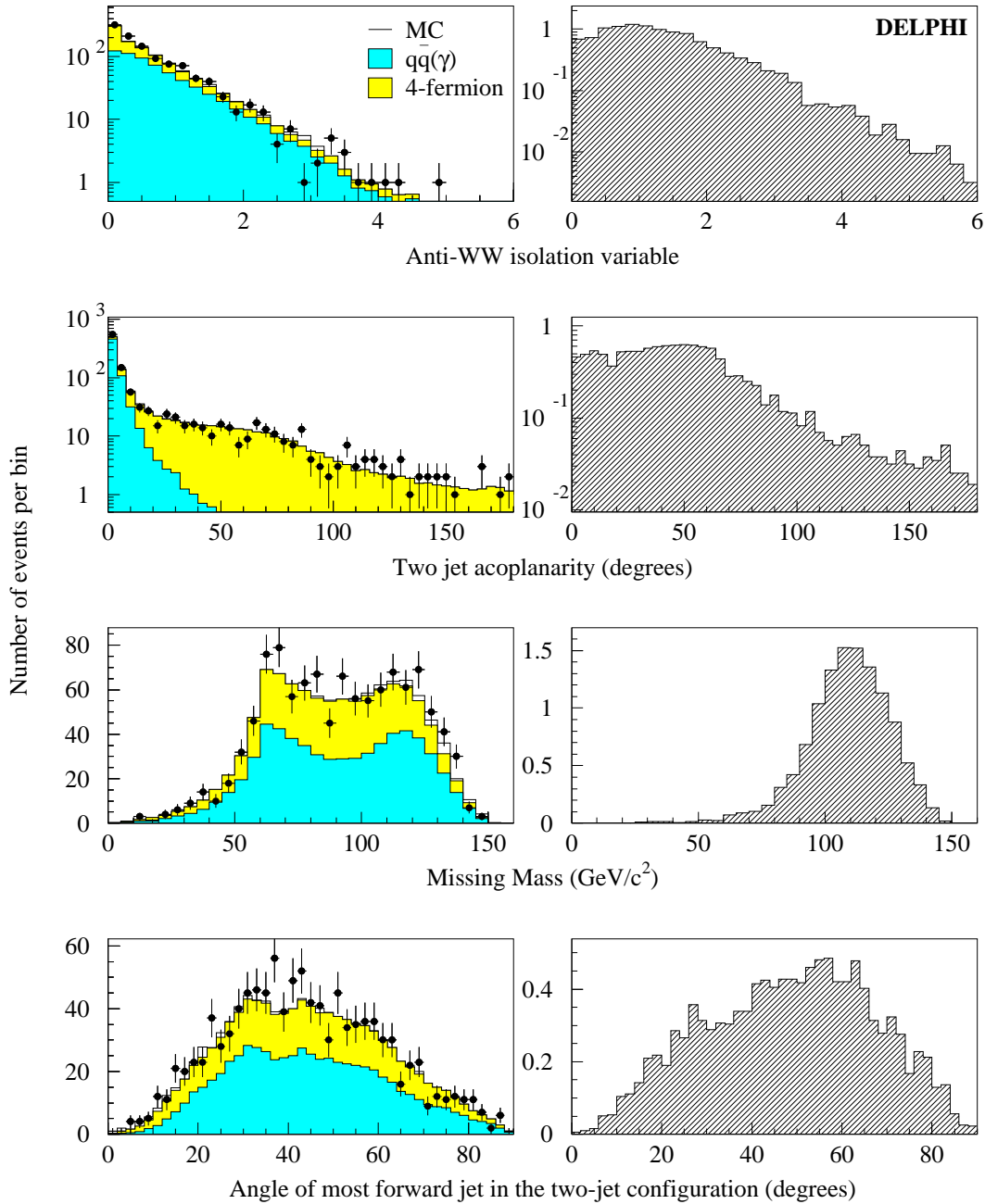,width=15.8cm}
\caption{ \hnn\ channel low mass analysis: 
  distributions of four analysis variables, as described in the text, 
at the preselection level. 
Data from the year 2000(dots) are compared with  {\sc SM} background process
  expectations (left-hand side histograms). The expected distributions
  for a 90~\GeVcc\ signal are shown in the right-hand side plots.
} 
\label{fig:hnunu_lm}
\end{center}
\end{figure}

\begin{figure}[htbp]
\begin{center}
\epsfig{figure=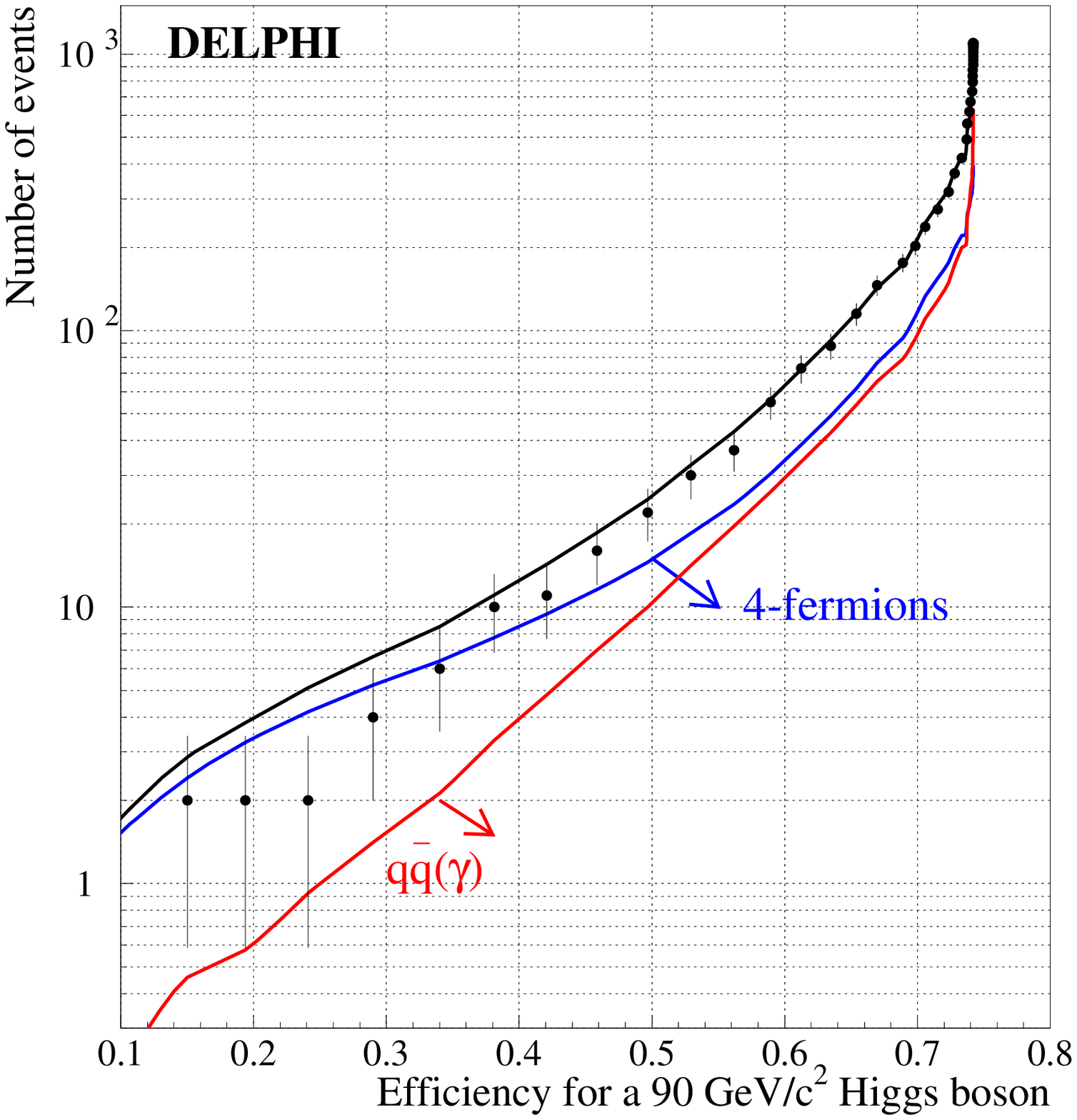,width=11cm} \\
\vspace{-0.5cm}
\caption{ \hnn\ channel, low mass analysis:  
  the observed rate  and that expected from the  {\sc SM} background processes
   as a 
  function of the efficiency for a 90~\GeVcc\ Higgs signal when
  varying the cut on the likelihood variable. 
  The total background 
and different background 
contributions are shown separately. 
  The dots represent the data. 
}
\label{fig:hnunu_disc_all}
\end{center}
\end{figure}

The distributions of four of the input variables are shown at preselection
level in Fig.~\ref{fig:hnunu_lm}. The comparison
between the observed rate and that  expected from the SM background processes
 in the signal-like tail of
the distribution of the likelihood discriminant variable is illustrated
in Fig.~\ref{fig:hnunu_disc_all}, 
which shows these rates 
as a function of the efficiency for a Higgs signal of 90~\GeVcc\ 
when varying the cut on the likelihood variable. 
To select the candidates, a loose cut of 1,0 is applied, 
leaving 90 events in data for a total expected background process rate of 
\mbox{$92.0 \pm 0.9 (stat.)$}.
The effect of the selections on  data and simulated samples  
for the two periods of operation are detailed in Tables~\ref{ta:hzsum2000e} 
and \ref{ta:hzsum2000u}. The signal  efficiencies for the first period
are shown as a function of Higgs boson mass in  Fig.~\ref{fig:hzeff} and for
both periods in 
Table~\ref{ta:hzeff}.
Even for very low masses, this analysis has non-negligible efficiencies.

\subsection{High mass Higgs bosons with missing energy}

The high mass analysis technique is essentially that outlined
in~\cite{ref:D2000}.
Both the preselection criteria and the final discriminating likelihood 
variable were optimised to achieve the maximum reduction of the
background to Higgs bosons with masses around 115~\GeVcc.

The general selection criteria to reject Bhabha, \gaga\ and beam-related 
background events are  described in~\cite{ref:pap99}.
Cuts were applied to reduce the \qqg\ channel with particular attention
to all cases where fake missing energy could be created.
In order to reject events coming from a radiative return to the Z with
photons emitted in the beam pipe, a two-dimensional selection criterion 
was set in the plane of 
 $\sqrt{s'}$  versus the  polar angle $\theta$ of the missing momentum.
This selection required 
that  $\sqrt{s'}$ (in \GeV)
 be greater than -0.6$\times\theta$+115 ($\theta$ in degrees)
 for $\theta<$40$^\circ$ (+0.6$\times\theta$+7 for $\theta >140^\circ$).
There is no selection on $\sqrt{s'}$ for $40^\circ < \theta <140^\circ$.
To reduce the contamination of radiative return events with photons in the
detector acceptance, events were rejected if the total electromagnetic energy 
within 30\mydeg\ of the beam axis was greater then $0.30\sqrt{s}$. 
Furthermore, events were rejected if the energy deposited
in the calorimeters  exceeded  $0.08\sqrt{s}$, $0.25\sqrt{s}$,
$0.35\sqrt{s}$, $0.4\sqrt{s}$ in the small angle luminosity monitor, the 
forward electromagnetic calorimeter, the barrel electromagnetic 
calorimeter and the hadronic calorimeter, respectively.
To reject events with photons crossing 
the small insensitive regions of the electromagnetic calorimeters, a
veto based on the hermeticity counters of DELPHI similar to that of
the low mass analysis~\cite{ref:pap99} was also applied.
To remove  background events with no missing energy it was required that
the effective centre-of-mass energy  $\sqrt{s'}$ was below $0.96\sqrt{s}$. 
Two-fermion events with jets pointing to the insensitive regions of the 
electromagnetic calorimeters or emitted close to the beam axis
are also a potential background due to mis-measurements of the jet 
properties. Events were thus rejected if 
the jet polar angles in the two-jet configuration were within $\pm$5\mydeg\
of 40\mydeg\ for one jet and of 140\mydeg\ for the other jet,
unless the acoplanarity was greater than 10\mydeg.
In addition, the acoplanarity in the two-jet configuration 
had to be larger than 6\mydeg\ when the transverse momentum of the event was 
below 6~\GeVc. The angle between either beam and the missing momentum of the
event had to be greater than 10\mydeg.
Both  jets in  the 
two-jet configuration had to be more than  12\mydeg\  away from
both  the beams,
 or 20\mydeg\  if the acoplanarity was less than 10\mydeg.

To reduce the semi-leptonic \WW\ background, which could fake the
high mass signal topology when the leptons (especially $\tau$ particles)
are hidden in the jets and thus increase the visible mass, the
following selection criteria were applied.
The energy of  the most energetic particle 
in the event was required to be less than 0.20$\sqrt{s}$.
At least one charged 
particle per jet was required for the events reconstructed in the 
free-jet configuration. 
Furthermore, when forcing the event into the two, three and free-jet 
configurations, there were upper limits on the
transverse momentum of a charged particle  with respect to its jet axis
of 10, 5 and 8~\GeVc\ respectively. These criteria were tightened to 
5, 3 and 4~\GeVc\ respectively when the charged particle was identified
as an electron or muon using the standard
criteria found in Ref.~\cite{ref:perfo}.

The final selection of signal-like events required that the total visible
energy 
was less than $0.75\sqrt{s}$. All the above criteria define the 
preselection. 
 
The multi-variable likelihood was constructed in the same way as
in the low mass analysis and combined the following ten variables:
the acoplanarity and 
the acollinearity in the two-jet configuration, 
the polar angle of the missing momentum, 
the  ${\mathrm b}$-tagging variable \xb,
the invariant mass in the transverse plane with respect to the beam axis, 
the anti-\WW\ isolation variable, 
the angle between the missing momentum and the closest jet in the free-jet 
configuration, 
the lowest charged multiplicity of any jet in the free-jet configuration,
the largest transverse momentum with respect to its 
jet axis of any charged particle in the free-jet configuration, 
and the visible mass. 

\begin{figure}[htbp]
\begin{center}
\epsfig{figure=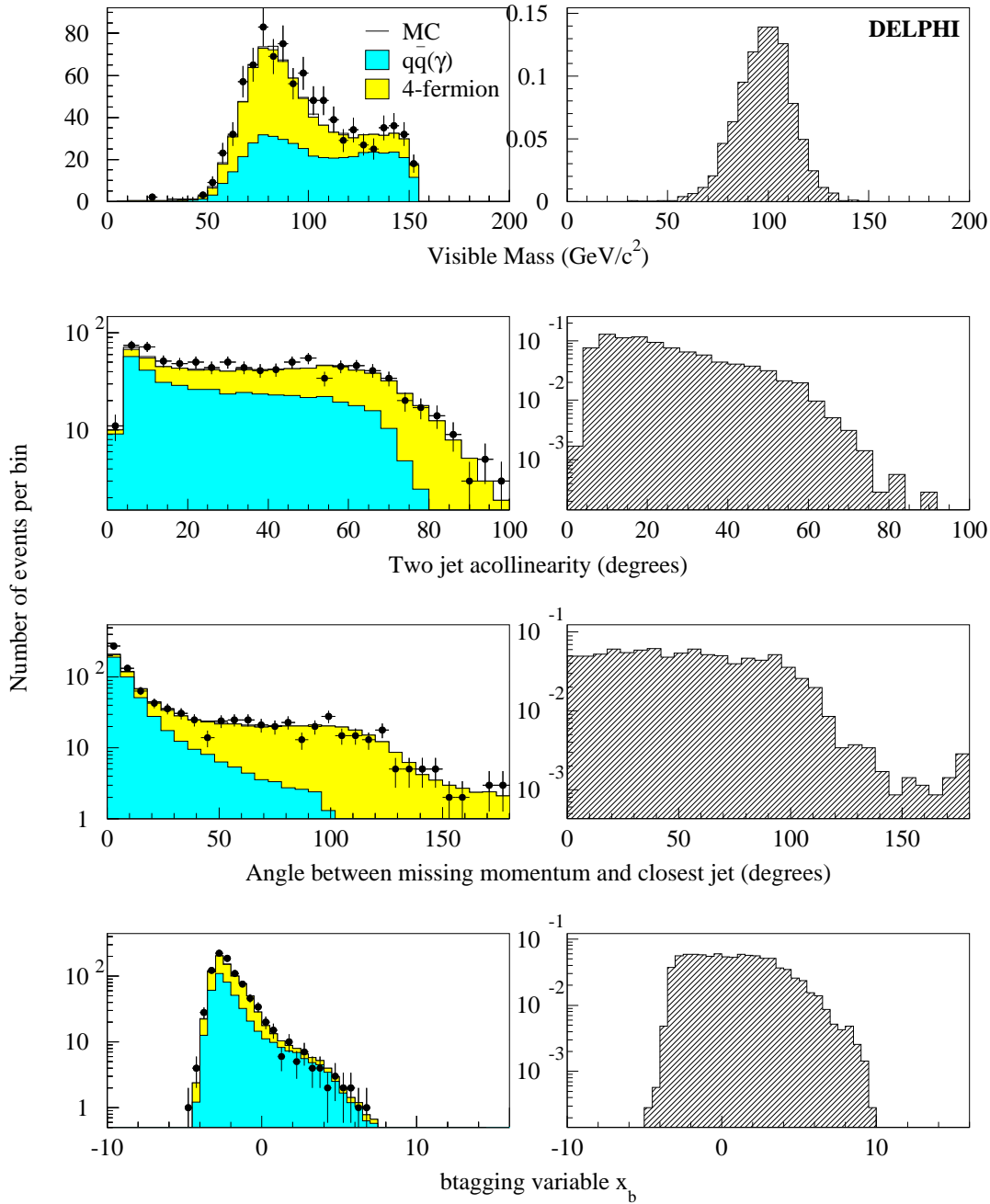,width=15.8cm}
\caption{ \hnn\ channel, high mass analysis: 
  distributions of four analysis variables, as described in the text, 
at preselection level. 
Data from the year 2000 (dots) are compared with  {\sc SM} background process 
  expectations (left-hand side histograms). The expected distributions
  for a 115~\GeVcc\ Higgs signal are shown in the right-hand side plots.
}
\label{fig:hnunu_hm}
\end{center}
\end{figure}

\begin{figure}[htbp]
\begin{center}
\epsfig{figure=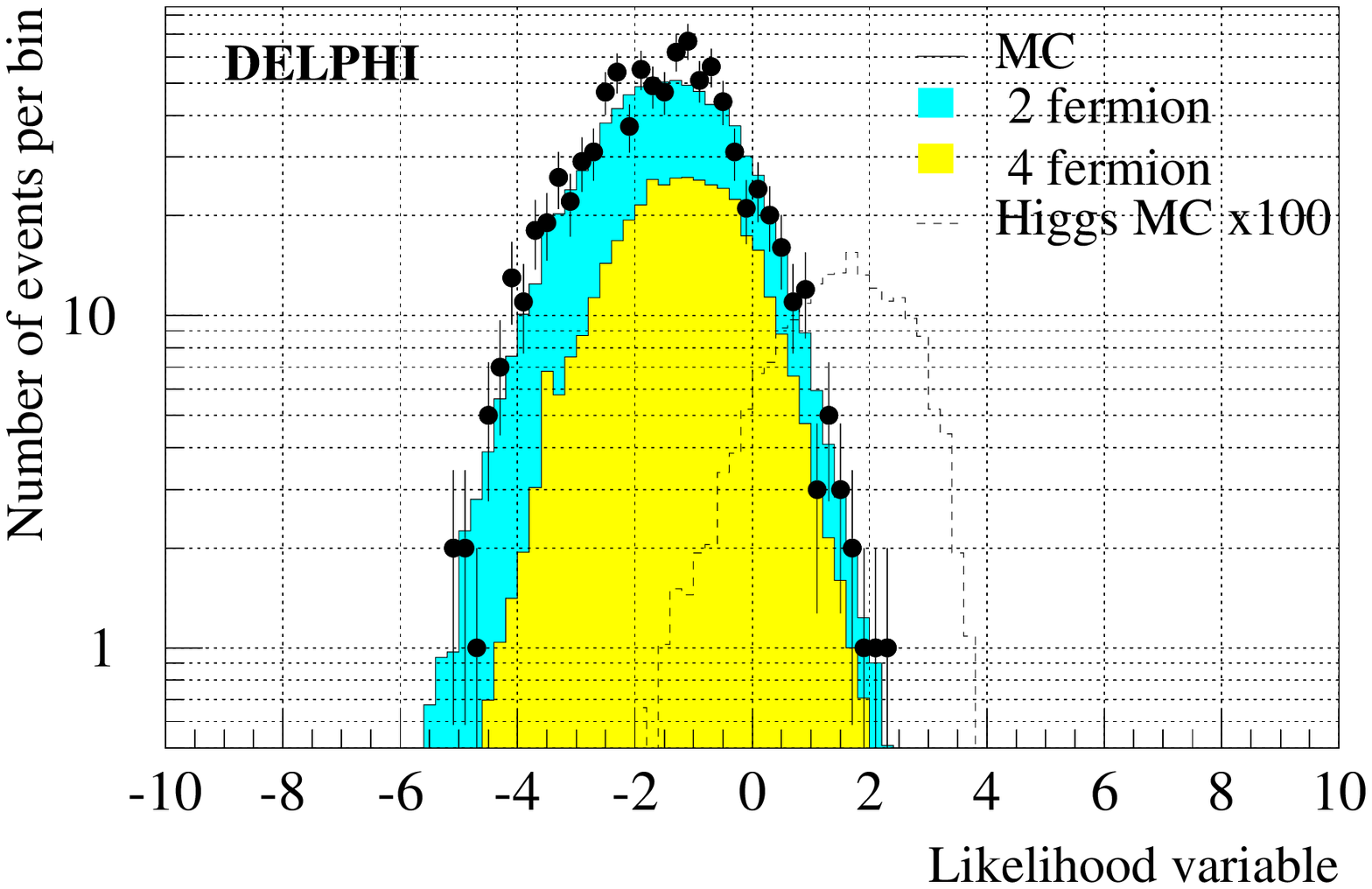, width=10.5cm} \\
\vspace*{-1.5cm}
\epsfig{figure=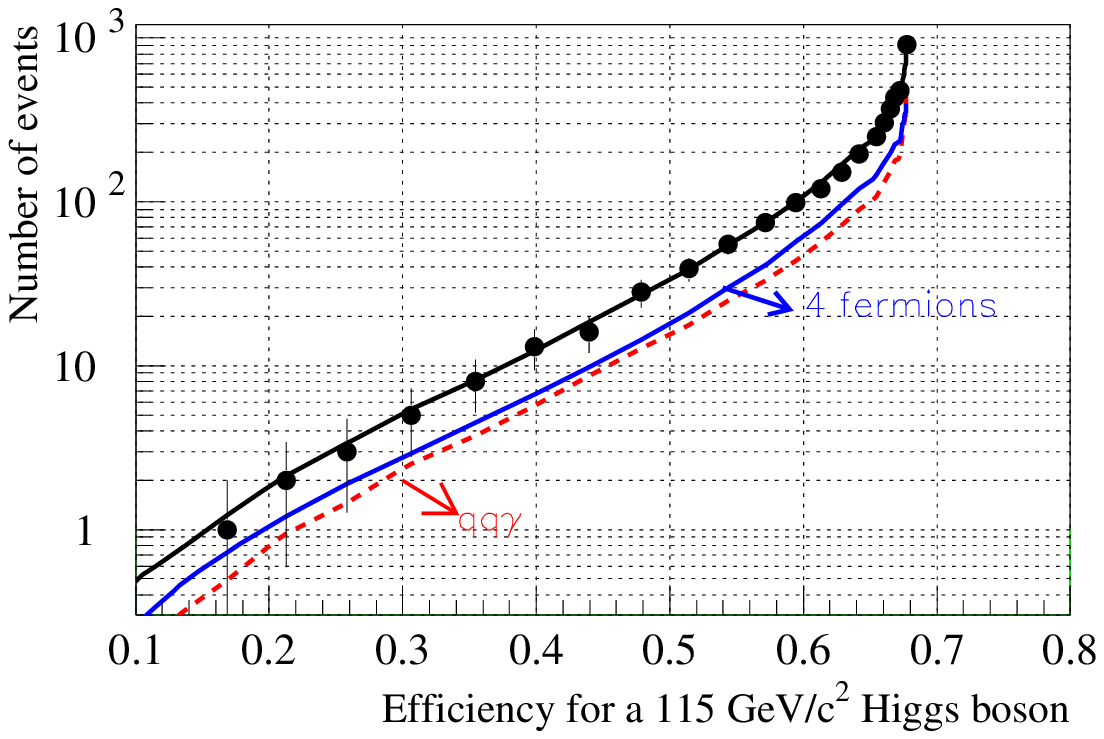,width=10.5cm}
\caption{ \hnn\ channel, high mass analysis: 
  Top: distributions of the likelihood variable 
  for the expected  {\sc SM} background processes (full histograms), 
  all data from the year 2000 (dots) and the 
  expected Higgs signal at 115~\GeVcc\ (dashed histogram, 
  normalised to 100 times the expected rate).
  Bottom: 
  The observed rate and that expected from the {\sc SM} background processes
  at \rs~=~206.5~\GeV\ as a 
  function of the efficiency for a 115~\GeVcc\ Higgs signal when
  varying the cut on the likelihood variable. 
  The different background contributions are shown separately. 
  The dots represent the data. 
}
\label{fig:hnunu_efb_all_hm}
\end{center}
\end{figure}

The distributions of four of the input variables are shown at preselection
level in Fig.~\ref{fig:hnunu_hm}. 
The top plot of Fig.~\ref{fig:hnunu_efb_all_hm} shows the 
distribution of the likelihood discriminant variable. 
The comparison between the observed rates and that expected from
SM background processes in the 
signal-like tail of this distribution is illustrated further 
on the bottom plot of  Fig.~\ref{fig:hnunu_efb_all_hm}, 
which shows these rates
as a function of the efficiency to detect a Higgs boson of mass 115~\GeVcc\ 
when varying the cut on the likelihood variable.  
To select the candidates, a minimal value of 0 is required,  
leaving 99 events in data for a total expected background rate of  
\mbox{$100.6 \pm 0.9 (stat.)$}.
The effect of the selections on  data and simulated samples  
for the two periods of operation are detailed in Tables~\ref{ta:hzsum2000e} 
and \ref{ta:hzsum2000u}. The signal  efficiencies for both periods
are shown as a function of the Higgs boson mass 
in  Table~\ref{ta:hzeff} and for the first period in
 Fig.~\ref{fig:hzeff},
where the efficiency in the low mass analysis can also be seen.
 The high mass analysis takes over from the low mass analysis at around 
105~\GeVcc\ and when the expected performances are calculated it
brings a gain  equivalent to at least 50\% more
luminosity for signal masses above 110~\GeVcc.

\subsection{Missing energy using Iterative Discriminant Analysis (IDA) }
\label{sec:AnaIDA}
A second analysis optimised for high  masses was made as a
cross-check.
This analysis used the iterative discriminant analysis (IDA)~\cite{ref:IDA}
method, which is a modification of Fisher's discriminant
analysis~\cite{ref:DA}. The IDA method introduces two elements, a non-linear
discriminant function (whereas the Fisher function is linear) and an
iterative procedure, to enhance the separation of signal events from
background.

\begin{figure}[htbp] 
\begin{center}
\epsfig{figure=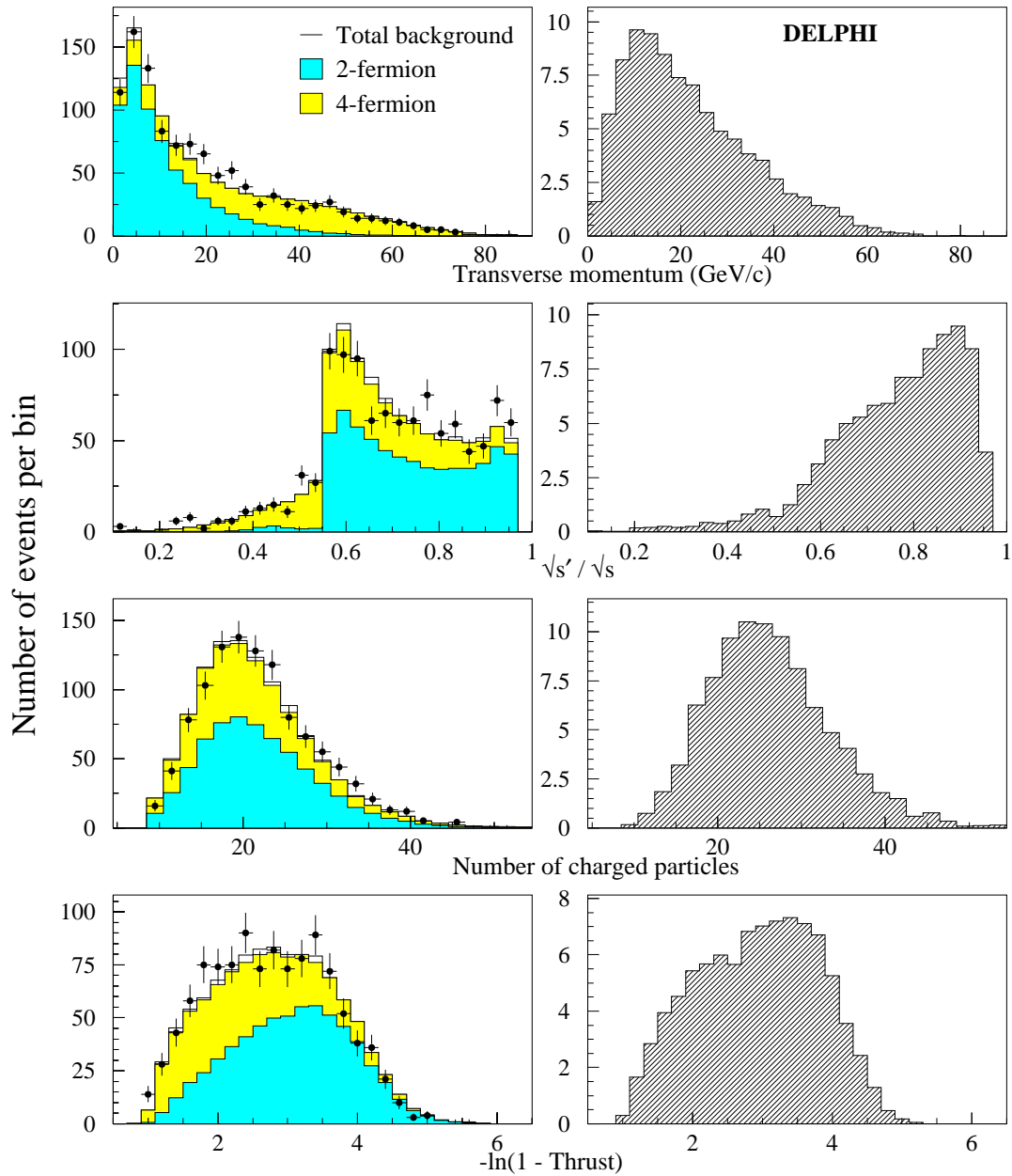,width=15.8cm}
\vspace{-1.2cm} 
\caption{\hnn\ channel, IDA analysis: distributions of four analysis
variables, as described in the text,  
 at the level of the common preselection for the IDA
analysis and low mass analysis.
Data (dots) are compared with {\sc SM} background expectations
(left-hand side histograms). The expected distributions for a
115~\GeVcc\ Higgs signal are shown in the right-hand side plots.
}
\label{fig:hnunu_ida}
\end{center}
\end{figure}

The same set of preselection criteria as in the low mass analysis was applied
to remove the bulk of the background events, before the IDA training. 
Ten variables were used to train the IDA:
the ratio of visible energy to centre-of-mass energy,
the energy around the most isolated particle in a double cone whose
inner and outer opening angles are normally 5$^\circ$ to 25$^\circ$
(but do depend upon energy),
the  ${\mathrm b}$-tagging variable \xb, 
the thrust in the rest frame of the visible system, 
the acoplanarity when forced to two jets 
 scaled by the sine of the minimum angle
between a jet and the beam axis, 
the transverse momentum, 
the anti-\WW\ isolation variable as explained in section~\ref{sec:hnn}, 
$\sqrt{s'}/\sqrt{s}$, 
the ${\mathrm b}$-tagging variable in the three-jet configuration, and 
the number of charged particles.
Fig.~\ref{fig:hnunu_ida} shows the distributions of four of the
 IDA variables at preselection level. 

As a next step, the event samples were reduced further by
imposing cuts in the tails of the signal distributions of the
variables used to train the IDA. For each variable in the combined
105 to 116~\GeVcc\ Higgs 
signal sample, the cuts removed about 0.5\% of the events in both the
upper and lower tails or about 1\% if 
only one tail was cut on.

The IDA consisted of two steps (iterations), keeping 85\% of the signal
in the first iteration. The training  samples were simulated
signal events with Higgs boson masses between 105
and 116~\GeVcc. Half of the available statistics, for both signal and
background samples, were used for the IDA training. The remaining
events were used to derive numbers for the background event rejection and
signal efficiencies, thus avoiding a statistical bias in these
estimates.

\begin{table}[htbp]
\begin{center}
\begin{tabular}{cccccc}     \hline
Selection & Data & Total  & \qqg  & 4 fermion & Efficiency (\%)\\
          &      & background        &       &           &    \\ 
\hline \hline
\multicolumn{6}{c} {Missing energy IDA analysis, first period, 157.8~\pbinv} \\ \hline
Anti \gaga       & 13038 & 12890$\pm$ 10 & 9669 &2929 & 85.6 \\
Preselection     &   787 & 786 $\pm$ 4 & 463  & 290  & 70.7 \\
eff(DA2)$ = 85\%$ &    21 &    16.6 $\pm$  0.6 &    7.9 &   8.6 & 47.8 \\ 
\hline
\hline
\multicolumn{6}{c} {Missing energy IDA analysis, second period, 57.5~\pbinv} \\ \hline
Anti \gaga       & 4475 & 4539 $\pm$ 6 & 3388 & 1060 & 85.1 \\
Preselection     &  303 & 288 $\pm$ 3 & 168  & 107  & 70.4 \\
eff(DA2)$ = 85\%$ &    9 &    6.76 $\pm$ 0.32 &    3.44 &    3.31 & 47.9 \\ 
\hline
\end{tabular}
\caption[]{
 \hnn\ channel, IDA analysis: effect of the selections on data
 and on simulated  background and signal events.
 Efficiencies are given for a signal with  \MH~=~115~\GeVcc, \sqrts=206.5~\GeV.
 The quoted errors are statistical only. For each period the
 first line shows the integrated luminosity used; the line labelled
 `eff(DA2)$ = 85\%$' shows the results when a cut on the IDA value
is applied such
that the efficiency for selecting a signal is 85\%; the cut
  value is different in the two periods.
}
\label{ta:hzsumIDA}
\end{center}
\end{table}

Table~\ref{ta:hzsumIDA} shows the 
effect of the selections on data and simulated samples.

\begin{figure}[htb]
\begin{center}
\epsfig{figure=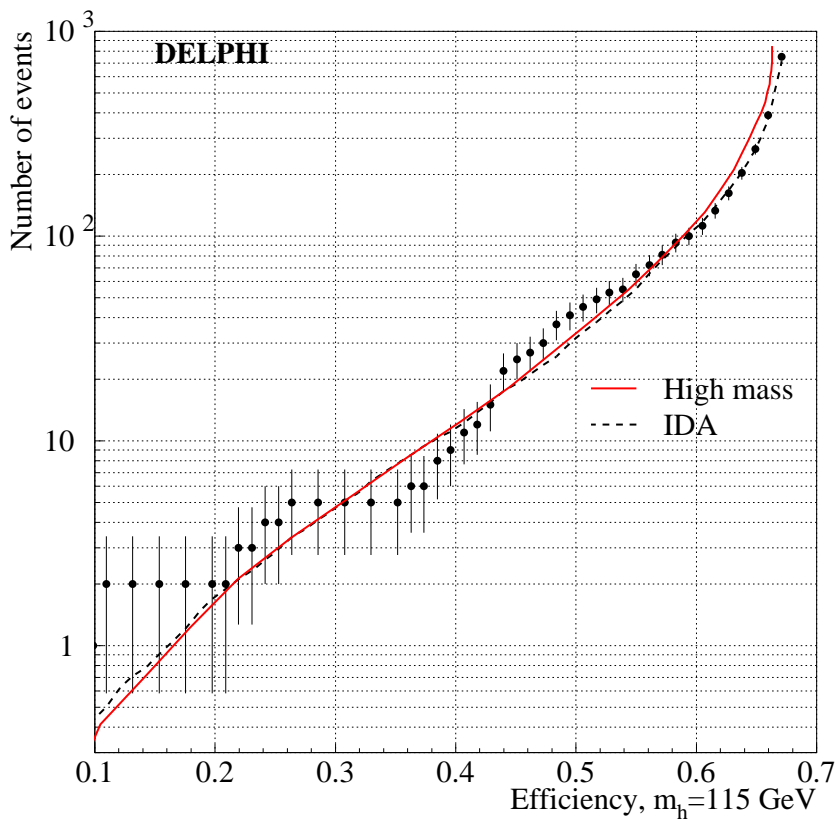,width=12cm} \\ 
\caption{\hnn\ channel, IDA analysis: the observed rate and that expected 
from the {\sc SM} background processes
as a function of the efficiency for a
 115~\GeVcc\ Higgs signal when varying the cut on the IDA variable.
The solid curve, from the high mass analysis, is included so that
the very similar performance of the two analyses can be seen.
The dots represent the data.}
\label{fig:hnunu_efb_all_ida}
\end{center}
\end{figure}

Fig.~\ref{fig:hnunu_efb_all_ida} shows the observed rate and 
that expected from SM 
background processes as a function of the efficiency for a Higgs signal of
115~\GeVcc.


The two high-mass analyses have different approaches, both in the
methods applied to extract the discriminant information and in the
treatment of the tails. 
The sensitivities and the results
 are very similar for the two analyses, as can be seen in
Figs.~\ref{fig:hnunu_efb_all_hm} and \ref{fig:hnunu_efb_all_ida}.

\section{Higgs boson searches in hadronic events}
\label{sec:4jet}

The analyses in the HZ and hA channels start from an  inclusive preselection,
after which all selection was performed by a Neural Network which 
removes some of the more distinct backgrounds. The output of the
Neural Network was then used
as the second variable in the CL calculations.
The  four-jet preselection, which eliminates $\gamma \gamma$ events and 
reduces the contributions from the  \qqg\ and $\Zz\gamma^*$ processes,
was not changed since the
previous analysis. The reader is referred to~\cite{ref:pap97,ref:pap98} 
for the exact description of the cuts, while only the
important features are briefly mentioned here.
After a selection of multi-hadron events
excluding those with  
an energetic photon in the calorimeters or lost in the beam pipe, 
topological criteria were applied to select multi-jet events. All 
selected events were then forced into a four-jet topology 
 and a minimal multiplicity and mass (1.5~\GeVcc) was required 
for each jet. After the preselection, different analysis 
procedures were applied in the \ZH\ and \hA\ channels.

\subsection{The HZ four-jet channel }
\label{sec:1}


Two analyses were applied to the whole range
of masses, and the results from the more powerful one were used.
One corresponds precisely to the account  published in Ref.~\cite{ref:pap99},
 and is not
described in this paper; the other has been optimised for high Higgs 
masses. The same automatic procedure as in the neutrino channel was applied to
select only the analysis with the better performance at each test point.
The range of
masses where the switches from the low mass to the high mass analyses occur 
lay between 99 and 110~\GeVcc\
 (with the majority of the 12 data sets changing at 105~\GeVcc).
The rest of this section describes the high-mass optimised analysis.

The final discriminant variable used in the four-jet channel was 
the output of an artificial neural network
(ANN) which combined thirteen variables. 
This is the same as was used in \cite{ref:D2000} without retraining.
The first of the variables was the global ${\mathrm b}$-tagging
variable \xb\ of the event. The next four  variables 
tested the compatibility of the event with the hypotheses of \WW\ and 
\ZZ\ production giving either 4 or 5 jets. 
Constrained fits were used 
to derive the probability density function measuring the compatibility 
of the event kinematics with the production of two objects of
hypothetical masses. 
This yielded a two-dimensional probability, the ideogram
probability~\cite{ref:ww183}.
To estimate compatibility with the \ZZ\ and \WW\ processes, the integral over
all boson masses of the ideogram probability
times the probability of obtaining that pair of masses from the process
in question was calculated.

The last eight  input variables, 
intended to reduce the \qqg\ contamination, were 
the sum of the second and fourth Fox-Wolfram moments,
the product of the minimum jet energy and the minimum opening angle between
any two jets,
the maximum and minimum jet momenta,
the sum of the multiplicities of the two jets with lowest multiplicity,
the sum of the masses of the two jets with lowest masses,
the minimum jet pair mass and
the minimum sum of the cosines of the opening angles of the two jet pairs
when considering all possible pairings of the jets.
The neural network was trained on independent samples, using signal
masses close to the kinematic limit.


The choice of the Higgs jet pair made use of both the kinematic 5C-fit 
probabilities 
(imposing four-momentum conservation and assigning the Z mass to
one pair of jets)
and the ${\mathrm b}$-tagging information 
in the event~\cite{ref:pap97}. 
The likelihood pairing function,
\[
  {\cal P}_b^{j_1} \cdot {\cal P}_b^{j_2} \cdot (
 (1-R_b-R_c) \cdot  {\cal P}_q^{j_3} \cdot {\cal P}_q^{j_4}
  +R_b         \cdot  {\cal P}_b^{j_3} \cdot {\cal P}_b^{j_4}
  +R_c         \cdot  {\cal P}_c^{j_3} \cdot {\cal P}_c^{j_4})
  \cdot P_{j_3,j_4}^{5C}
\]
was calculated for each of the six possibilities to combine the jets
$j_1$, $j_2$, $j_3$ and $j_4$ and assign the jet pairs to a \Hz\ or \Zz\ 
hypothesis.
${\cal P}_b^{j_i},{\cal P}_c^{j_i},{\cal P}_q^{j_i}$ are the 
probability densities of getting 
the observed ${\mathrm b}$-tagging value for the jet $j_i$ 
when originating from 
a $b$, $c$ or light quark, estimated from simulation.
$R_b$ and $R_c$ are the hadronic branching fractions
of the \Zz\ into $b$ or $c$ quarks~\cite{ref:rpp2000}, 
and $P_{j_3,j_4}^{5C}$ is the probability 
of the kinematic 5C-fit with the jets $j_3$ and $j_4$ 
assigned to the \Zz.
The pairing that maximised this function was selected
and the reconstructed Higgs boson mass was the result of the 5C-fit for
that pairing.
The proportion of right matchings for the Higgs jet pair, 
estimated in simulated signal events with 115~\GeVcc\ mass, 
was around 53\% at preselection level, 
increasing to  73\% after a cut on ANN of 0.81, as used later for
figure~\ref{fig:mass_pl1}.
This technique is better than using just the probability 
of the kinematic 5C-fit, both for \ZH\ and \ZZ\ events.

\begin{figure}[htbp]
\begin{center}
\begin{tabular}{cc}
\epsfig{figure=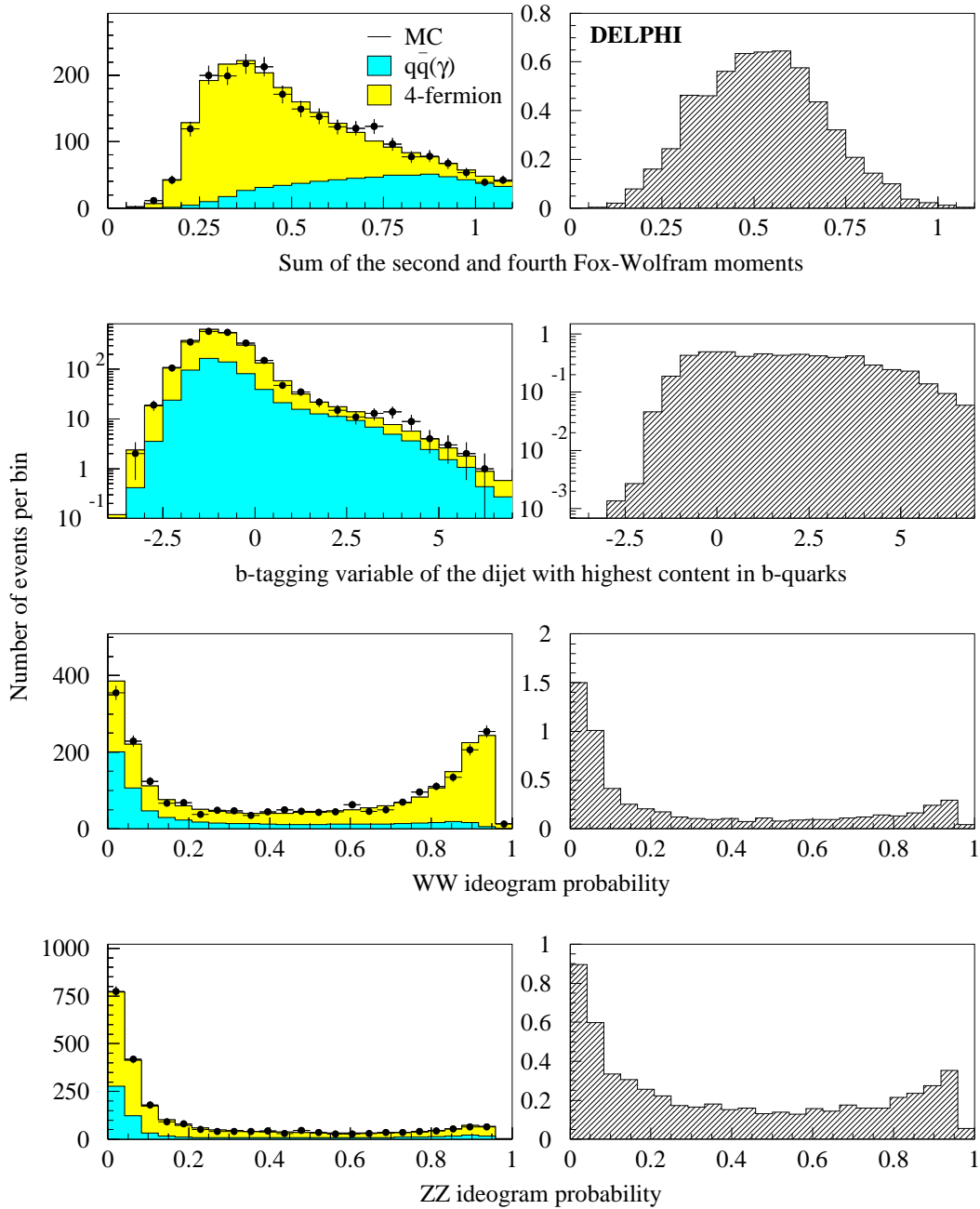,width=15.7cm} 
\end{tabular}
\caption[]{\hqq\ channel: 
  distributions of four analysis variables,
  as described in the text, at preselection level. Data from the year 2000
  (dots) are compared with  {\sc SM} background process 
  expectations (left-hand side histograms). The  expected distributions
  for a 115~\GeVcc\ Higgs signal are shown in the right-hand side histogram.
}
\label{fig:hqq}
\end{center}
\end{figure}

\begin{figure}[htbp]
\begin{center}
\epsfig{figure=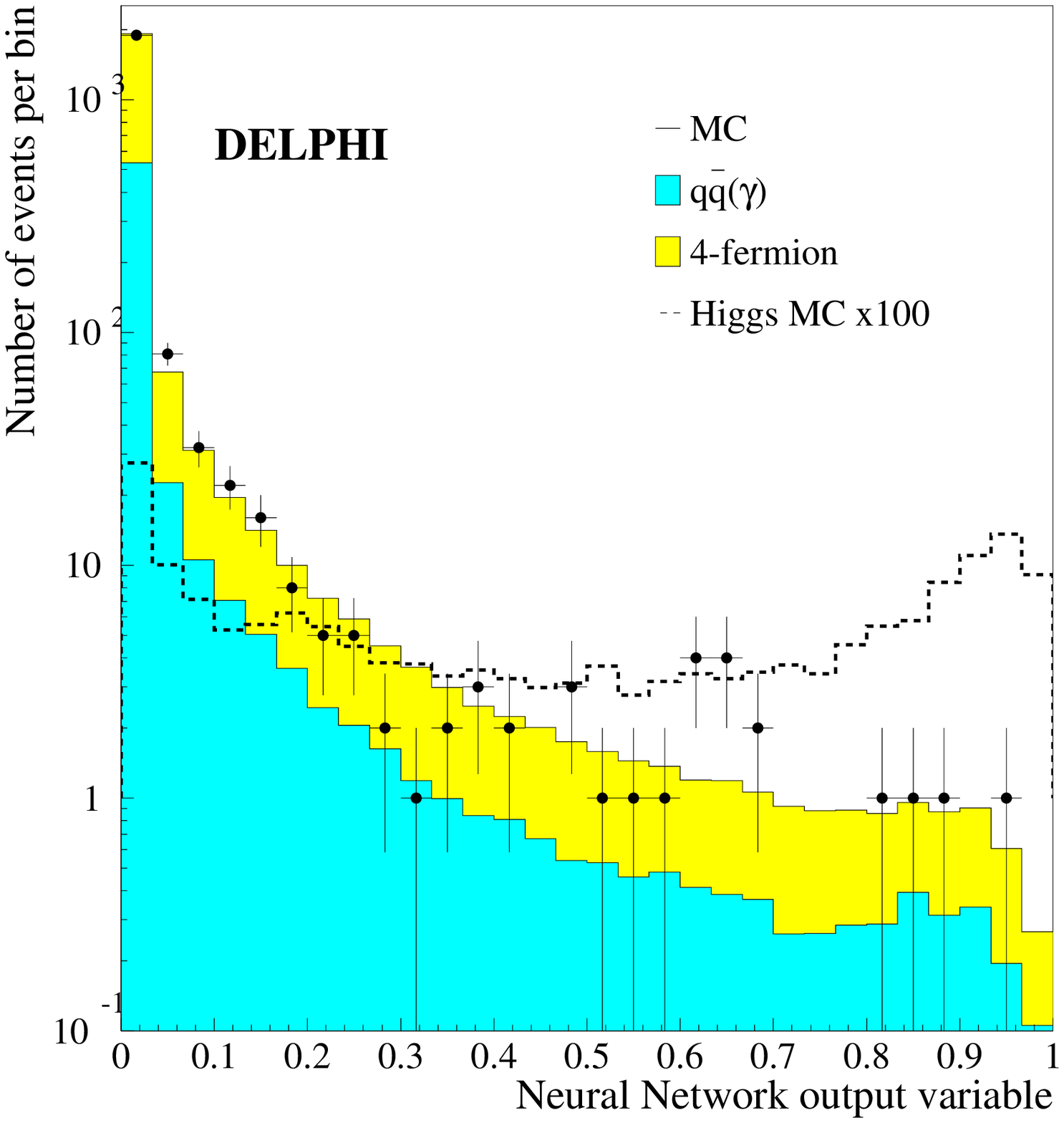,width=10cm} \\
\epsfig{figure=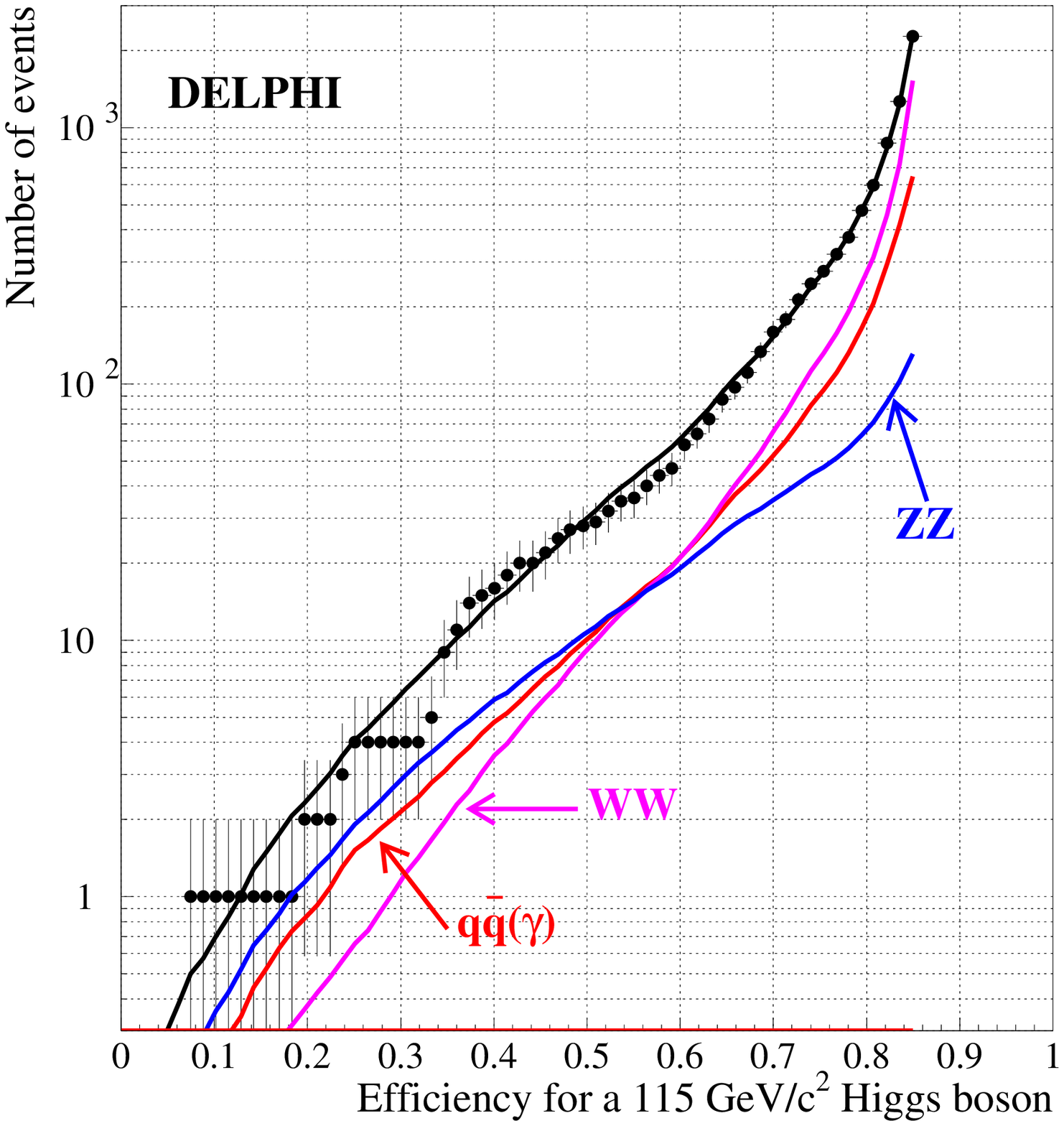,width=10cm} \\
\vspace{-0.5cm}
\caption{\hqq\ channel: 
Top:
  the distribution of the neutral network variable used to select
  Higgs candidates in the 2000 data. The signal expectation is shown
 for a 115~\GeVcc\ Higgs, normalised to 100 times the expected rate. The
  dots represent the data.
Bottom:
  curve of the expected  {\sc SM} background rate in the 2000 data as a 
  function of the efficiency for a 115~\GeVcc\ Higgs signal when
  varying the cut on the neural network variable. 
  The different background  contributions are shown separately. 
  The dots represent the data. 
}
\label{fig:hqq_disc}
\end{center}
\end{figure}

The agreement between data and background process simulations after the
four-jet
preselection is illustrated 
in  Fig.~\ref{fig:hqq}, which shows the distributions of 
the sum of the second and fourth Fox-Wolfram moments
as an example of the kinematic variables,
the global ${\mathrm b}$-tagging variable,
and the WW and ZZ ideogram probabilities
for the configuration with 4 jets.
Fig.~\ref{fig:hqq_disc} shows the performance  of the final discriminating
variable in terms of the background  rate as a function of 
the efficiency  for a 115~\GeVcc\ Higgs signal, and the agreement between
simulation and data.
The effect of the selections on  data and simulated samples  
for the two periods of data taking are detailed in Tables~\ref{ta:hzsum2000e} 
and \ref{ta:hzsum2000u}. The signal  efficiencies for the first period
are shown as a function of the Higgs boson mass
in  Fig.~\ref{fig:hzeff} and for both periods in 
Table~\ref{ta:hzeff}.
The data are also analysed by the low mass analysis, which
 selected 180 candidates 
for the confidence level calculation
while 173 were expected from SM background processes.

 The final CL calculations were made in the plane of the ANN value versus 
the Higgs boson mass estimator, using only events where the ANN was  greater
than 0.2.
This gives 47.7$\pm$0.3(stat.) events expected from background  processes,
 whilst 40 are observed.


\subsection{The hA four-b channel}

This channel benefited most from the data reprocessing and
improved ${\mathrm b}$-tagging.
The analyses 
  include not only data from the year 2000, 
but also the reprocessed  1999 data.
  After the common four-jet preselection, events were preselected further,
requiring a visible energy greater than 120 $\times
\sqrt{s}/189$~\GeV,
 \mbox{$ \sqrt{s'}$} greater than 150 $\times
\sqrt{s}/189$~\GeV, a missing momentum component along the beam
direction lower than 30~\GeVc\ and at least two charged particles
per jet. A four-constraint kinematic fit requiring energy and
momentum conservation was then applied, and the two jet-pair masses
were calculated for each of the three different jet pairings. As
the possible production of {\sc MSSM} Higgs bosons through the
\hA\ mode dominates at large \tbeta, where the two bosons are
almost degenerate in mass, the pairing defining the Higgs boson
candidates was chosen as that which minimizes the mass difference
between the two jet pairs
and the reconstructed Higgs boson masses were taken from the 4C-fit for
that pairing.
The final discrimination between
background and signal events was then based on a multidimensional variable
which combined the following twelve variables  as the output of an
artificial neural network: the event thrust, the sum of the second
and fourth Fox-Wolfram moments,
the product of the minimum jet energy and the minimum opening
angle between any two jets,
the minimal $y_{cut}$ values for which the event is clustered into
4 jets ($y_{34}$) and into 5 jets ($y_{45}$),
 the maximum and
minimum jet momenta,
the sum of the multiplicities of the two jets with lowest
multiplicity,
the minimum jet pair mass, the production angle of the
Higgs boson candidates,
the sum of the four jet ${\mathrm b}$-tagging variables and the minimum
jet pair ${\mathrm b}$-tagging variable.
The neural network was trained using signal masses between 80 and
95~\GeVcc, and about 10\% of the simulated background events, and
this one training applied to all data sets.

\begin{table}[htbp]
\begin{center}
\begin{tabular}{cccccc}     \hline
Selection & Data & Total  & \qqg  & 4 fermion & Efficiency (\%)\\
          &      & background        &       &           &    \\ 
\hline \hline
\multicolumn{6}{c} {hA four-jet channel 228~\pbinv\ 1999 } \\ \hline
Tight preselection     & 2224 & 2211.4$\pm$ 2.5   &   650.0 &  1561.4 &    91.6 \\
Candidate selection   &  217 & 191.6 $\pm$ 0.8&  81.4&  110.2&    89.0 \\
\hline
\multicolumn{6}{c} {hA four-jet channel 163.7~\pbinv\ 2000 1$^{\rm st}$ period} \\ \hline
Tight preselection     & 1459 & 1500.2$\pm$ 2.1   &   406.9 &  1093.3 &    91.2 \\
Candidate selection   &  127 & 129.3 $\pm$ 0.7&  50.6&  78.5&    89.4 \\
\hline
\multicolumn{6}{c} {hA four-jet channel 60.1~\pbinv\ 2000  2$^{\rm nd}$ period} \\ \hline
Tight preselection     & 495 & 547.2$\pm$ 1.1   &   148.1 &  399.1 &    90.8 \\
Candidate selection   &  48 & 45.2 $\pm$ 0.3&  17.4&  27.8&    88.2 \\
\hline
\end{tabular}
\caption[]{
\hA\ hadronic channel: effect of the selections on 
  data and  simulated  background events. 
 Efficiencies are given for a signal with 
 \MA = \mh = 90 \GeVcc.
 The quoted errors are statistical only. For each case, the
 first line shows the integrated luminosity used; the line labelled
`candidate selection' shows the events used for calculating the confidence
levels.
}
\label{ta:ha_all}
\end{center}
\end{table}

\begin{figure}[htbp]
\begin{center}
\begin{tabular}{cc}
\epsfig{figure=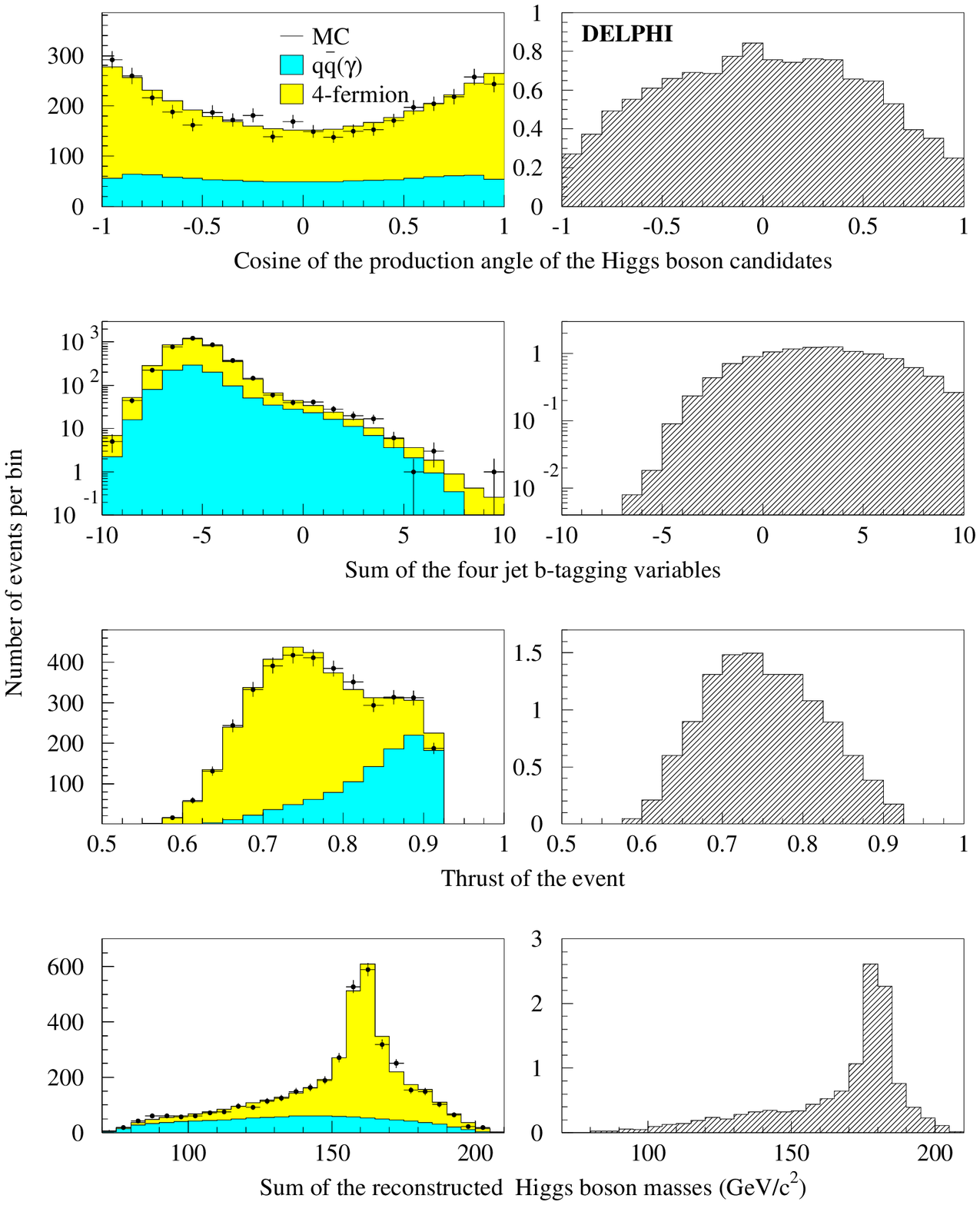,width=15.7cm}
\end{tabular}
\caption[]{hA hadronic channel:
  distributions of four analysis variables,
  as described in the text, at preselection level. Data from the years
  1999 to 2000
   (dots) are compared with  {\sc SM} background process
  expectations (left-hand side histograms). The expected distributions
  for a 
 \MA = \mh = 90 \GeVcc\ signal are shown in the right-hand side plots.
}
\label{fig:4b}
\end{center}
\end{figure}

\begin{figure}[htbp]
\begin{center}
\epsfig{figure=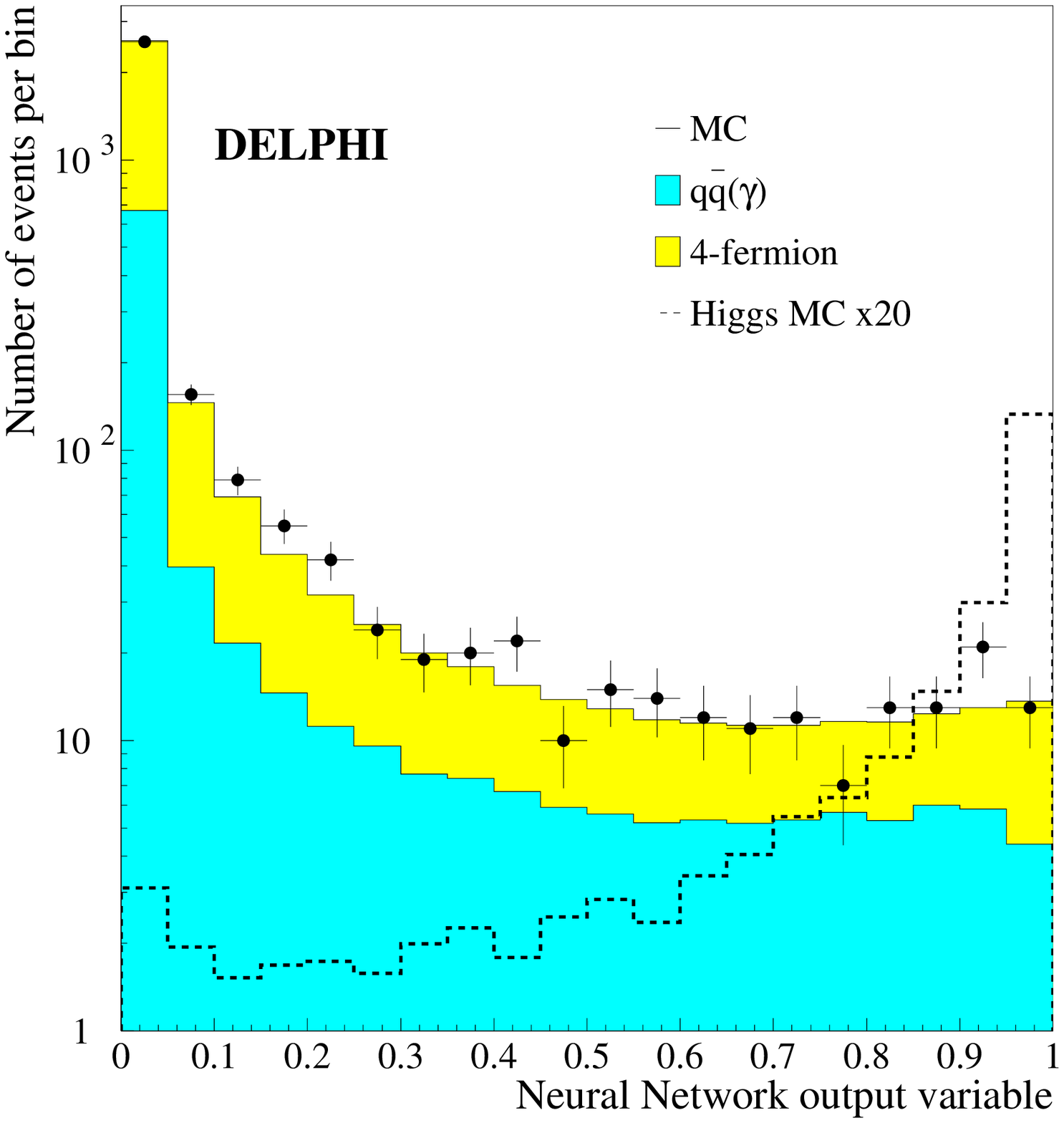,width=10.cm} \\
\epsfig{figure=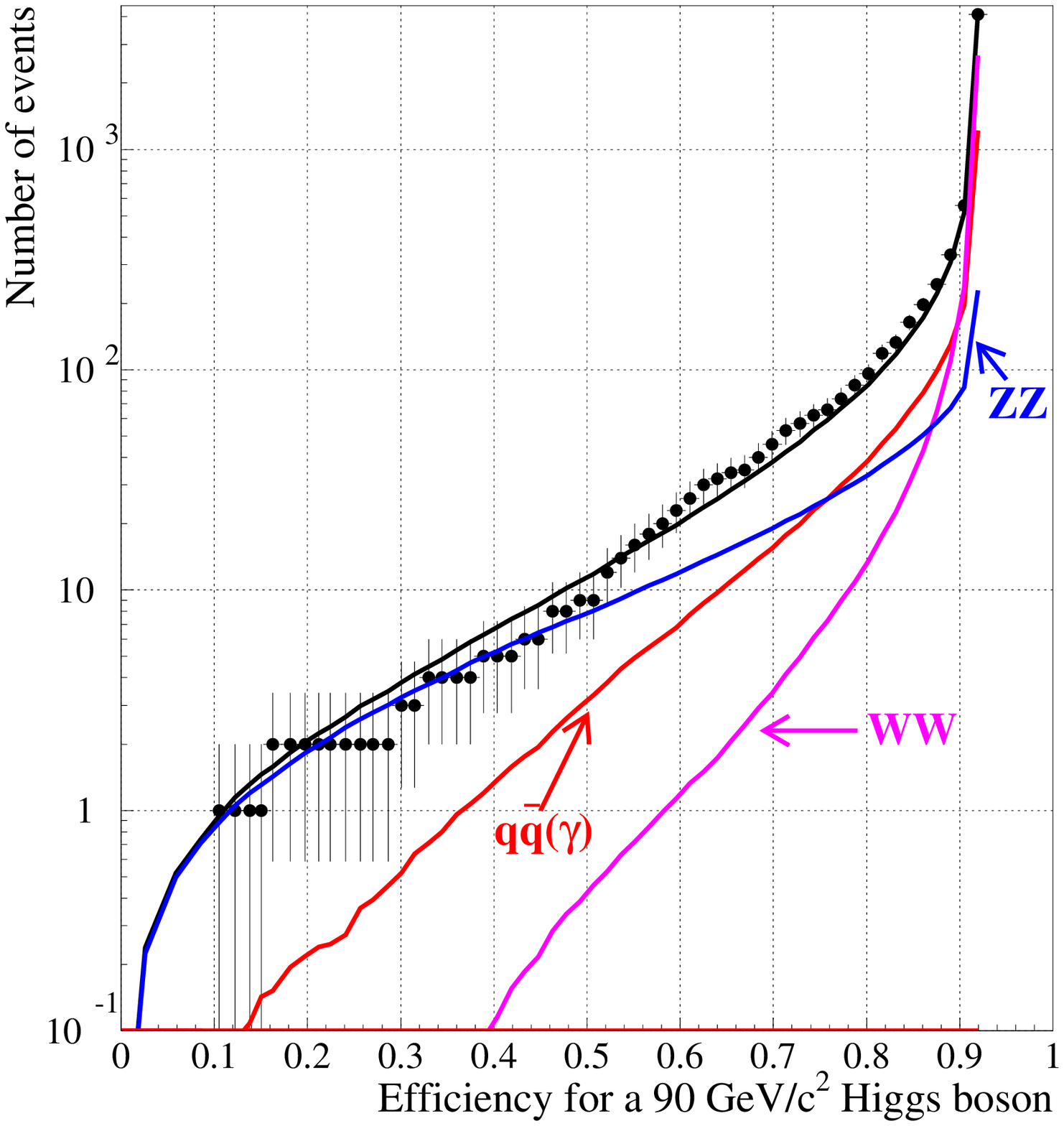,width=10.cm} \caption{hA hadronic
channel: Top: distributions of the ANN variable
  for the expected  {\sc SM} background processes (full histograms),
  the data from the years 1999 and 2000
  (dots) and the   expected 
 \MA = \mh = 90 \GeVcc\ Higgs signal 
  (dashed histogram, normalised to 20 times the expected rate).
  Bottom: curve of the expected  {\sc SM} background rate for 
  all data from 1999 and 2000
  as a function of the efficiency for the same signal
   when varying the cut on the ANN variable.
  The different background contributions are shown separately.
  The dots show the data. }
\label{fig:4b_disc}
\end{center}
\end{figure}

The agreement between data and background channel simulations after the
preselection is illustrated in  Fig.~\ref{fig:4b}, which shows the
distributions of three input variables and of the sum of the
reconstructed Higgs boson masses.
 Fig.~\ref{fig:4b_disc} shows the
distribution of the final discriminant variable and, as an
example, the total expected background rate and the data 
from 1999 and 2000
as a function of the efficiency for a
signal with 
 \MA = \mh = 90 \GeVcc, when varying the cut
on the discriminant variable. As a final selection, a minimal
value of 0.1 is required, leading to 392 events in data, with
 \mbox{$366.2 \pm 1.1 (stat.)$} expected from background processes.
  The effect of the selections on data and simulated samples
 are detailed in Table~\ref{ta:ha_all}
while representative efficiencies at the end of the analysis are reported 
as a function of Higgs boson masses in
Tables~\ref{ta:scaneff} and~\ref{ta:haeff}
and in Fig.~\ref{fig:4b_eff}.

 The two-dimensional calculation of the confidence levels uses the
 ANN variable and the sum of the reconstructed Higgs boson masses.


\subsection{Additional {\sc MSSM} results}\label{sec:admssm}

  In the purely hadronic final state, which is the dominant topology
in the {\sc MSSM}, additional signals were considered. 

  In a small region of the parameter space where the \hZ\ production
process is dominant, the decay \hAA\ opens. As 
the low mass \lhqq\ analysis proved to perform  reasonably  on that 
signal, no dedicated procedure was set up and that analysis
was applied as such on the two simulated 
(\hAA )(\Zqq ) channels. The corresponding
efficiencies are shown in Table~\ref{ta:aaqqeff} at \sqrts=206.5~\GeV\ for
both  data-taking periods. The efficiencies and PDFs
obtained for the ($ {\mathrm{h}} \rightarrow {\mathrm{AA}} \rightarrow {\mathrm{c\overline{c}c\overline{c}}} $) (\Zqq) channel were conservatively applied
to the two channels where one A boson decays into b's while the other
decays into c's.
  
  The two four-jet final states expected in the {\sc MSSM} have common 
features. As a consequence, the two analyses developed specifically for 
each of them perform rather well on the other signal. As an example, the
efficiencies of the low mass \lhqq\ analysis applied on the four-b signal 
and that of the \hA\ four-b analysis applied on the \lhqq\ signal are
given in Tables~\ref{ta:ad4beff} and~\ref{ta:adhqeff} 
at 199.6~\GeV\  and at 206.5~\GeV\ for both data-taking periods. 
Thus, when combining the results in the \hZ\ and \hA\ channels to 
derive confidence levels in the {\sc MSSM}, 
both selected signals are included in the 
results of these two analyses at all energies above 191.6~GeV. This
leads to a gain in sensitivity of around 1~\GeVcc\ on the masses
of {\sc MSSM} Higgs bosons in regions of
the parameter space where both the \hZ\ and \hA\ production processes
contribute.

\section{Systematic errors}
\label{sec:systematics}

The systematic errors for each channel are discussed below.

\subsection{Systematic errors in the \hee\  search}

  The systematic uncertainties on background rates and signal efficiency
estimates
are mainly due to the imperfect simulation of the detector response
and were estimated as described in~\cite{ref:pap97}:
each cut in turn was adjusted until the fraction of events accepted
in simulation matched that found in data. The corresponding changes 
in background and signal rates were summed in quadrature.
 
The relative error on the efficiencies is typically $\pm 2\%$
while that on the background rate estimates is $\pm 5\%$.

\subsection{Systematic errors in the \hmm\ search}

The imperfect simulation of the detector response leads to
systematic errors in background process rate and signal efficiency
evaluation. As explained
in~\cite{ref:pap97}, each of the momentum and angular cuts was varied in a 
range
given by the difference between the mean values of the simulated and
real data distributions of the corresponding variable at preselection 
level.
This method cannot be used for the muon pair identification tag, which is 
a discrete variable. To estimate the effect of possible differences in
muon pair tagging between the data and the simulated samples, 
each muon candidate in the simulated samples was allowed to migrate
randomly from its original level of identification to one of the two
neighbouring ones, with a
probability of 1.5\%, with migrations to non-physical levels ignored.
 This probability corresponds to 
the difference observed in muon identification results 
between real data and simulation.
For the relative error on the expected number of background events 
we also include a 2\% uncertainty on the \ZZ\ cross-section, 
so  the total systematic error is estimated to be $\pm$2.8\% on the
background. 
For the efficiencies, an overall  relative systematic uncertainty of $\pm 1\%$
can be quoted, independent of \MH.


\subsection{Systematic errors in  the \ttqq search } 

Systematic uncertainties from the imperfect modelling of the detector
response were estimated by moving each selection cut according to
the resolution in the corresponding variable. The main contributions arise
from the cuts on the \toto\ invariant mass and electromagnetic energy.
The total relative systematic uncertainties amount to $\pm$6\% 
on signal efficiencies and $\pm$11\% on the background process
 estimates.

\subsection{Systematic errors in the \hnn\  search  } 

Systematic uncertainties in the low mass  analysis due to the imperfect
 modelling of 
the detector response were derived by rescaling, bin by bin, the
 contents of each 
PDF from simulation to those in data, restricting to bins that
contained at least one percent of the total statistics. 
The analysis was then repeated with the rescaled PDF for each
variable in turn and the largest difference with respect to the 
initial result was taken as systematics.
These systematics amount to $2\%$  for the efficiencies and to $4\%$ 
for the background in the first data taking period and come from the
PDF of the acoplanarity distribution. These numbers become $3\%$ and
$9\%$ for the second data taking period.
Systematic
uncertainties due to the use of non-independent samples in the
definition of the likelihood PDFs and in the final result amount to
$\pm2\%$ for the efficiencies and to $\pm1\%$ for the background processes
in the first operational period.  These numbers become $\pm3\%$ 
and $\pm4\%$
for the second operational period.  
Thus, the overall uncertainties amount to
$\pm2.8\%$ ($\pm4.1\%$) for the signal efficiencies and $\pm4.1\%$ 
($\pm9.8\%$) for
the background channels, for the first (second) period of data taking.

The same method was used to 
derive the systematic uncertainties in the high mass analysis.
This was done
separately for the two operational periods. 
Systematic uncertainties
due to the use of non-independent samples in the definition of the
likelihood PDFs and in the final result amount to $<\pm1\%$ for both the
signal efficiencies and the background for the two operational
periods. These uncertainties are well below the statistical errors.
Systematic uncertainties due to the imperfect modelling of the
detector response (coming from the PDF of the visible mass
distribution) amount to $\pm1\%$ for the efficiencies and to $\pm7\%$ for
the background processes in the first operational period. These
numbers become $\pm2\%$ and $\pm7\%$ for the second operational period.

\subsection{Systematic errors in the four-jet  searches}

In the \HZ\ search the systematic uncertainties from the imperfect
modelling of the detector response were estimated by repeating the
selection procedure on the distribution of the neural network variable
obtained by smearing, in turn, each of the distributions of the
thirteen input variables according to the resolution in the variable.
This leads to relative uncertainties of $\pm$5\% related to b-tagging.
 Uncertainties in the anti-QCD variables are $\pm$2\% in the
background process
estimations and $\pm$1\% in the signal efficiencies. Systematic
uncertainties related to the ideogram probabilities are $\pm$3\% for
the background and $\pm$2\% for the efficiencies.  This results in
overall relative uncertainties of $\pm6.2\%$ in the background rates
and $\pm5.5\%$ in efficiency estimates for each period.

   Systematic uncertainties in the \hA\ search 
due to the use of non-independent samples
in the training of the ANN and in the final result
derivation were estimated at the level of  $\pm 1.0\%$ relative,
by repeating the whole procedure with two independent samples of
smaller size. Systematic uncertainties due to the imperfect modelling
of the detector response were derived as 
above.
The uncertainty related to ${\mathrm b}$-tagging amounts to  $\pm 10\%$ on
background and  $\pm 1.2\%$ on signal, while  
that related to shape variables is  $\pm 5.0\%$ in the
background rate and $\pm 1.2\%$  in signal efficiency estimates.
Combining all these results in 
overall relative uncertainties of  $\pm 11.2\%$ and $\pm 1.7\%$  on background 
processes and signal efficiency.

\subsection{Summary of systematic errors}

\begin{table}[htbp]
\begin{center}
\begin{tabular}{clcc}     \hline
\HZ\ selection         & period & background & signal          \\ 
\hline \hline
\hee\ channel          & both   &   $\pm$5\%      & $\pm$2\%   \\
\hmm\ channel          & both   &   $\pm$2.8\%    & $\pm$1\%   \\
Tau channel            & both   &  $\pm$11\%      & $\pm$6\%   \\
\hnn\  (low mass )     & first  &   $\pm$4.1\%    & $\pm$2.8\%   \\
\hnn\  (low mass )     & second &  $\pm$9.8\%     & $\pm$4.1\%   \\
\hnn\  (high mass)     & first  &   $\pm$7\%      & $\pm$1\%   \\
\hnn\  (high mass)     & second &   $\pm$7\%      & $\pm$2\%   \\
\hqq\ channel          & both   &  $\pm$ 6.2\%    & $\pm$5.5\% \\
\hline \hline
\hA\ selection         & period & background & signal          \\ 
\hline \hline
Four-jet channel       & both   &   $\pm$11.2\%   & $\pm$1.7\% \\
\hline
\end{tabular}
\caption[]{
The systematic error estimates for  the individual channels. The
missing energy channels are somewhat more sensitive to the condition of the
detector, and result in the larger errors during the second period.}
\label{ta:systematics}
\end{center}
\end{table}

The error estimates obtained in each channel are shown in
Table~\ref{ta:systematics}. In principle there might be small
correlations between the errors in the different channels, from
for example the cross-section of the \ZZ\ process and the b-tagging
procedure. However, both these enter in significantly different ways
in the different channels, and the correlated component is therefore rather
small, and has been neglected.


\section{Results}
\label{sec:results}

The results of the searches presented in the previous
sections 
are used to calculate the consistency of the data with
signal and background hypotheses, and derive 
confidence levels as a function 
of the masses of the 
neutral Higgs bosons in the {\sc SM} and {\sc MSSM}.

\subsection{Reconstructed mass spectra}

\begin{figure}[htbp]
\begin{center}
\epsfig{figure=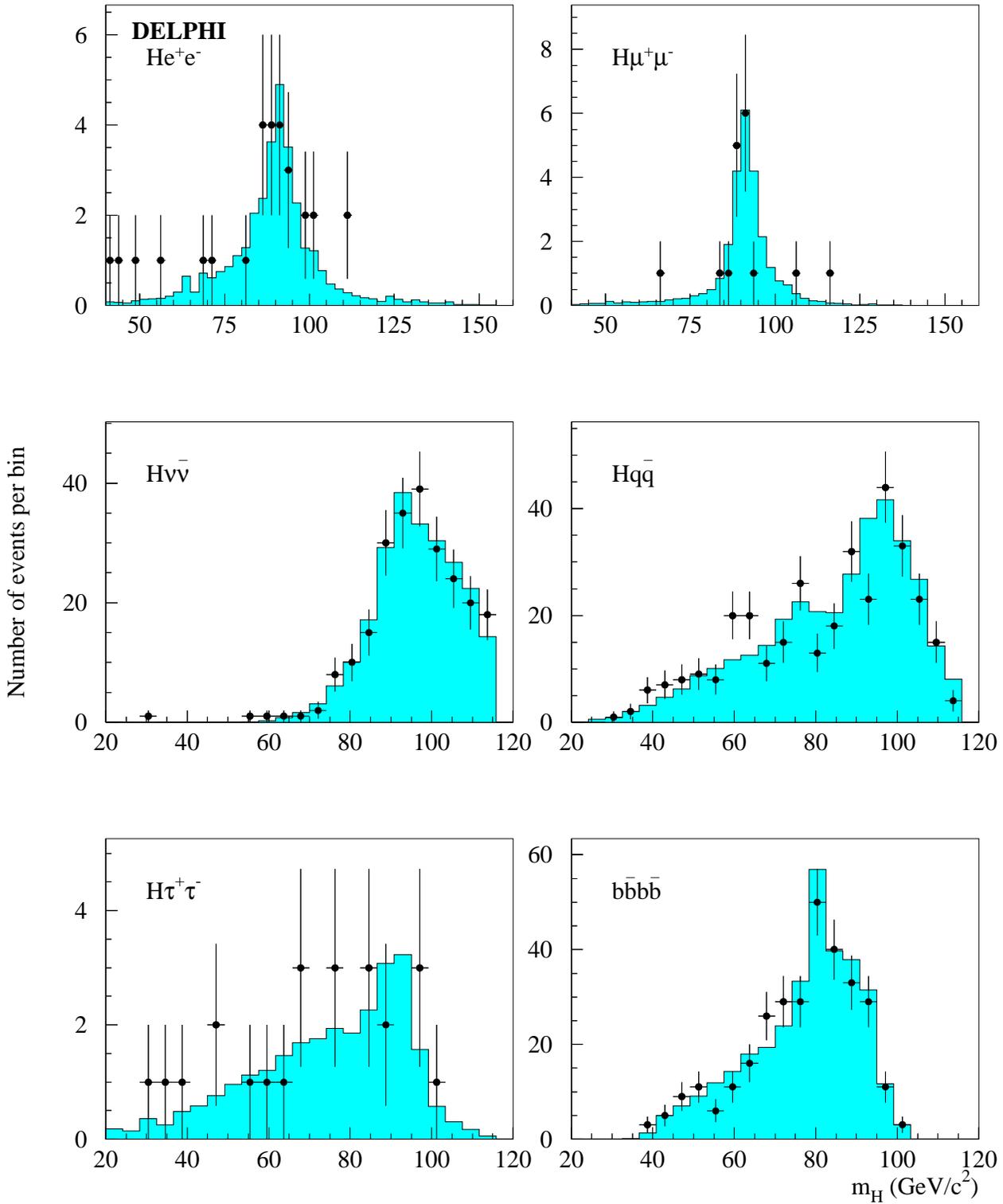,width=15.9cm} \\
\vspace{-0.1cm}
\caption[]{
Distributions of the reconstructed Higgs boson mass(es)
in each channel when all the data from 183 to 209~\GeV\ are
combined at the  candidate selection level.
This plot contains almost all the LEP2 data set used for the
extraction of the final limits.
Data (dots) are compared with  {\sc SM} background 
expectations (full  histograms).
}
\label{fig:mass_pl}
\end{center}
\end{figure}

The 
distributions of the reconstructed Higgs boson mass(es) at the level of
the selected candidates
are presented in Fig.~\ref{fig:mass_pl}. These plots include our
previous LEP2 results above 183~\GeVcc.
The data are consistent with the simulated background events.

\begin{figure}[htbp]
\begin{center}

\epsfig{figure=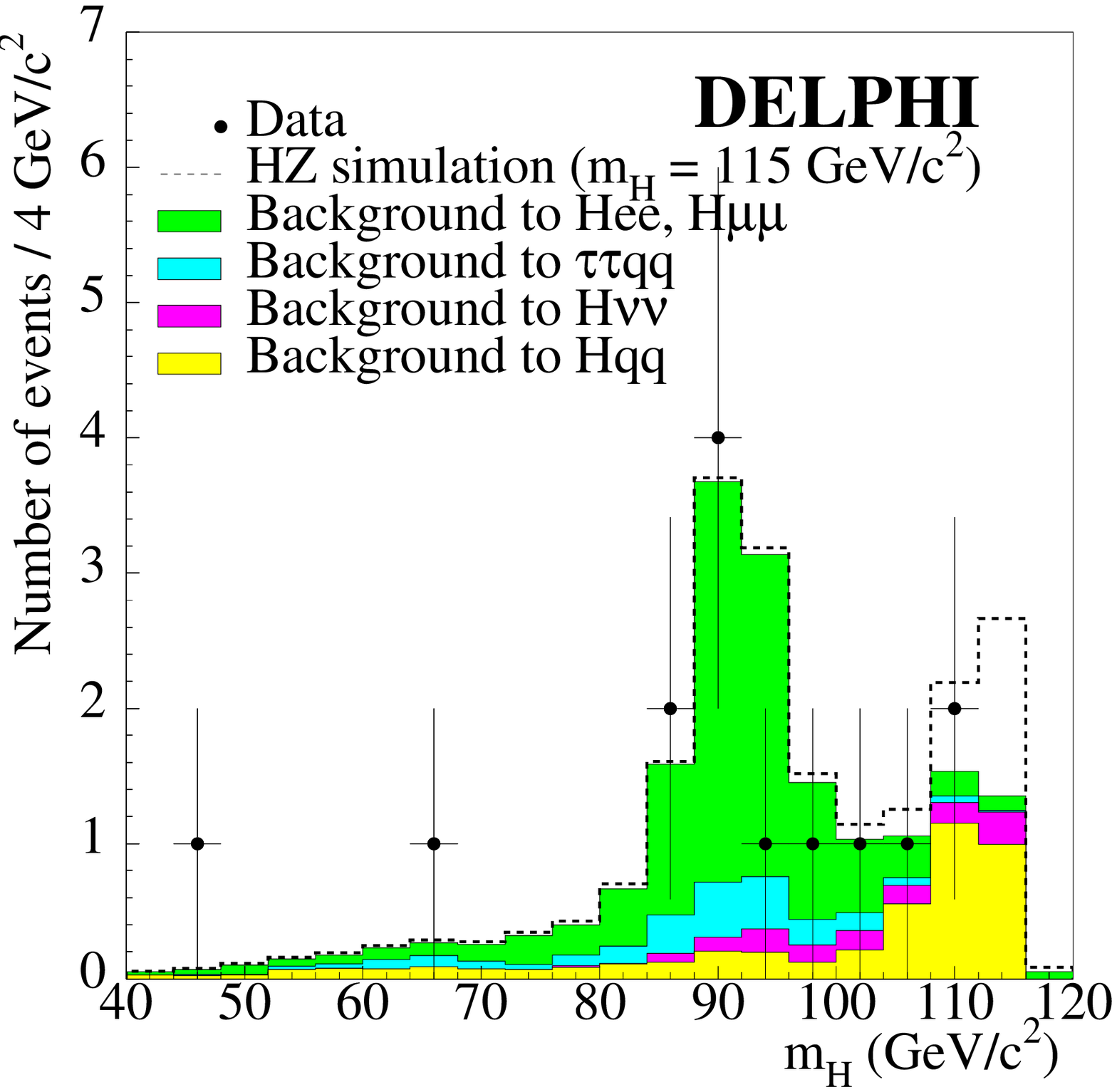,width=11.cm} \\
\vspace{-0.1cm}
\caption[]{
Distribution of the  reconstructed {\sc SM} Higgs boson masses
for the tightly selected candidates
in each channel from the 2000 data.
Data (dots) are compared with  background process expectations (full 
histograms)
and with the normalised signal spectrum added to the background channel 
contributions (dashed histogram). The mass hypothesis for the simulated signal
spectrum is \MH~=~115~\GeVcc.
}
\label{fig:mass_pl1}
\end{center}
\end{figure}

The results for only the year 2000 {\sc SM} analysis
are shown in Fig.~\ref{fig:mass_pl1}.
For this figure there is an extra  selection:  a 
cut on the second discriminant variable was applied such that a signal
might be more  apparent. This cut is fixed for all channels 
so that, for  events  with mass greater
than 110~\GeVcc\ in data taken at 206.5~\GeV, the signal to background ratio 
should be 0.75.
The selections required a minimal 
${\mathrm b}$-tagging value of 0.49 in the \hee\ and
 -2.5 in the \hmm\ channels, minimal
likelihood values of 0.58, 2.22  in the \ttqq\ and \hnn\ channels,
 respectively, and a minimal neural network output of 0.81 in the
\hqq\ channel. 
The corresponding observed and expected rates in each period
are given in Table~\ref{ta:tight}, which can be compared with
Tables~\ref{ta:hzsum2000e} and \ref{ta:hzsum2000u}.

\begin{table}[htbp]
\begin{center}
\begin{tabular}{cccccc}     \hline
Channel           & Data & Total        & \qqg & 4 fermion & Efficiency (\%)\\
                  &      & background   &      &           &    \\ 
\hline
\multicolumn{6}{c}{First period } \\ 
\hline \hline
\hee          &   1  & 1.28$\pm$ 0.04& 0.09 & 1.19      &   33.4  \\ 
\hline
\hmm              &   5  & 6.38$\pm$0.12& 0.04 & 6.34      &   64.3  \\ 
\hline
Tau               &   1  & 1.53$\pm$0.10& 0.08 & 1.45      &   10.0    \\
\hline
\hnn\ (High mass) &   1  & 0.87$\pm$0.08& 0.34 & 0.53      &   17.4  \\ 
\hline
\hqq          &   3  & 3.03$\pm$ 0.1 & 1.08 & 1.95      &   25.6  \\
\hline
\multicolumn{6}{c} {Second period } \\ 
\hline \hline
\hee          &   0  & 0.42$\pm$0.03& 0.04 & 0.38      &   33.4 \\ 
\hline
\hmm              &   2  & 2.36$\pm$0.05& 0.02 & 2.34      &   64.4 \\ 
\hline
Tau               &   0  & 0.53$\pm$0.03& 0.03 & 0.50      &   12.1  \\
\hline
\hnn\ (High mass) &   0  & 0.30$\pm$0.05& 0.14 & 0.17     &   16.5  \\ 
\hline
\hqq          &   1  & 1.11$\pm$0.05& 0.40 & 0.71      &   25.3 \\
\hline
\end{tabular}
\caption[]{
Candidates selected for the {\sc SM} channels by tight cuts  on data and
 simulated  background processes. The last column gives the efficiencies at
a mass \MH = 115~\GeVcc.
}
\label{ta:tight}
\end{center}
\end{table}


There are three events with a signal to background ratio higher
than 0.2 for the hypothesis
\MH~=~115~\GeVcc; all are four-jet Higgs boson candidates.
Two of them have reconstructed Higgs boson mass above 105 \GeVcc\
while the other has mass $97.4$ \GeVcc. The  value from the ANN is higher
than 0.8  for all three events, which were all  collected
at a centre-of-mass energy of 206.6 \GeV.
The first candidate 
was reconstructed 
with a Higgs boson mass of  110.7 \GeVcc\ and with an ANN value of 0.85.
The pairing selected corresponds to jet pair masses after the 4C fit 
(i.e. before the Z mass is fixed to its central value) 
of 105.1 and 96.8 
\GeVcc. In the high mass jet pair one of the jets has a
high  ${\mathrm b}$-tagging value
with a clear secondary vertex; in the low mass 
jet pair there is also a jet with a high ${\mathrm b}$-tagging value.
The third and fourth 
jets, ordered by ${\mathrm b}$-tagging,  have values  which have 
probabilities of 25$\%$ and 67$\%$ to occur in a four-b event. 
The second candidate 
 was reconstructed 
with a Higgs boson mass of  108.2 \GeVcc\ and with an ANN of 0.83.
The pairing selected corresponds to jet pair masses of 113.1 and 87.1 
\GeVcc. The two jets of the low mass jet pair have high ${\mathrm b}$-tag
values
with a secondary vertex in each  jet.
The third candidate
 was reconstructed with 
a Higgs boson mass of  97.4 \GeVcc\ and with an ANN of 0.96, the highest
of all the events collected in year 2000.
The two pairings with lowest $\chi^2$, after a 5C fit with the
Z mass fixed,   
had $\chi^2$ values of 3.58 and 3.97; they correspond to Higgs boson mass
estimators  of 
113.4 \GeVcc\ and 97.4 \GeVcc.
After applying the algorithm described in section~\ref{sec:1}
the second one was selected. The jet pair masses for such a pairing 
were 101.3 and 87.0 \GeVcc. The two jets of the high mass pairing 
had  high ${\mathrm b}$-tag values with one clear secondary vertex in each jet,
while the jets of the low mass jet pair had ${\mathrm b}$-tagging values 
which have 
probabilities of 34$\%$ and 56$\%$ to occur in a four-b event.

\begin{figure}[htbp]
\begin{center}

\epsfig{figure=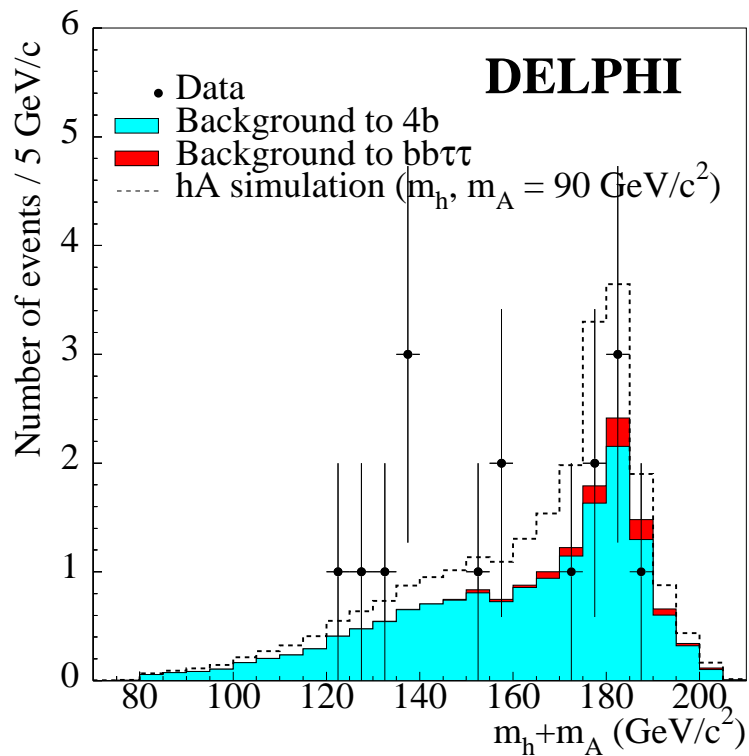,width=11.cm} \\
\vspace{-1.5cm}
\caption[]{
Distribution of the sum of the reconstructed Higgs boson masses
for the tight candidates
in the \hA\ channels in the 1999 and 2000 data.
Data (dots) are compared with  {\sc SM} background process expectations (full 
histograms)
and with the normalised signal spectrum added to the background channel 
contributions (dashed histogram). The mass hypothesis for the simulated signal
spectrum is
\mh=\MA~=~90~\GeVcc.
}
\label{fig:mass_pl2}
\end{center}
\end{figure}

The events selected by a  tight cut in the {\sc MSSM} 
Higgs boson search are shown in Fig.~\ref{fig:mass_pl2}.
The cut has been made on the second discriminant variable such that a
signal might be more apparent. 
The selections  require a minimal 
likelihood value of 0.90  in the \ttqq\  channel,
and a minimal neural network output of 0.95 in the
\bbbar\bbbar\ channel.
Both the 2000 and 1999 results are included.
The analysis of the \ttqq\ channel for 1999 is taken from our previous
publication~\cite{ref:pap99} for completeness.
The corresponding observed and expected rates in each period
are given in Table~\ref{ta:tight-ha}, which can be compared with
Table~\ref{ta:ha_all}.

\begin{table}[htbp]
\begin{center}
\begin{tabular}{cccccc}     \hline
Channel           & Data & Total  & \qqg  & 4 fermion & Efficiency (\%)\\
                  &      & background        &       &           &    \\ 
\hline
\multicolumn{6}{c}{1999 data, 191.6 - 201.7~\GeV } \\ 
\hline \hline
Tau               &   0  &   0.78$\pm$0.06& 0.13 &  0.65 &   12.8 \\
\hline
Four-jet          &   6  &   7.82$\pm$0.2&   1.7&    5.7&    55.0 \\
\hline
\multicolumn{6}{c}{First period 2000 } \\ 
\hline \hline
Tau               &   0  &   0.4$\pm$0.05& 0.04 &  0.38 &   11.5 \\
\hline
Four-jet          &   7  &    5.7$\pm$0.2&   1.7&    4.1&    55.0 \\
\hline
\multicolumn{6}{c} {Second period 2000} \\ 
\hline \hline
Tau               &   0  &  0.11$\pm$0.03&  0.01&  0.11&    11.0 \\
\hline
Four-jet          &  3  &    1.8$\pm$0.06& 0.41&     1.36&    50.3 \\
\hline
\end{tabular}
\caption[]{
Candidates selected for the {\sc MSSM} channels by tight cuts  on data and
 simulated  background processes.
The last column gives the efficiencies at
a mass \mh = \MA = 90~\GeVcc.
The tau results from 1999 are taken  from our previous publication.
}
\label{ta:tight-ha}
\end{center}
\end{table}

The three events with highest significance for the 
hypothesis \MA=90~\GeVcc\
at \tbeta~=~20.6 (corresponding to \mh~$\sim$90~\GeVcc)
are described below. The three events are from the 4b channel.

 The first event, collected at a centre-of-mass energy of 206.6~\GeV,
is reconstructed with a sum of Higgs boson masses of 120.9~\GeVcc\ and 
with an ANN of 0.99.
Its content in b-quarks is high, the jet with the 
lowest ${\mathrm b}$-tag value having a probability of 70\% to have such a
value in a four-b event.  
However, in the three possible jet pairings, the mass difference 
between any two jet pairs is well within the resolution
expected on this variable. The differences are 8.8, 4.4, and 5.1~\GeVcc,
corresponding to a sum of masses of 178.2, 120.9 and 202.3~\GeVcc, 
respectively.  
If instead of the minimal mass difference between the two jet pairs, 
the $\chi^2$ of a five-constraint fit
imposing equal masses of the two jet pairs 
is used as a criterion to pair the jets, the high mass solution would be 
selected with a $\chi^2$ of 5.7, while the low mass solution
has a $\chi^2$ of 7.0 and the third combination, the one closest to the
$ZZ$ hypothesis, has a $\chi^2$ of 7.9. 
Although this event 
appears to be a good 4b candidate, it has
an ambiguity in the mass estimation which allows an almost
equally good interpretation as an on-shell \ZZ\ candidate, or as a 
\ZZ$^*$ candidate with the Z$^*$ far below or above its nominal mass.

The second event, collected at a centre-of-mass energy of 205.1~\GeV,
is reconstructed with a
sum of Higgs boson masses of 180.5~\GeVcc\ and with an ANN of 0.97.
The jet pairings which are not selected have a much larger $\chi^2$
for an equal mass hypothesis than that which is used, suggesting that
the event is likely to be due to \ZZ\ production.

The third event, taken at a centre-of-mass energy of 206.6~\GeV\
is reconstructed with a
sum of Higgs boson masses of 178.6~\GeVcc\ and with an ANN of 0.96.
It has three well reconstructed secondary vertices, which explains the
high value of the jet ${\mathrm b}$-tagging variables and hence that of ANN.
However, as in the first candidate, it has two pairings almost equally 
probable, with mass differences of 
9.3 and 19.4~\GeVcc, corresponding to a sum of masses of
178.6 and 201.9~\GeVcc, respectively. The $\chi^2$ 
of a five-constraint fit is 14.3 for the pairing close
to the \ZZ\ hypothesis, and 14.5 for the high-mass solution.

\subsection{The  {\sc \bf SM} Higgs boson}
\label{sec:smresults}

Confidence levels as a function of the {\sc SM} Higgs boson mass are derived,
 combining
the data analysed in the previous sections with those taken at energies
from 161.0 to 202~\GeV~\cite{ref:pap96,ref:pap97,ref:pap98,ref:pap99}.
The expected cross-sections and branching ratios are taken 
from the database provided by the LEP Higgs working group,
using the {\tt HZHA}~\cite{ref:hzha} package
with the top mass set to 174.3~\GeVcc. 
As noted earlier, the \hnn\ and \hqq\ channels each use two analyses, one
for most of the range and the other optimised for the kinematic limit.
The selection between these is done independently for each energy window
and each mass hypothesis under consideration.

\begin{figure}[htbp]
\begin{center}
\epsfig{figure=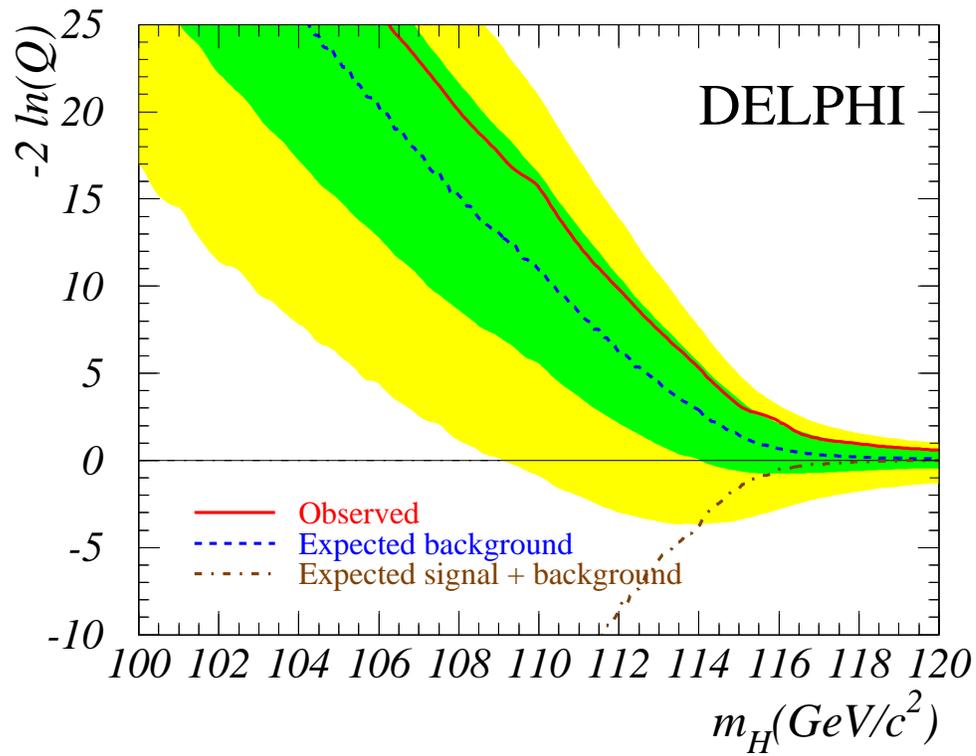,width=14.cm} \\
\caption[]{
 {\sc SM} Higgs boson: test-statistic $-2\ln \likear$ for each \MH\ hypothesis
in data (solid) and its
expected median value in background-only experiments (dashed). 
The bands correspond to the 68.3\% and 95.0\% confidence intervals
from experiments with only background processes.
The dash-dotted curve shows the expected mean value if a signal were
present; the error bands on this would be rather similar in width
to those on the background-only curve at the same mass.
}
\label{fig:xi_sm}
\end{center}
\end{figure}

\begin{figure}[htbp]
\begin{center}
\vspace*{-0.2cm}
\epsfig{figure=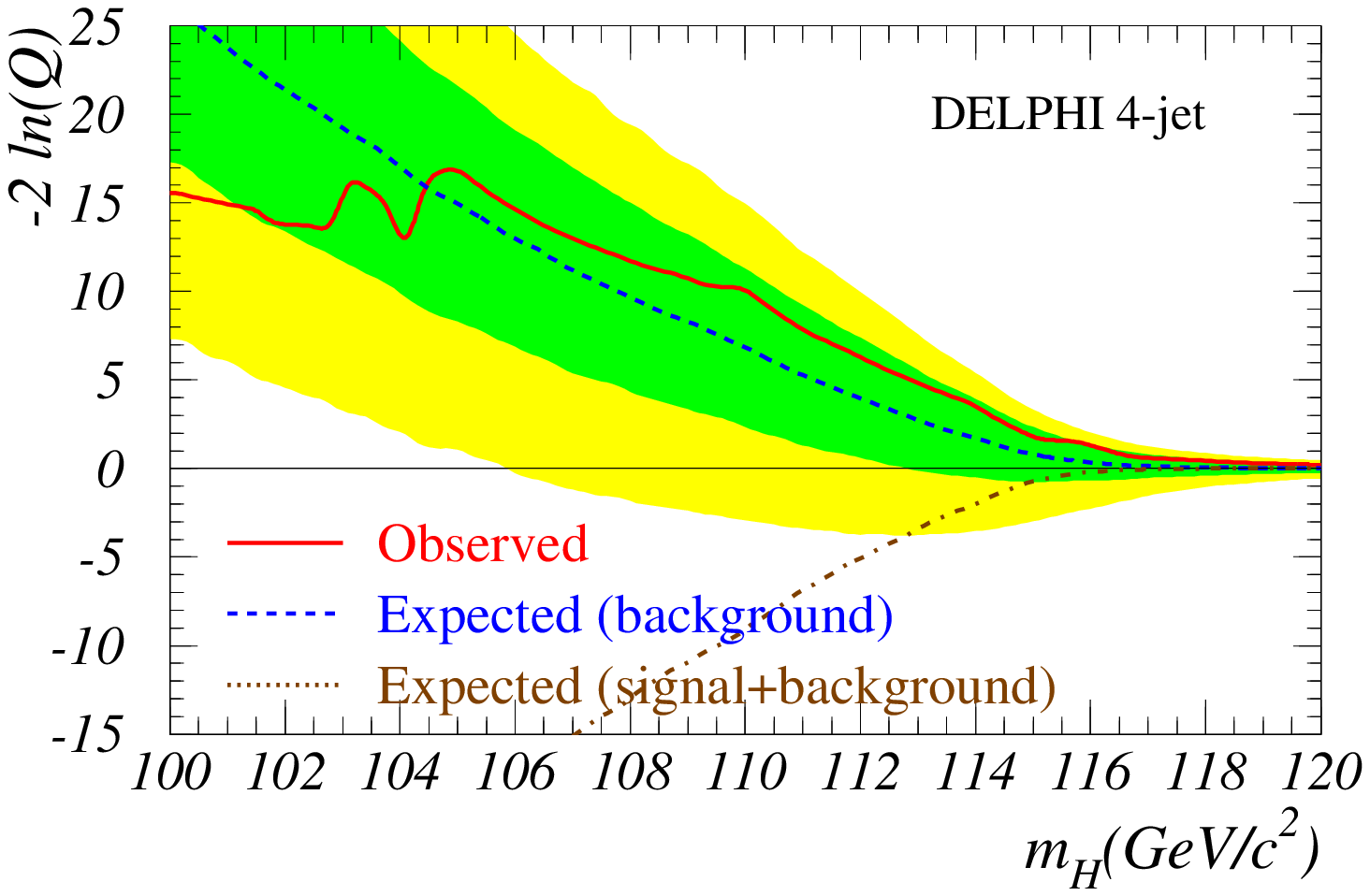,width=7.5cm}
\epsfig{figure=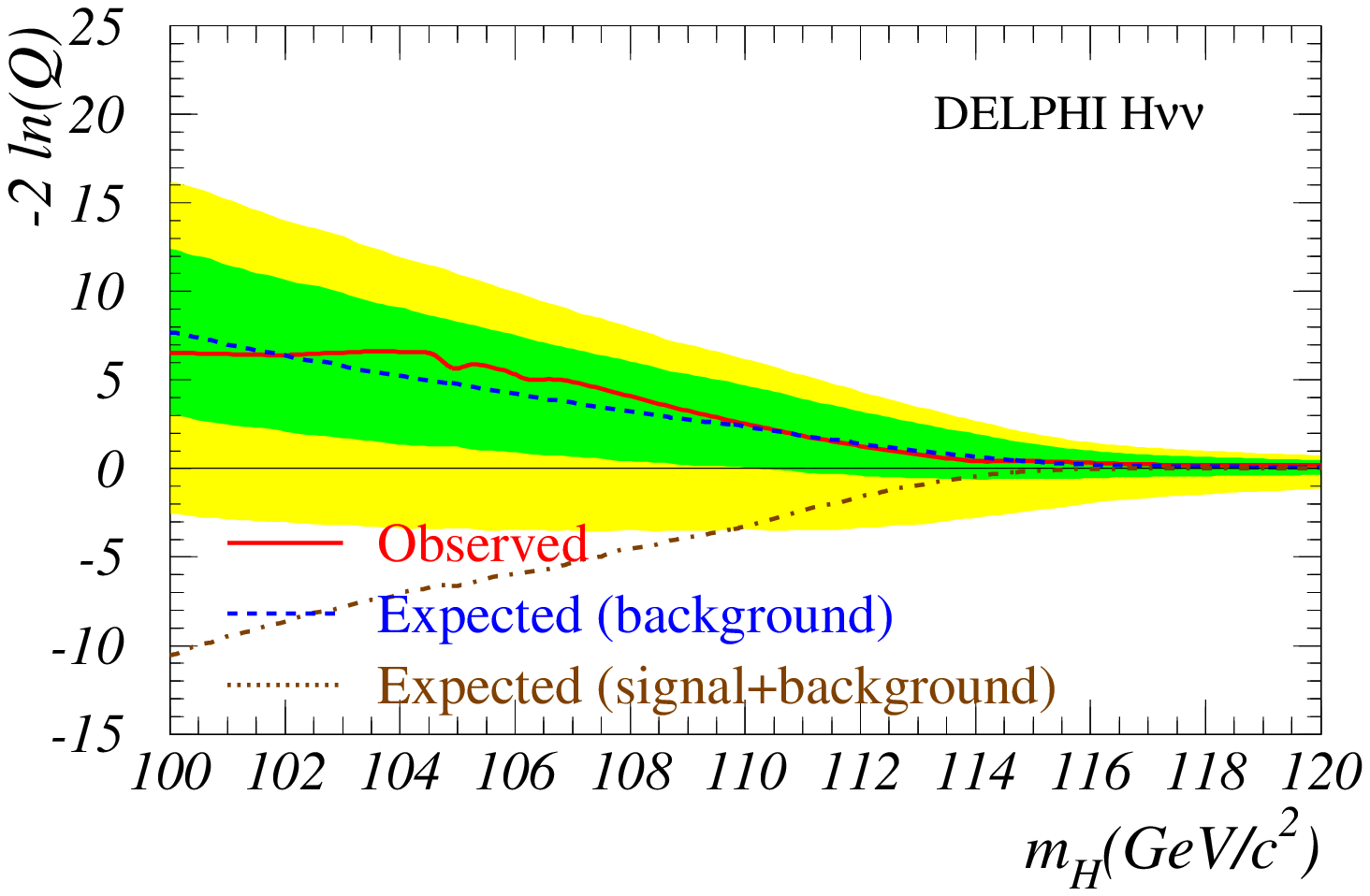,width=7.5cm}
 \\
\epsfig{figure=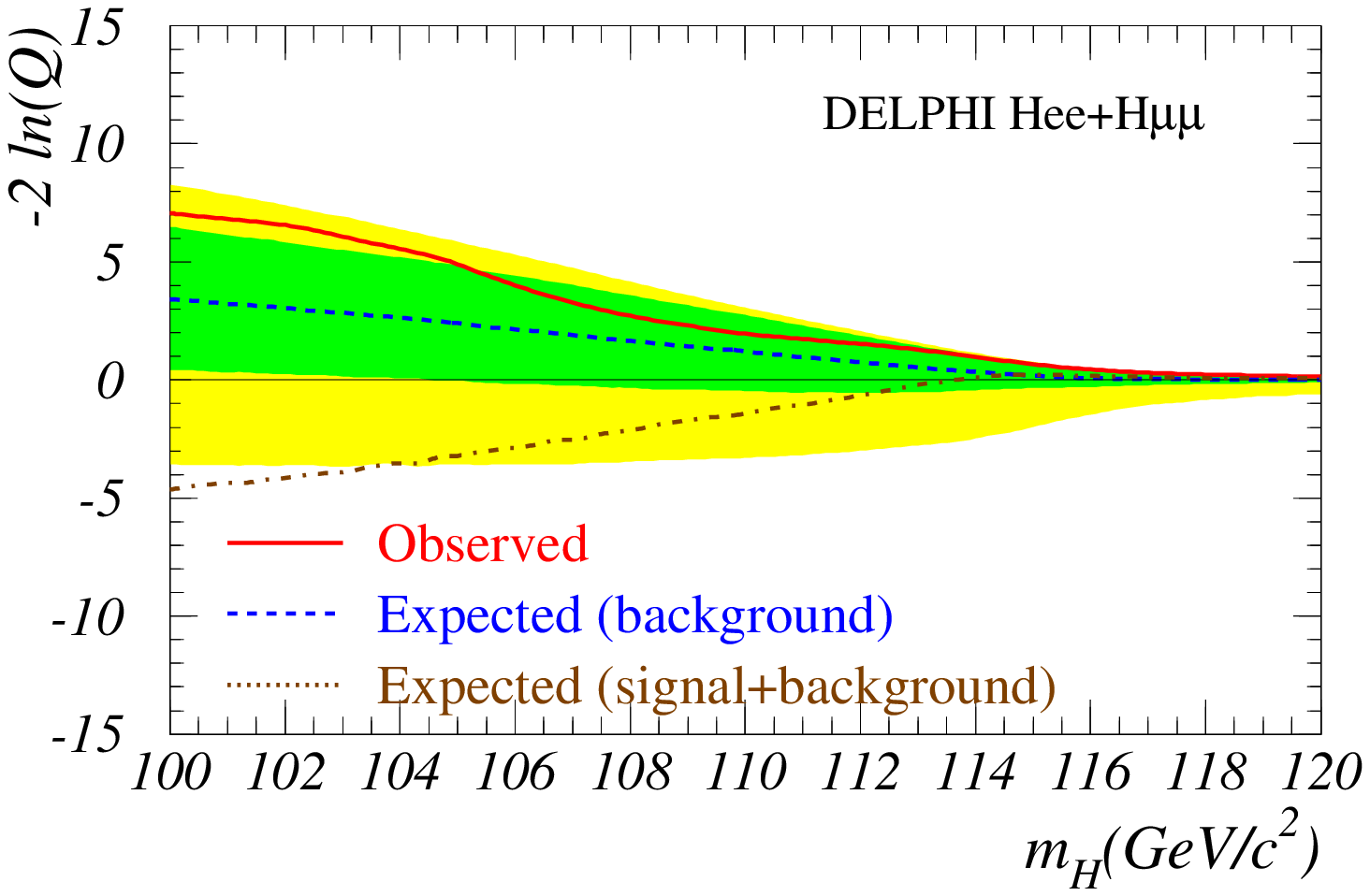,width=7.5cm}
\epsfig{figure=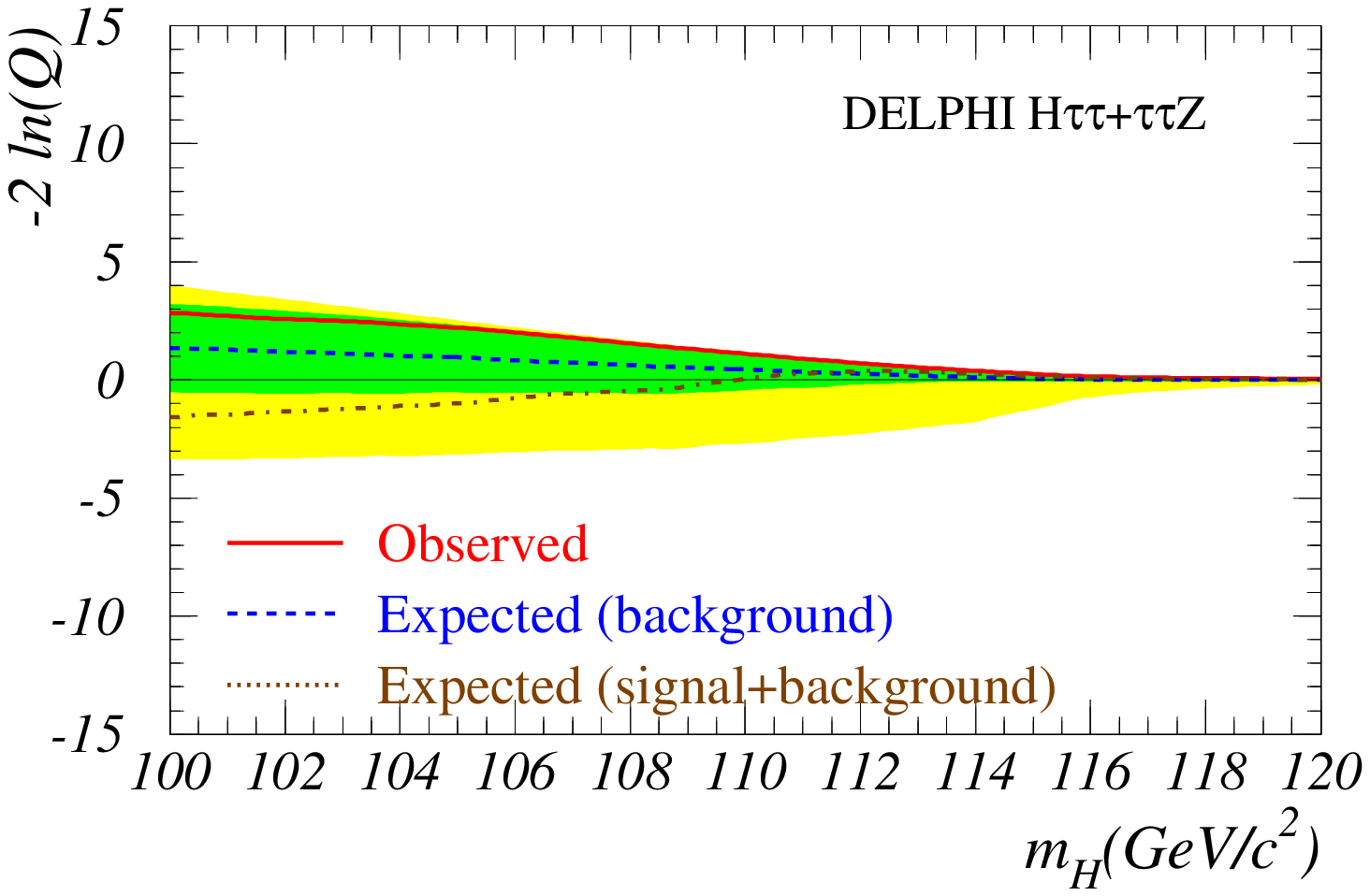,width=7.5cm}
 \\
\vspace*{-0.4cm}
\caption[]{
 {\sc SM} Higgs boson: test-statistic $-2\ln \likear$ for each \MH\ hypothesis
in four channels: hadronic, \hnn, {\Hz}ll and channels with tau leptons.
The conventions are as in figure~\ref{fig:xi_sm}.
No channel produces a signal-like result at any mass.
}
\label{fig:xi_sm4}
\end{center}
\end{figure}

  The curve of the test-statistic \likear\ as a function of 
the mass hypothesis is 
shown in Fig.~\ref{fig:xi_sm}, where the observation is compared with the
xpectations from experiments with only background processes  and from
experiments where both signal and background channels exist.
Over the whole range of masses, 
the test-statistic remains positive, while in the event of a discovery  
it would be negative for mass hypotheses close to the actual mass of the 
signal.

The same curve is shown in Fig.~\ref{fig:xi_sm4}, except that 
the test-statistic \likear\ is broken into 4 channels,
four jets, \hnn, \hee\ plus \hmm\ and tau channels.
In the region abouve 110~\GeVcc\ the \hnn\ channel has a result 
which is median for background, while the other three have a
slight deficit 
compared with the background, of the order of one sigma.

\begin{figure}[htbp]
\begin{center}
\epsfig{figure=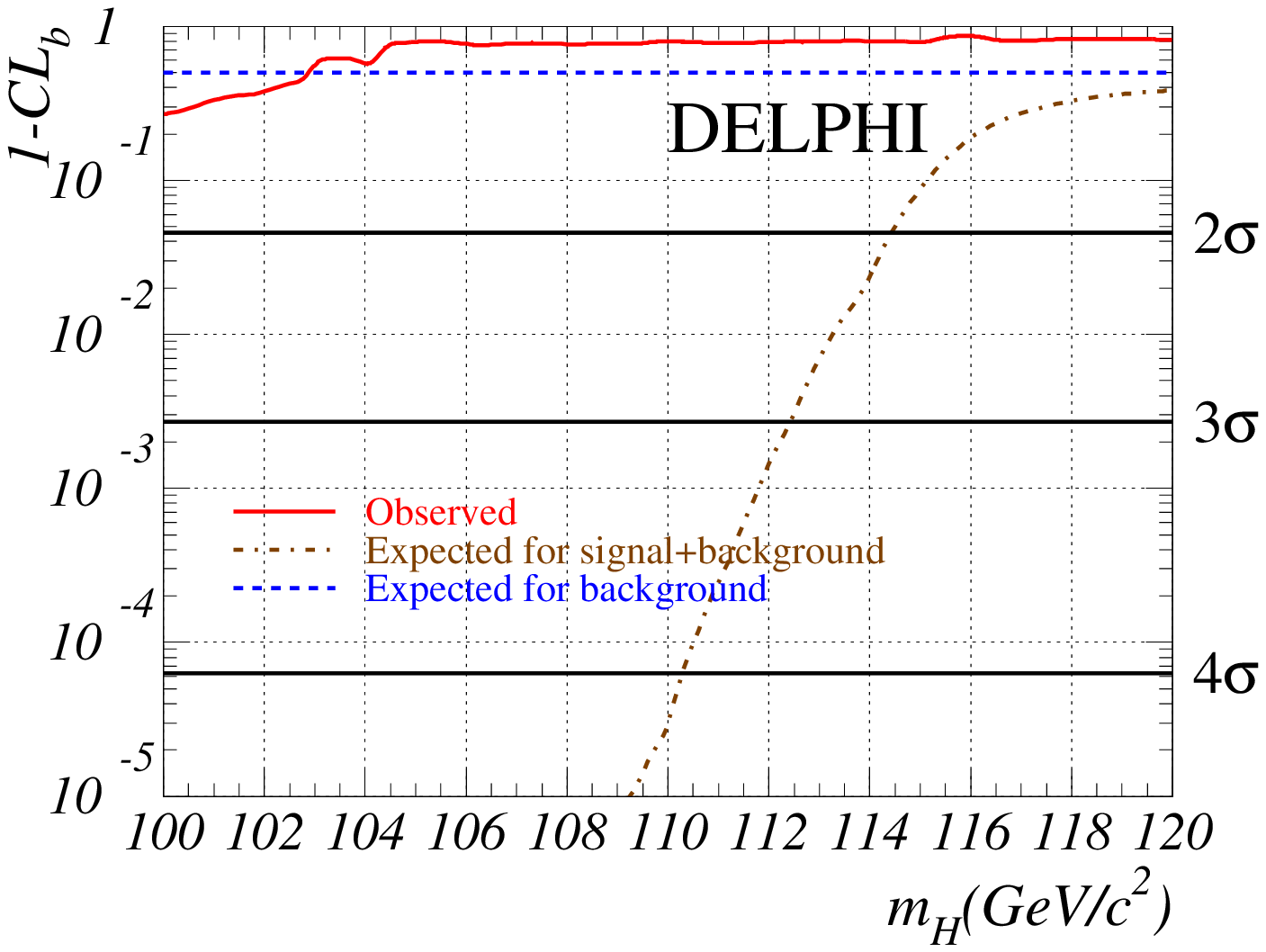,width=14.cm}\\
\epsfig{figure=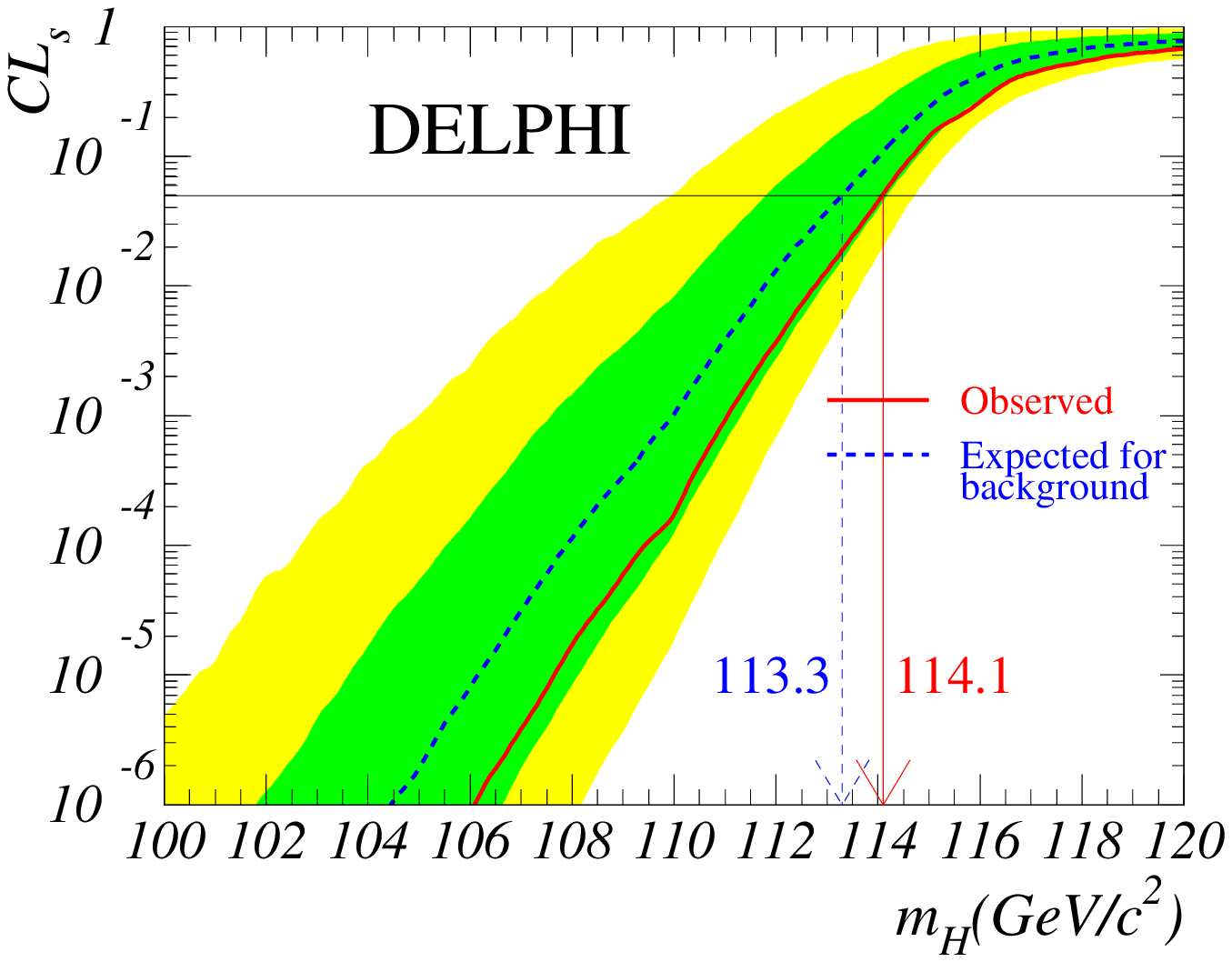,width=14.cm}
\caption[]{
 {\sc SM} Higgs boson: confidence levels as a function of \MH.
Top:  1-\CLb\  for the background hypothesis. 
The full curve is the observation, 
the dashed curve is the median expected for background only, 
and the dash-dotted curve is the median expected at a particular
\MH\ value when tested for that \MH value.
A signal would appear
as a  downward deviation.
Bottom: \CLs, the pseudo-confidence level 
for the signal hypothesis. 
Curves are the observed (full) and expected median 
(dashed) confidences from experiments with only background channels
while the bands correspond to the 68.3\% and 95.0\% confidence intervals 
for the hypothesis of only background processes.
The intersections of the 
curves with the horizontal line at 5\% define the 
expected and observed 95\% CL lower limits on \MH.
}
\label{fig:cl_sm}
\end{center}
\end{figure}

Curves of the confidence level \CLb\ and \CLs\ (as defined in
section~\ref{sec:limits}) as a function of
the test mass \MH\ are shown in Fig.~\ref{fig:cl_sm}.
In the presence of a sizable Higgs signal,
the value of the observed \CLb\ (top of Fig.~\ref{fig:cl_sm}) 
would approach one, since it measures the fraction of 
experiments with only background processes which are more background-like
than the observation. 
Here the compatibility between the observation and the expectation 
from background processes is well within one standard deviation over 
the range of masses tested. 
The pseudo-confidence level in the signal is shown in Fig.~\ref{fig:cl_sm} 
(bottom).
The observed 95\% {\sc CL} lower limit on the mass is 114.1~\GeVcc\
while  the expected median limit is $113.3~\GeVcc$.
If the  Higgs boson mass was below  107.5~\GeVcc\ this search
would produce an  expected 5$\sigma$ discovery; all such masses are 
instead excluded at 99.999\% {\sc CL} or better.

It is possible to  calculate the Bayesian credibility of these
(essentially frequentist) 95\% CL limits. This is the probability that 
our results are correct, that is to say that the true Higgs boson mass is
greater
than our limit.
Like any Bayesian probability this needs
a prior belief. Working within the framework of the Standard Model
two interesting
priors are to take a probability flat in $\log m_H $ up to 1~TeV, or
the same but modified by the electroweak fits results as they are 
currently known~\cite{ref:Amsterdam2002} {(i.e. $m_H =  81^{+52}_{-33}$).}
The posterior probability density function is obtained by 
   multiplying the prior by the likelihood distribution from this experiment
and normalising. 
The credibilities when integrating  from the quoted limit to 1~TeV 
 are 99.97\% for the flat case and
99.8\% when the electroweak fit results are considered.
These probabilities, of course, assume that there is exactly one Standard
Model Higgs. By construction the \CLs\ method will always produce 
large credibilities.

\subsection{Limits on the H coupling to Z(W)}
\label{sec:sm-cross-limit}

\begin{figure}[htbp]
\begin{center}
\epsfig{figure=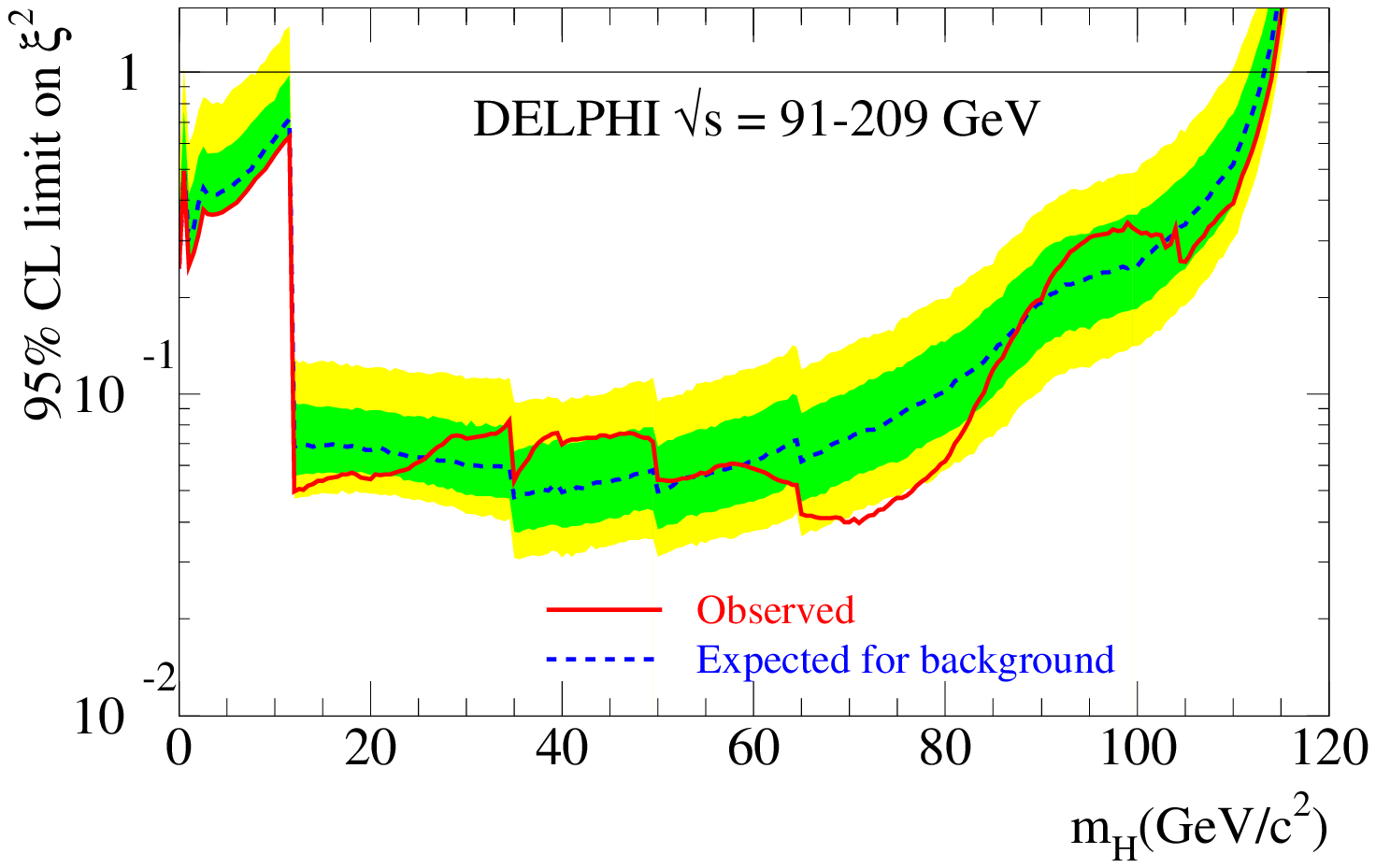,width=13cm}
\caption[]{
95\% CL upper bound on $\xi^2$, 
where $\xi$ is the HVV (V=\W\ or \Zz) coupling normalised to that
in the {\sc SM}, assuming {\sc SM} branching fractions for the Higgs 
boson. The limit
observed in data (full curve) is shown together with the expected median limit
in background process experiments (dashed curve).  The bands correspond to the 
68.3\% and 95.0\% confidence intervals from background-only experiments.
The limits are significantly less stringent below 12~\GeVcc, the \bbbar\
threshold, where only results obtained on subsets of the LEP 1 data are used.
}
\label{fig:ghvv}
\end{center}
\end{figure}

  In a more general approach, the results of the searches for a 
{\sc SM} Higgs boson can be used to set a 95\% CL upper bound 
on the Higgs boson production as a function of its mass.
Here it is assumed that the Higgs boson decay properties are identical to
those in 
the {\sc SM} but that the Higgs boson couplings to pairs of 
\Zz\ and \W\ bosons (the latter arising in the \WW\ fusion production
mechanism) may be smaller. To achieve the best sensitivity over
the widest range of mass hypotheses, the results described in this paper 
are combined consistently with those obtained at lower energies at 
LEP2~\cite{ref:pap99,ref:pap98,ref:pap97,ref:pap96}, as well as with those
obtained at 
LEP1~\cite{ref:paplep1} which covered masses up to 60~\GeVcc.
For each mass hypothesis, the production 
cross-section is decreased with respect to its {\sc SM} value
until a pseudo-confidence level \CLs\ of 5\% is obtained.  

The coupling  $\xi$ is introduced as the HVV (V=\W\ or \Zz) coupling
normalised to that
in the {\sc SM}, assuming {\sc SM} branching fractions for the
Higgs.\footnote{The Higgs Lagrangian could include a term of the form:
${\cal L}= g^3_{HZZ}HZ_\mu Z^\mu ,$ 
where $g^3_{HZZ} = 0.5gm_W\delta_Z$ and   $\xi\equiv(1+\delta_Z)$, which
vanishes in the case $\delta_Z = 0$.}
In practice, $\xi$ is dominated by the HZZ coupling.
The  95\% CL upper bound on $\xi^2$ is shown in Fig.~\ref{fig:ghvv} for
masses of the Higgs boson from 0 to 120~\GeVcc. 
The {\sc SM} result described in the previous section corresponds to a
ratio of 1.

\subsection{Neutral Higgs bosons in the  {\sc MSSM}}

  The results in the \hZ\ and \hA\ channels reported in
the previous sections are combined with the same statistical 
method as for the {\sc SM}, also using earlier results at LEP2 
energies~\cite{ref:pap99,ref:pap98,ref:pap97,ref:pap96,ref:pap95},
to derive confidence levels in scans of the {\sc MSSM} parameter
space. The exclusion limits
obtained at LEP1~\cite{ref:mssmlep1} (\mh$>$44~(46)~\GeVcc\ when \mh\ 
is above (below) the ${\mathrm{AA}}$ threshold) are
used as external constraints to limit the number of points in
the scans.

\subsubsection{The benchmark scenarios}
\label{sec:benchmark}

   At tree level, the production cross-sections and the 
Higgs branching fractions in the  {\sc MSSM}
depend on two free parameters, \tbeta\ and one Higgs 
boson mass, or, alternatively, two Higgs boson masses, e.g. \MA\ and \mh. 
Radiative corrections introduce additional parameters related
to supersymmetry breaking.
Hereafter, the usual assumption that some of them
are identical at a given energy scale is made: hence, 
the SU(2) and U(1) gaugino mass terms are assumed to be
unified at the so-called GUT scale, 
while the sfermion mass terms or the squark trilinear
couplings are assumed to be unified at the EW scale.
Within these assumptions, the parameters beyond tree level are:
the top quark mass, the Higgs mixing parameter, $\mu$, 
the common sfermion mass term at the EW scale, $M_{\rm susy}$,
the SU(2) gaugino mass term at the EW scale, $M_2$,
the gluino mass, $m_{\tilde{g}}$,
and the common squark trilinear coupling at the EW scale, $A$.
The U(1) gaugino mass term at the EW scale, $M_1$, is 
related to $M_2$ through the GUT relation
$M_1 = (5/3) {\rm \tan}^2\theta_W M_2$.
The radiative corrections affect the relationships between the masses of 
the Higgs bosons, with the largest contributions arising from the 
top/stop loops. 
As an example, the h boson mass, which is below that of the Z boson
at tree level, increases by a few tens of \GeVcc\ in some regions
of the {\sc MSSM} parameter space due to radiative corrections.

\begin{table}[htbp]
{\small
\begin{center}
\begin{tabular}{c|cccccc}     \hline
scenario & \mtop\   & $M_{\rm susy}$ & $M_2$ & $m_{\tilde{g}}$ & $\mu$ 
         & $X_t$ \\
& (\GeVcc) & (\GeVcc) & (\GeVcc) & (\GeVcc) & (\GeVcc) & (\GeVcc)\\
\hline
\mbox{$ m_{\mathrm h}^{\rm max}$} scenario &
     174.3 & 1000 & 200 & 800 & -200 & 2 $M_{\rm susy}$ \\
no-mixing &
     174.3 & 1000 & 200 & 800 & -200 & 0 \\
large $\mu$ &
     174.3 & 400 & 400 & 200 & 1000 & -300 \\
\hline
\end{tabular}
\caption[]{
Values of the underlying parameters for the three 
representative {\sc MSSM} scenarios scanned in this paper.
Note that  $X_t$ is $ A - \mu \cot \beta$.
}

\label{ta:benchmarks}
\end{center}
}
\end{table}

 In the following, three benchmark scenarios are considered, as suggested 
in Ref.~\cite{ref:new_pres}. These are quite representative 
since the limits obtained in these schemes with earlier results
were only slightly reduced in more general parameter scans~\cite{ref:pap99}.
The first two scenarios, called the
\mbox{$ m_{\mathrm h}^{\rm max}$} scenario and the no-mixing scenario,
rely on radiative corrections computed at partial two-loop order 
as in~Ref.\cite{ref:FDradco}. The values of the underlying parameters are
quoted in Table~\ref{ta:benchmarks}.
The two scenarios differ only by the value of $X_t = A - \mu \cot \beta$, 
the parameter which controls the mixing in the stop sector, and hence
has the largest impact on the mass of the h boson.
The \mbox{$ m_{\mathrm h}^{\rm max}$} scenario leads to the maximum
possible h mass as a function of \tbeta. The no-mixing
scenario is its counterpart with vanishing mixing, leading to upper
bounds on \mh\ which are at least 15~\GeVcc\ lower than in the
\mbox{$ m_{\mathrm h}^{\rm max}$} scheme.

 The third scenario, called the large $\mu$ scenario, predicts at least 
one scalar Higgs boson with a mass within kinematic reach at LEP2 
in each point of the MSSM parameter space. However, there are regions 
for which the Higgs bosons cannot be detected because of  
vanishing branching fractions into b-quarks.
In this scenario, the radiative corrections are computed  
as in Ref.~\cite{ref:radco}.
The values of the underlying parameters are given in 
Table~\ref{ta:benchmarks}. The main difference with the
two previous schemes is the large and positive value of $\mu$
and the relatively small value of $m_{\tilde{g}}$.

 It must be noted that, with respect to the calculations 
of Ref.~\cite{ref:FDradco,ref:radco} used in this paper, 
recent theoretical improvements exist that include
more complete two-loop order radiative corrections and a redefinition
of the underlying parameters of the benchmark scenarios, that lead
in particular to an extended allowed range of the h boson 
mass Ref.~\cite{ref:new_calc}. 
These changes will probably reduce the excluded region in $\tan\beta$.

\subsubsection{The procedure} \label{procedure}

  In the three benchmark scenarios, a scan was performed over the {\sc MSSM} 
parameters \tbeta\ and \MA. The range in \MA\ spans from 12~\GeVcc, 
the minimal value which has been searched for at LEP2 in the DELPHI
analyses, up to the maximal value allowed by each scenario~\cite{ref:new_pres},
that is up to $M_{\rm susy}$, which is 1~TeV/$c^2$ in the 
\mbox{$ m_{\mathrm h}^{\rm max}$} 
and no-mixing schemes, and 400~\GeVcc\ in the large $\mu$ scenario
(see Table~\ref{ta:benchmarks}).
The range in \tbeta\ extends from the minimal value allowed in each 
scenario \footnote{The minimal value of \tbeta\ is 0.7 in the large
$\mu$ scenario and 0.4 in the other two  schemes. For lower values, some 
parameter combinations give rise to unphysical negative mass squared values.}
 up to 50, a value chosen in the vicinity of the ratio of 
the top- and b-quark masses, which is an example of the large \tbeta\
hypothesis favoured in some constrained {\sc MSSM} models~\cite{ref:susygut}.
The scan steps were 1~\GeVcc\ in \MA\ and 0.1 in \tbeta\ in the regions 
where \mh\ varies rapidly with these parameters. 

  At each point of the parameter space, the \hZ\ and \hA\ cross-sections 
and the Higgs branching fractions were taken from  databases provided by
the LEP Higgs working group, Ref.~\cite{ref:hwg}, on the basis of the
theoretical calculations in Refs.~\cite{ref:FDradco,ref:radco}.
The signal expectations in each channel were then derived from the 
theoretical cross-sections and branching fractions,
the experimental luminosity and the efficiencies. A correction
was applied to account for different branching fractions of the
Higgs bosons into \bbbar\ and \toto\ between the test point 
and the simulation (e.g.~for the \hZ\ process, the simulation was done in the 
{\sc SM} framework). 
   For the hA channels, to account for non-negligible widths of the h and 
A bosons at large \tbeta\ 
the set of efficiencies from the \mh, \MA\ 
simulations was applied for \tbeta~$<$ 30. Above that value, efficiencies
were linearly interpolated in \tbeta\ between the efficiencies from the 
\mh, \MA\ simulations and those from the simulations at \tbeta~=~50.
As the Higgs boson widths grow approximately linearly with 
\tbeta\ above 30, a linear interpolation is valid. 
The same holds for the discriminant information, for which 
the same interpolation software was used as discussed in
section~\ref{sec:limits}
for the PDF interpolation in mass or centre-of-mass energy. 

  Finally, when combining the results in all channels to derive confidence
levels, only independent channels must be included, which requires some 
special treatment for a few non-independent cases. 
As already mentioned in section~\ref{sec:htau},
the four \tautauqq\ signals, which were covered by the same analysis, were
thus combined into one global \tautauqq\ channel prior to the confidence
level computation. The same applies to the four signals selected by the
low mass \hZ\ four-jet analysis  - the \lhqq\ signal, the two (\hAA )(\Zqq )
signals and the 4b signal (see section~\ref{sec:admssm}) - 
or to the two signals selected by the \hA\ four-b analysis - 
the 4b signal and the \lhqq\ signal (see section~\ref{sec:admssm}). 
Moreover, in three cases, there was also a large overlap in the  
events selected by two different analyses: 
the low and high mass analyses in the missing energy channel or 
in the four-jet \hZ\ channel and the four-jet \hZ\ and \hA\ analyses.
In each case, only one of the two analyses was selected at each input point 
and at each centre-of-mass energy, 
on the basis of the smallest expected \CLs\ from experiments with no signal
(that is, on the basis of the strongest average exclusion if no signal 
is present).
This ensures that the channels which are then
combined in the global confidence level computations are independent.

\subsubsection{Consistency tests in the \hA\ channels}
 
\begin{figure}[htbp]
\begin{center}
\vspace{-0.9cm}
\epsfig{file=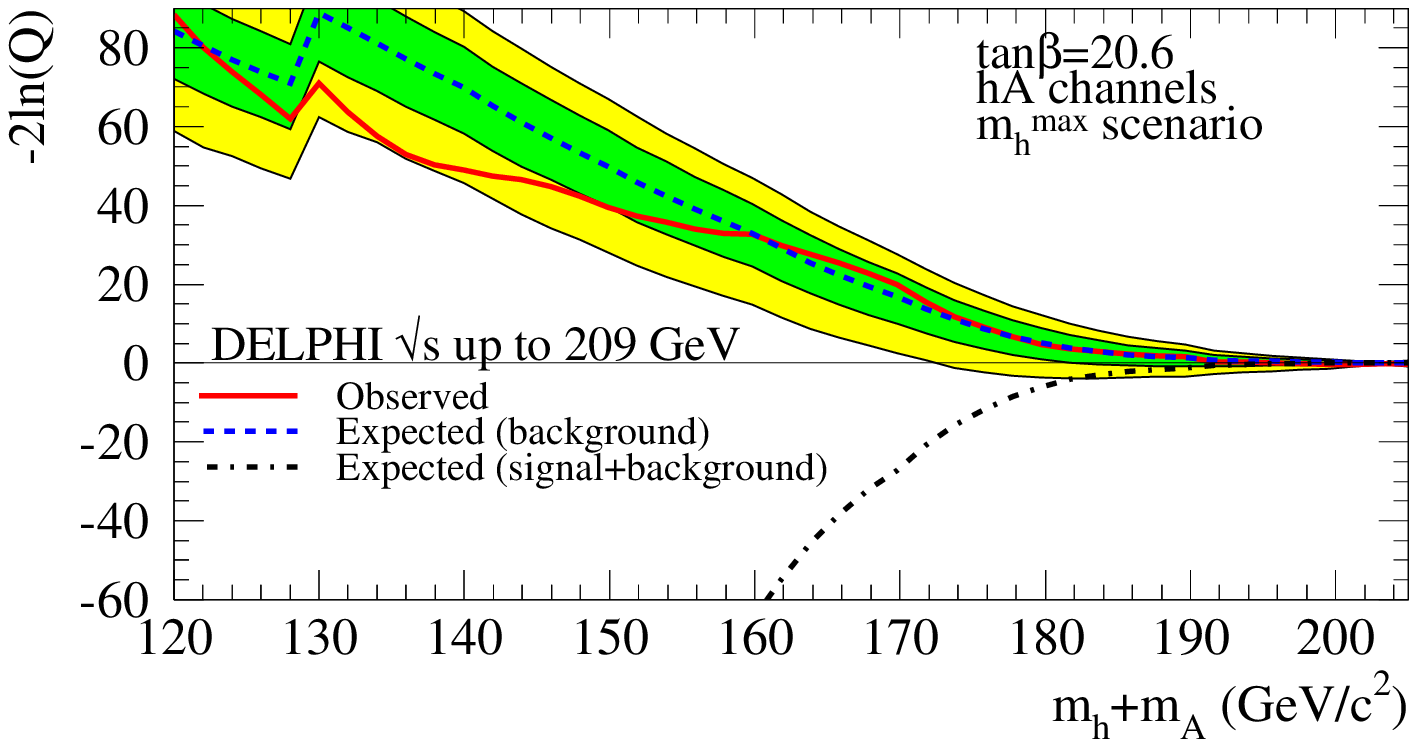,height=7.5cm} \\
\vspace{-0.1cm}
\epsfig{file=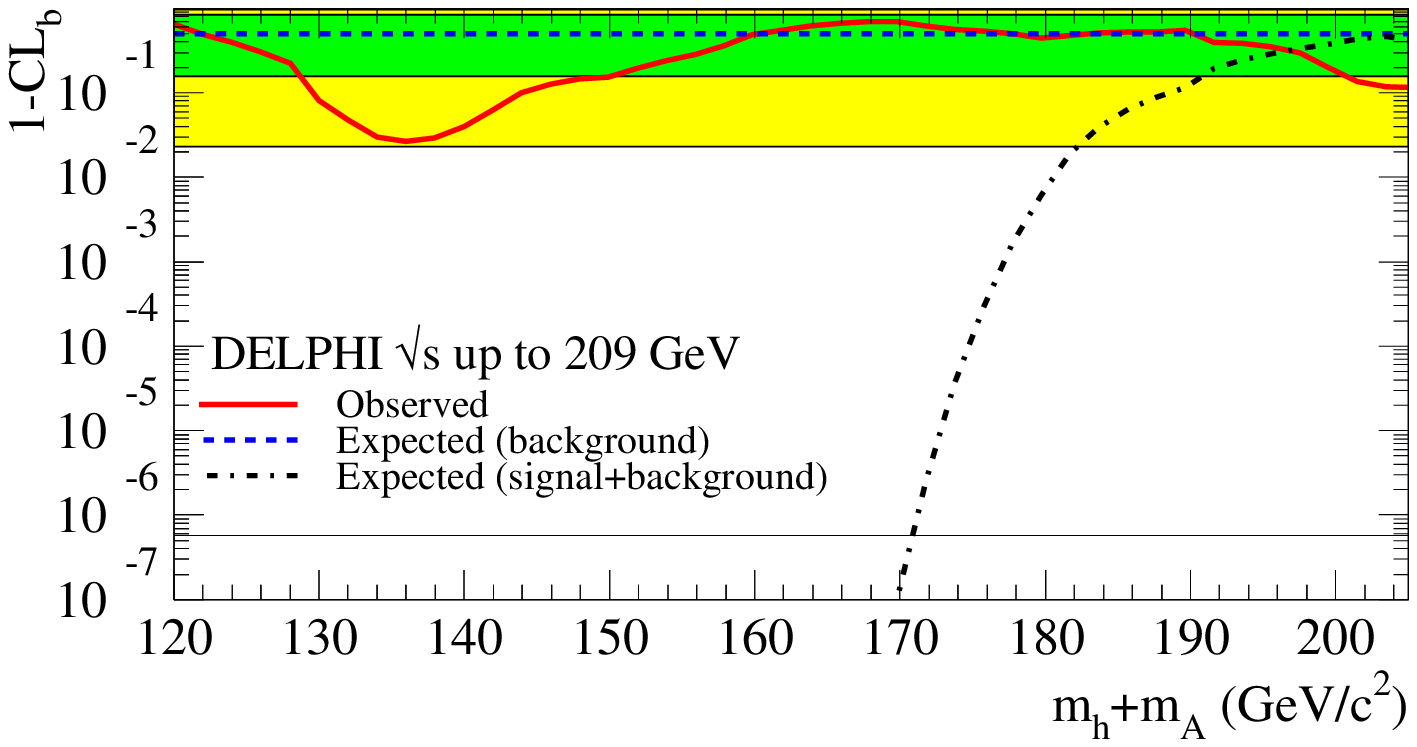,height=7.5cm} \\
\vspace{-0.1cm}
\epsfig{file=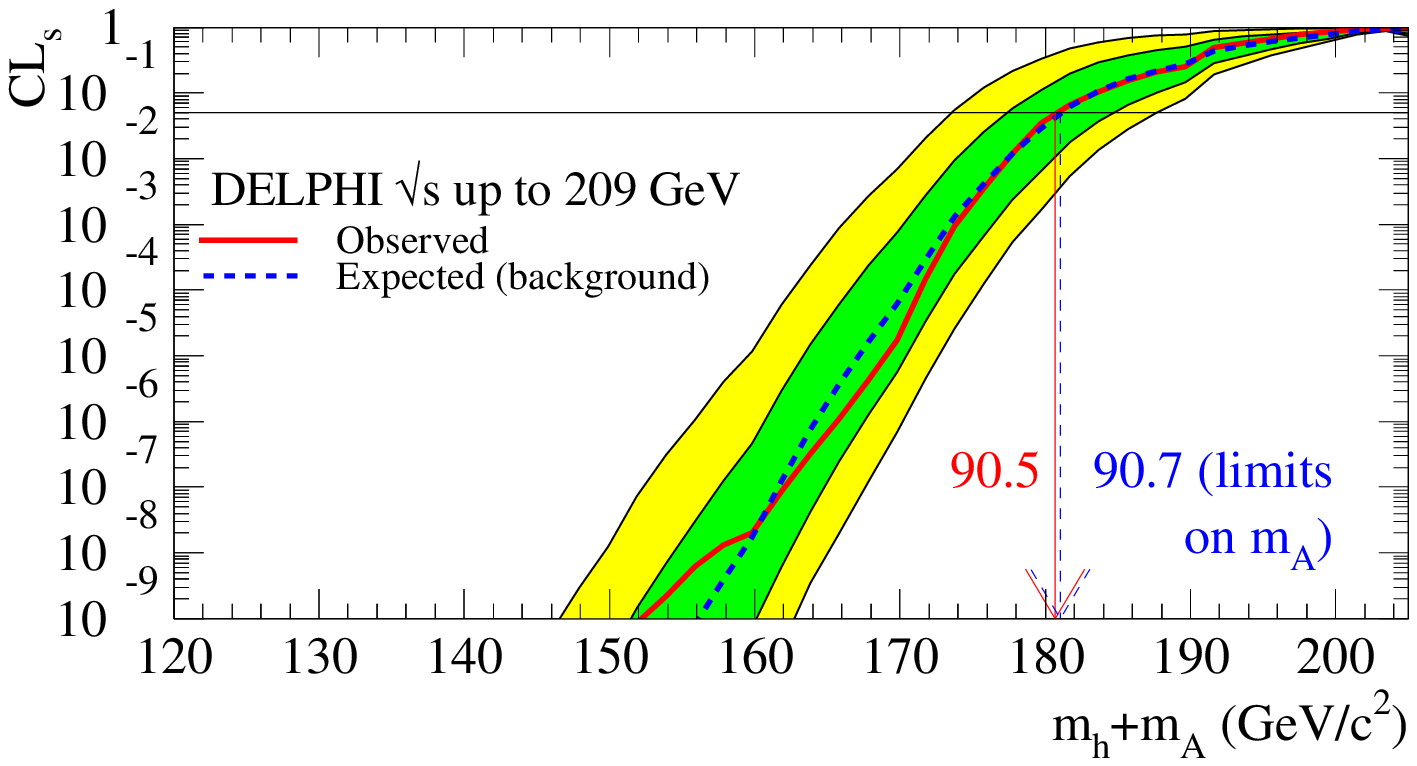,height=7.5cm}
\caption[]{
hA analyses: test-statistic (top) and 
confidence levels in the background-only hypothesis (middle) and
in signal hypothesis (bottom) as  functions of \mh+\MA.
Curves are the observed (full) and  median expected 
(dashed) results from background-only experiments
while the bands correspond to the 68.3\% and 95.0\% confidence intervals 
from the latter.
The dash-dotted curves are the expected mean values 
from experiments where a signal of mass given in the abscissa is
added to the background.
The limited range of the results derived at 188.7~\GeVcc\
explains the hook at 130~\GeVcc\ in the top plot.
}
\label{fig:cls}
\end{center} 
\end{figure}

Fig.~\ref{fig:cls} shows the curves of the test-statistic 
\likear\ and of the confidence levels \CLb\ and \CLs\ as a function of 
the test mass \mh+\MA, when using only the results of the 
two \hA\ analyses applied onto the two \hA\ signals. 
The signal cross-sections are from the \mbox{$ m_{\mathrm h}^{\rm max}$} 
scenario at \tbeta~=~20.6.
Over the whole range of test masses,
data are in reasonable agreement with the background process expectations.
For test masses \mh+\MA\ around 135~\GeVcc\ 
a two standard deviation effect is observed which is due to
the small excess of events in the 4b channel with reconstructed 
masses in that region, as seen in Fig.~\ref{fig:mass_pl2}. 
The mass resolution is around 5~\GeVcc, and therefore the region
120 to 205~\GeVcc\ shown contains some 15 or 20 effectively independent
points, and therefore it is not surprising  that there is a deviation
of this magnitude at some mass.
There is a small excess, only just over one sigma, at 200~\GeVcc, which
is not apparent in Fig.~\ref{fig:mass_pl2}.
It involves only 2 significant
events with reconstructed mass greater than 200~\GeVcc, and these have NN
values of 0.94 and 0.93 respectively, and so do not appear in the mass
plot.

Furthermore, the \CLs\ curves show that, in the particular
scenario under study, the exclusion 
limits on \mh+\MA, both observed and expected, are around 181~\GeVcc. The 
exclusion and discovery potentials rise fast when testing mass 
hypotheses below that value. As an example, the 5 sigma discovery potential 
reaches 171~\GeVcc\ in \mh+\MA (see 1-\CLb\ curves), only 10~\GeV\ below the limit
at 95\% {\sc CL}, and this sum of masses is experimentally excluded with a 
\CLs\ of $10^{-4}$ (see \CLs\ curves).

\subsubsection{Exclusion regions}

\begin{figure}[htbp]
\begin{center}
\begin{tabular}{cc}
\hspace{-1.2cm}
\epsfig{figure=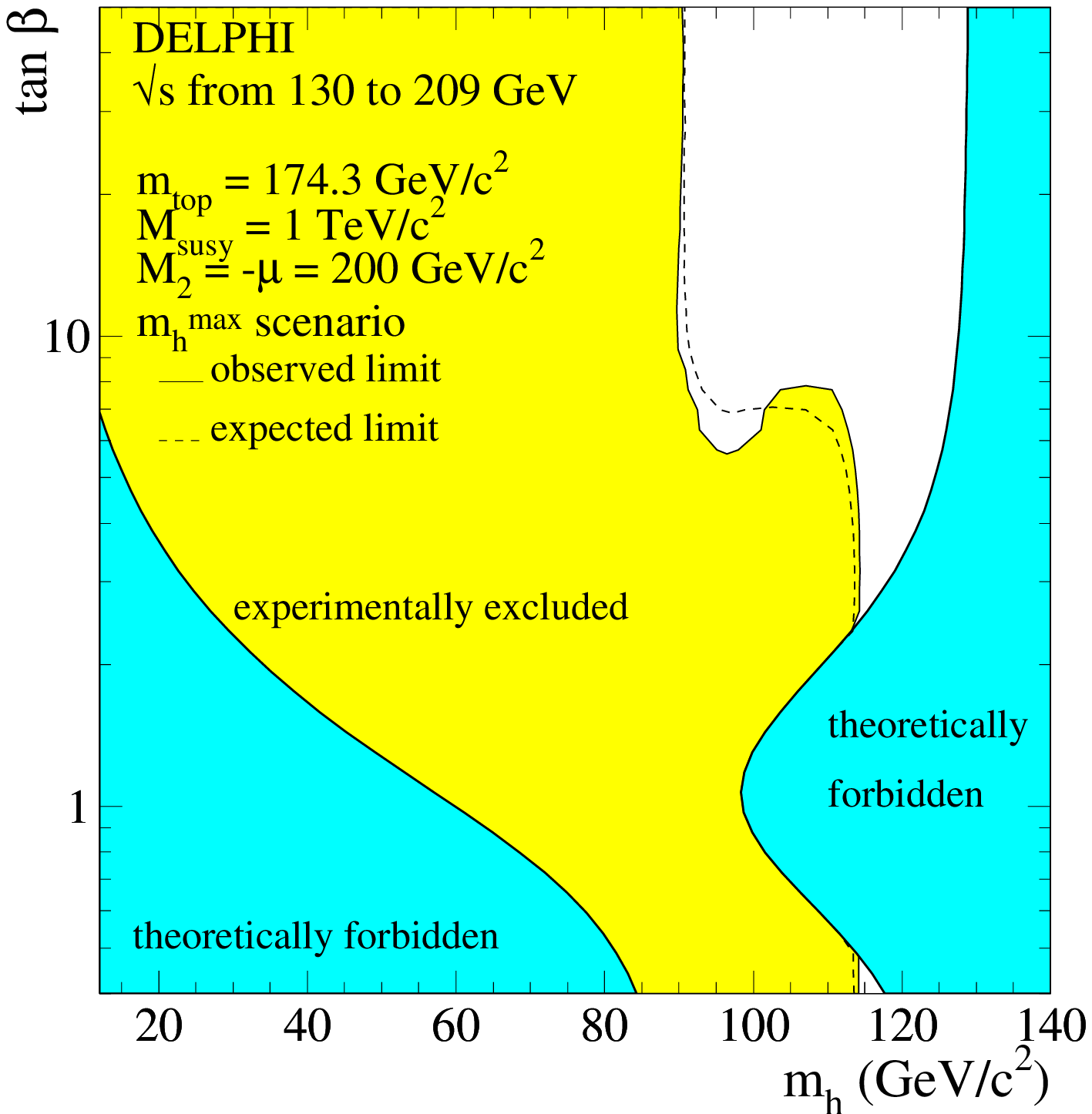,height=9cm} &
\hspace{-1cm}
\epsfig{figure=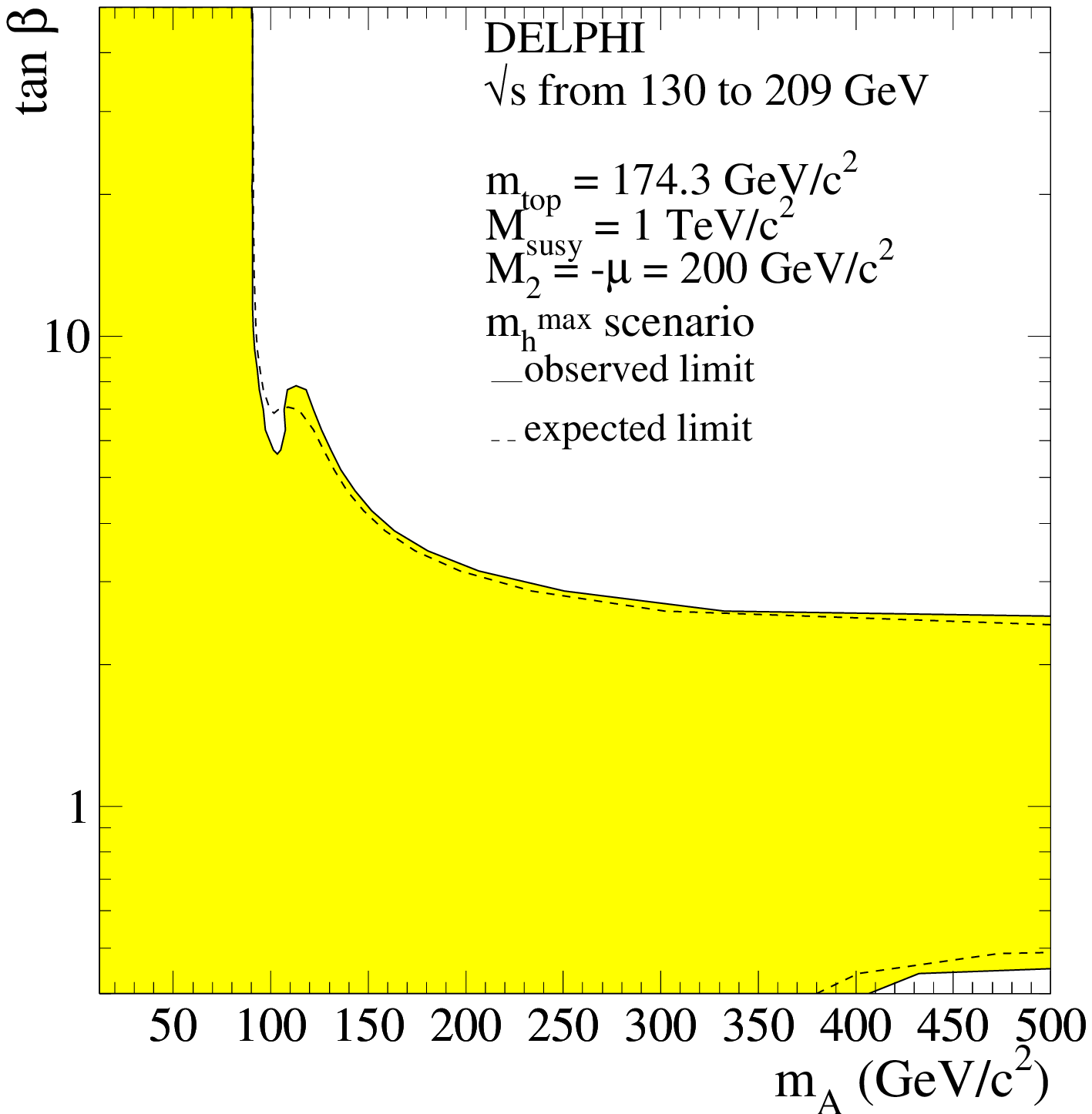,height=9cm} \\
\hspace{-1.2cm}
\epsfig{figure=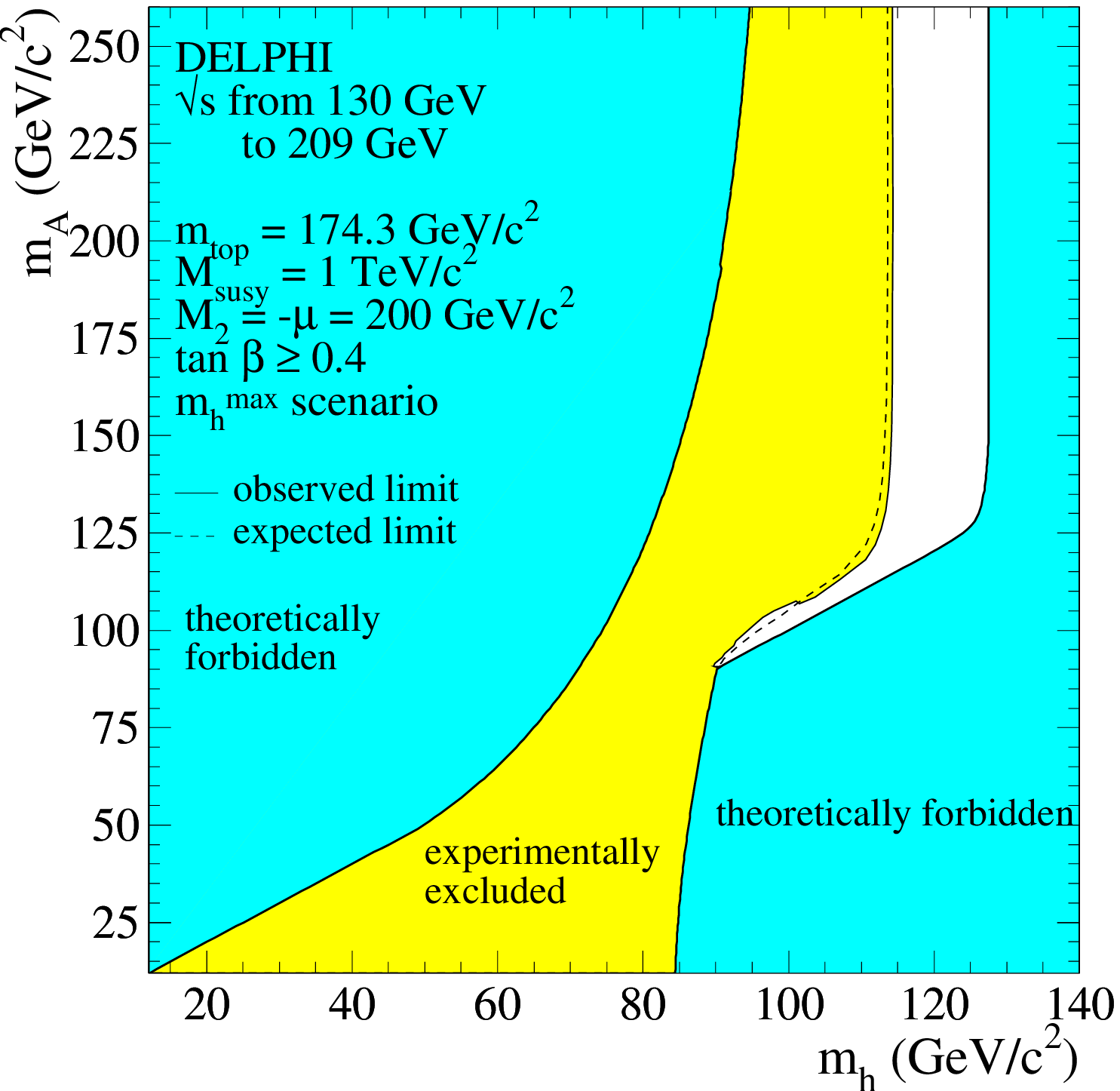,height=9cm}
\end{tabular}
\caption[]{
    {\sc MSSM} Higgs bosons: regions excluded at 95\% CL
   by the searches in the combined \hZ\ and \hA\ channels,
   in the \mbox{$ m_{\mathrm h}^{\rm max}$} scenario.
   The dark shaded areas are the regions not allowed by the 
    {\sc MSSM} model in this scenario.
   The dashed curves show the median expected limits.}
\label{fig:limit_max}
\end{center}
\end{figure}

\begin{figure}[htbp]
\begin{center}
\begin{tabular}{cc}
\hspace{-1.2cm}
\epsfig{figure=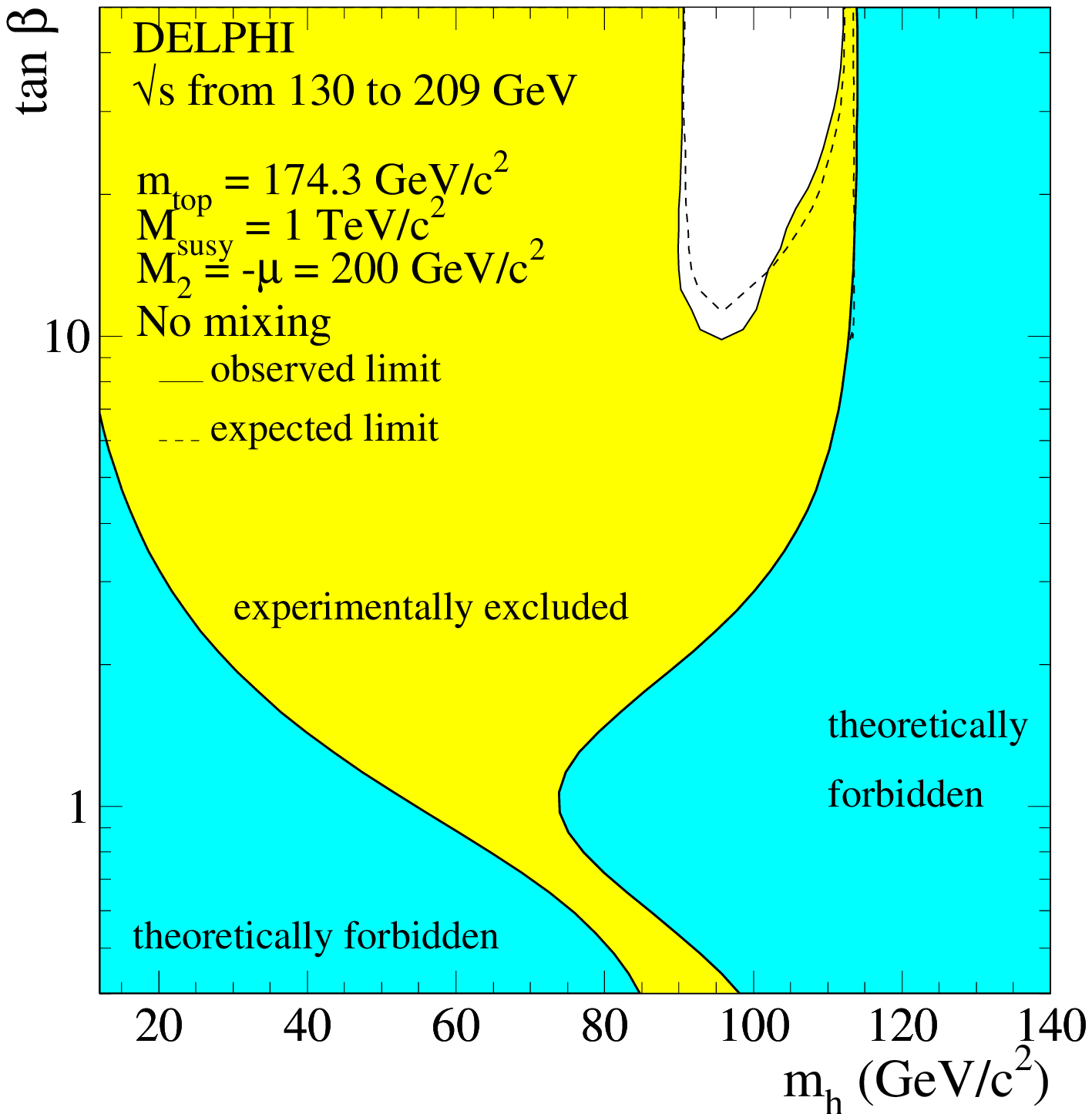,height=9cm} &
\hspace{-1cm}
\epsfig{figure=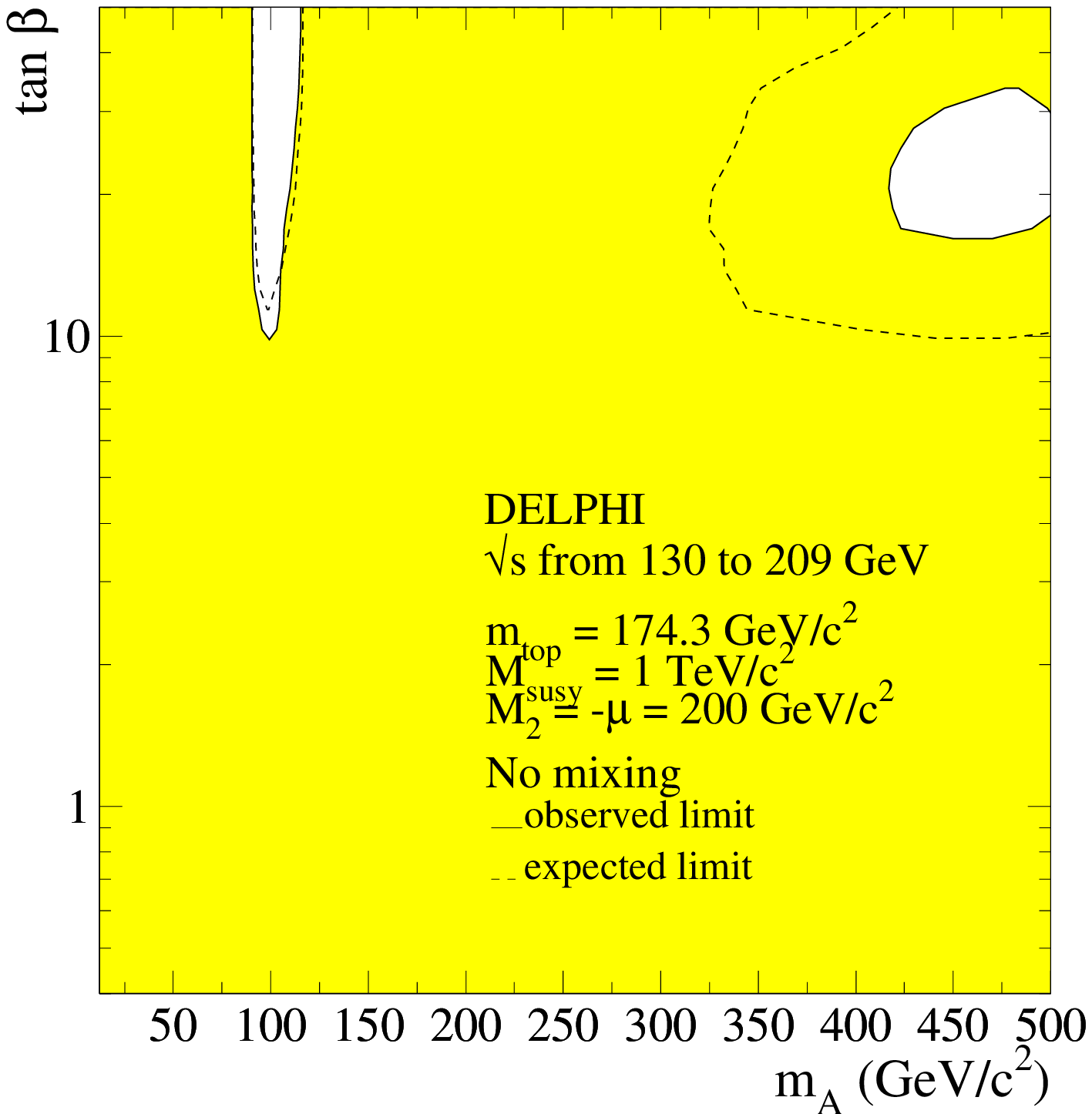,height=9cm} \\
\hspace{-1.2cm}
\epsfig{figure=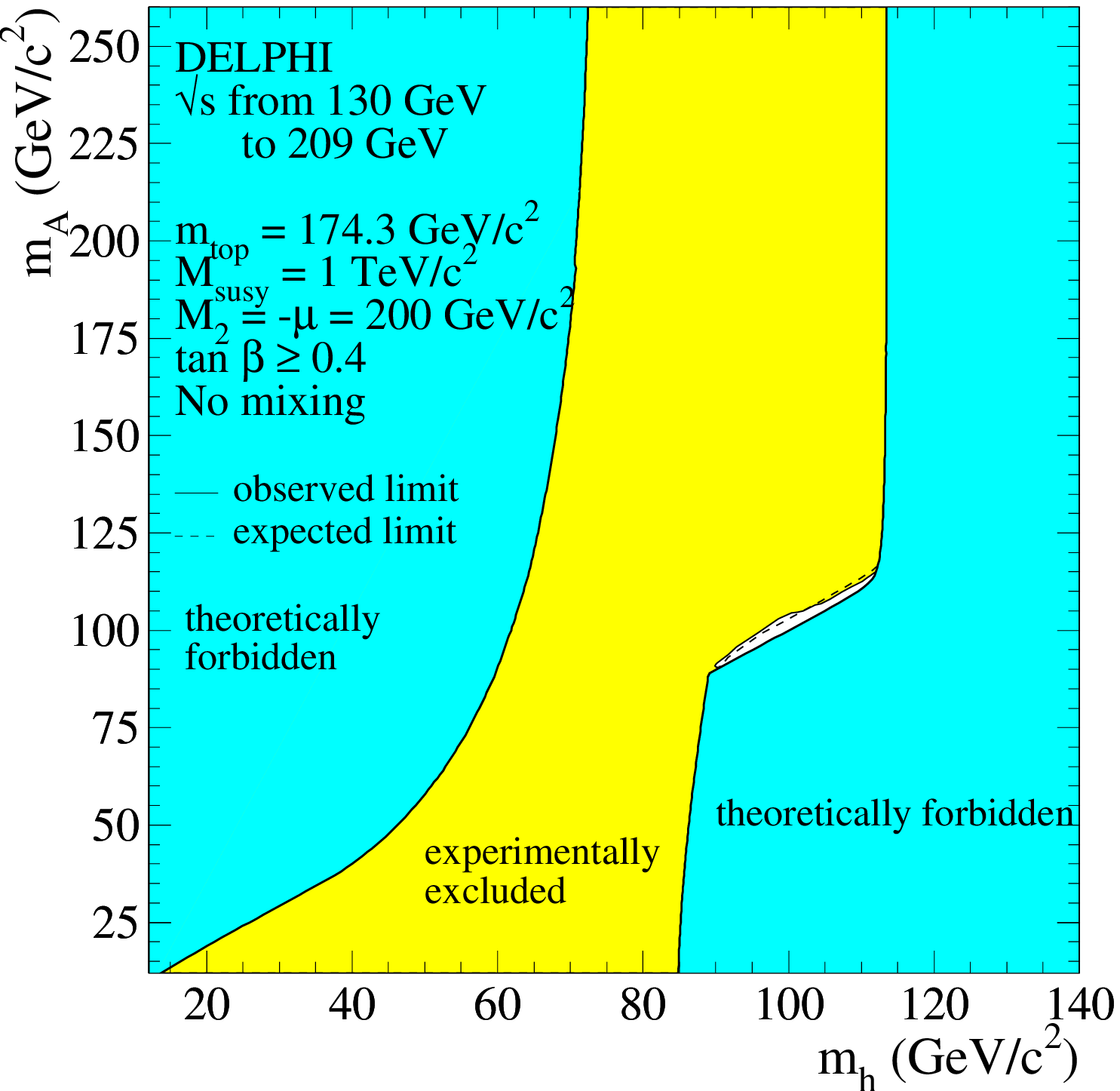,height=9cm}
\end{tabular}
\caption[]{
    {\sc MSSM} Higgs bosons: regions excluded at 95\% CL
   by the searches in the combined \hZ\ and \hA\ channels,
   in the no-mixing scenario.
   The dark shaded areas are the regions not allowed by the 
    {\sc MSSM} model in this scenario.
   The dashed curves show the median expected limits.}
\label{fig:limit_no}
\end{center}
\end{figure}

  Combining the results in the \hZ\ and \hA\ channels gives
regions of the {\sc MSSM} parameter space which are excluded at 
95\% CL or more. The excluded regions in the (\mh, \tbeta), 
(\MA, \tbeta) and (\mh, \MA) planes are presented 
in Fig.~\ref{fig:limit_max} for the \mbox{$ m_{\mathrm h}^{\rm max}$} scenario 
and in Fig.~\ref{fig:limit_no} for the no-mixing scenario.
For \MA\ below the kinematic threshold \mh~=~2\MA, which occurs at
low \tbeta\ only, the decay h$\rightarrow$AA opens, in which case it
supplants the h$\rightarrow$\bbbar\ decay.
However, in most of the
region, the A$\rightarrow$\bbbar\ and A$\rightarrow$\ccbar\ branching fractions
are  large
which explains why the results in the (\hAA) \qqbar\ channels 
reported in section~\ref{sec:admssm}, 
combined with studies of the \hAA\ decay
at lower energies~\cite{ref:pap99,ref:pap97,ref:pap96}, 
exclude  this region in both scenarios.

The above results establish 95\% {\sc CL} lower limits on \mh\ and \MA,
for either assumption on the mixing in the stop sector and
for all values of \tbeta\ above 0.4:

\[ \mh > 89.7~\GeVcc \hspace{1cm}
   \MA > 90.4~\GeVcc  .\]

\noindent
The expected median limits are 90.6~\GeVcc\ for \mh\ and 90.5~\GeVcc\ 
for \MA. The limit in \MA\ is reached in the no-mixing scenario at \tbeta\
around 30, that is in a region where Higgs bosons have non-negligible widths,
while the limit in \mh\ is obtained in the \mbox{$ m_{\mathrm h}^{\rm max}$} 
scenario at \tbeta\ around 10, in a region where both the \hZ\ and 
\hA\ processes contribute.
Furthermore, there are excluded ranges in \tbeta\ between 
0.4 and 9.36 (expected [0.4-9.36]) in the no-mixing case 
and between 0.54 and 2.36 (expected [0.54-2.14]) in the
\mbox{$ m_{\mathrm h}^{\rm max}$} scenario. 
Note that in the case of no-mixing, while the observed and expected
 lower limits on \tbeta\ are the same, they appear at quite different
 A and h masses.

\begin{figure}[htbp]
\begin{center}
\begin{tabular}{c}
\epsfig{figure=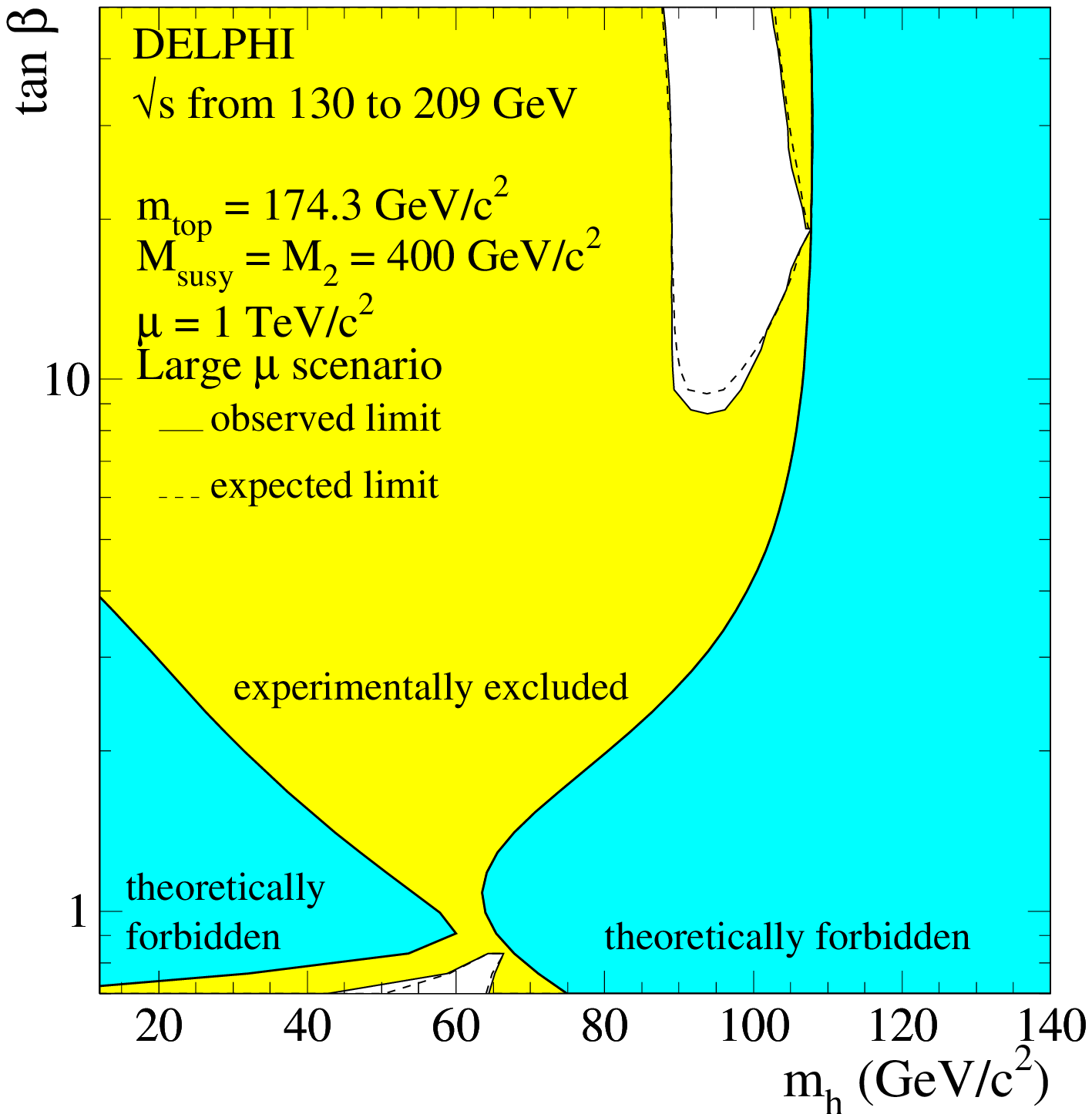,height=11.5cm}\\
\vspace{-0.2cm}
\epsfig{figure=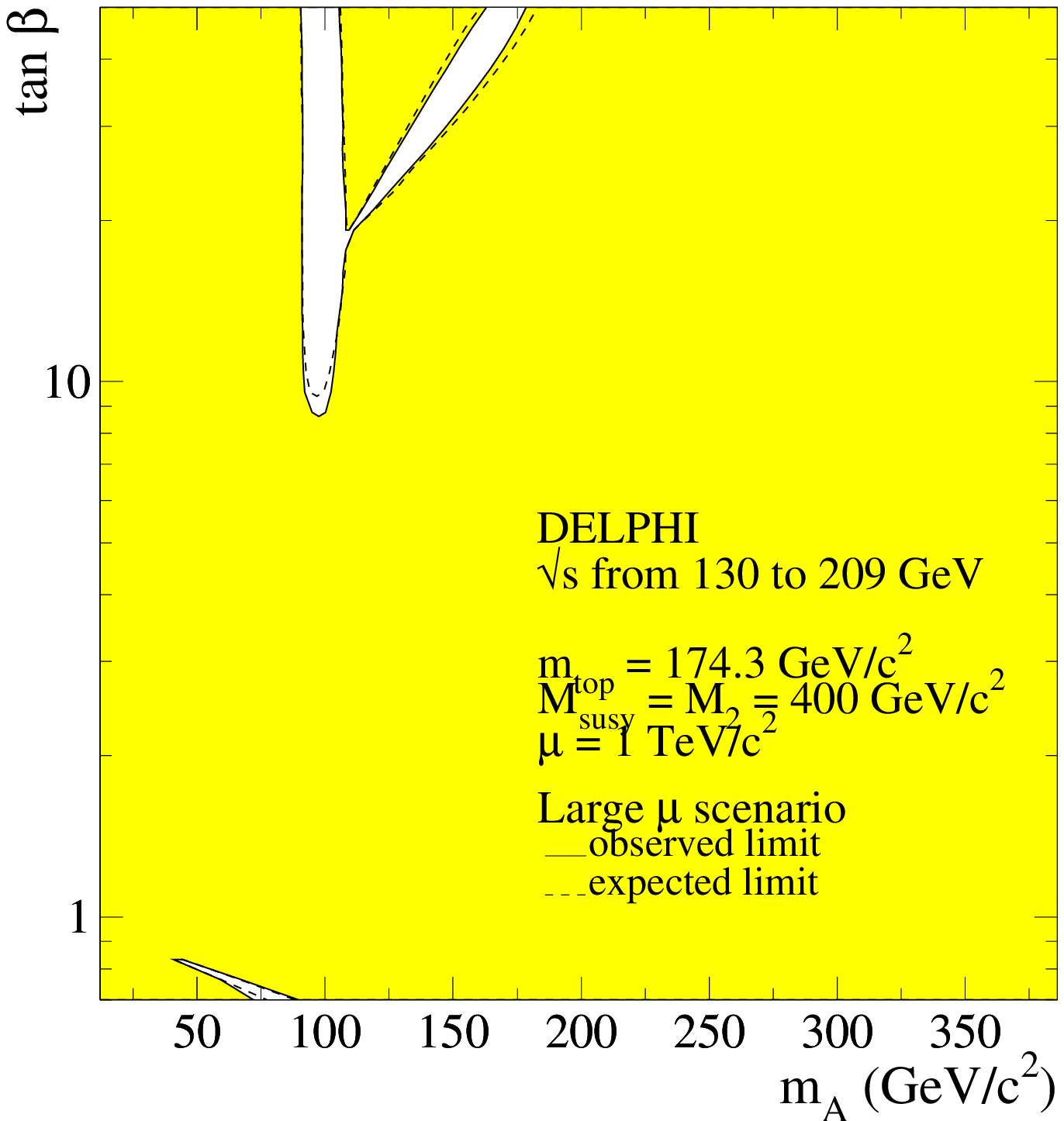,height=11.5cm}
\end{tabular}
\caption[]{
    {\sc MSSM} Higgs bosons: 
   regions excluded at 95\% CL, by the searches in the combined \hZ\ and \hA\ 
   channels, in the large $\mu$ scenario.
   The dark shaded areas are the regions not allowed 
   by the  {\sc MSSM} model in this scenario.
   The dashed curves show the  median expected limits.}
\label{fig:limit_mu}
\end{center}
\end{figure}

  The excluded regions in the large $\mu$ scenario are presented in the 
(\mh, \tbeta) and (\MA, \tbeta) planes in Fig.~\ref{fig:limit_mu}.
A large fraction of the allowed domain is excluded by the present results 
in the \hZ\ and \hA\ channels. In particular, given that the 
theoretical upper bound
on the h boson mass in that scenario is slightly above 107~\GeVcc, 
the sensitivity of the \hZ\ channels is high even at large \tbeta, which
explains why the excluded region reaches the theoretically forbidden
area for large values of \tbeta. On the other hand, there is an unexcluded 
hole in the low \tbeta\ region at \mh\ around 60~\GeVcc\ which is due to a 
loss of sensitivity because of vanishing h$\rightarrow$\bbbar\ branching 
fractions in that region.
The unexcluded area at large \tbeta\ is mostly due to low expected rates
in the \hZ\ and \hA\ channels 
(the \hA\ kinematic limit is close and the ZZh coupling is low) 
rather than to vanishing branching fractions into b's. 
At these unexcluded points the second scalar boson, H, is
kinematically accessible and has a large branching fraction into b-quarks. 
Allowing for its production in the scans should lead to an improved 
sensitivity. 
There are also points with vanishing branching 
fractions of the h boson into b-quarks, and it is expected that improvements 
could be made by allowing for other decay modes, such as gluons or c quarks.
 These  possibilities, and the theoretical scenarios referred to in
section~\ref{sec:benchmark}, could profitably
 be explored in a future publication.


\subsection{Limits on the coupling between h, A and Z}
\label{sec:mssm-cross-limit}

\begin{figure}[htbp]
\begin{center}
\epsfig{figure=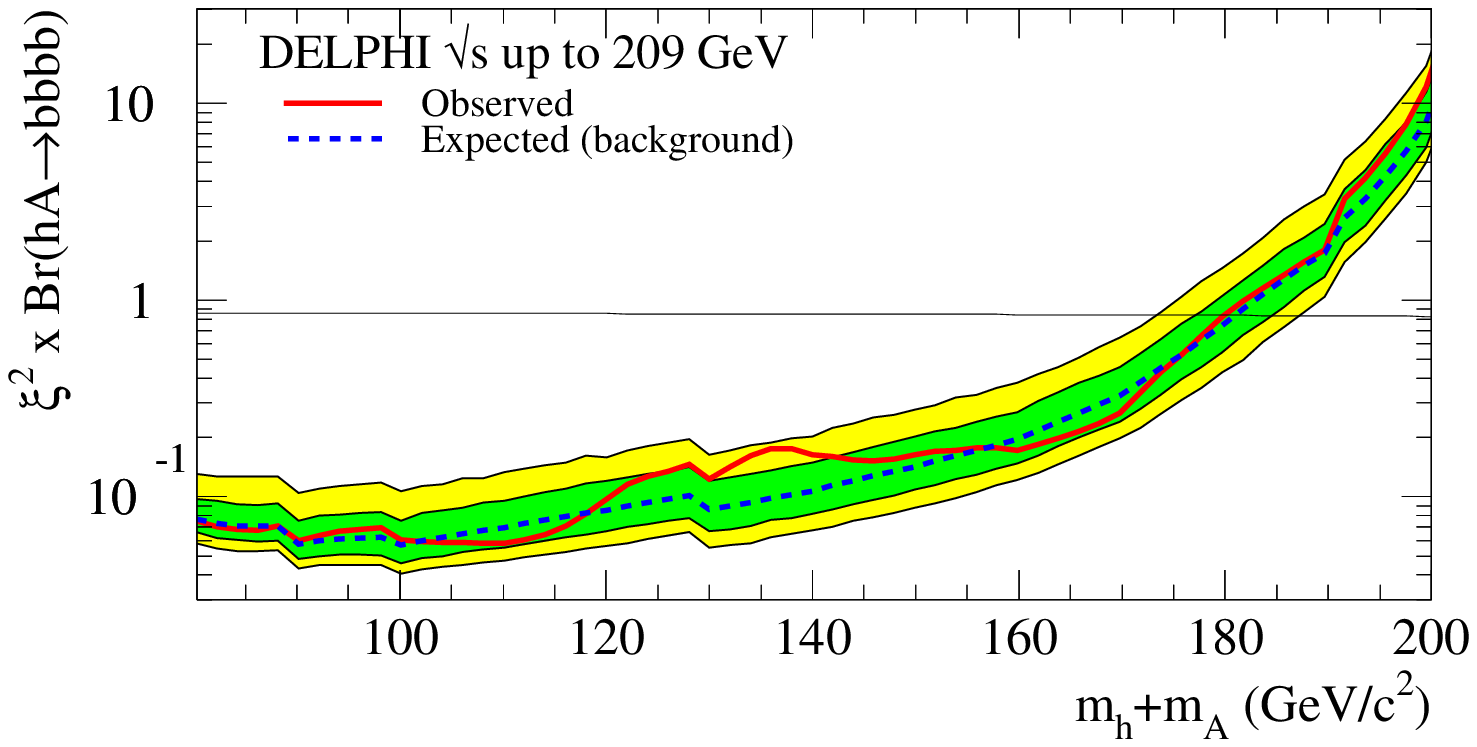,height=9cm} \\
\epsfig{figure=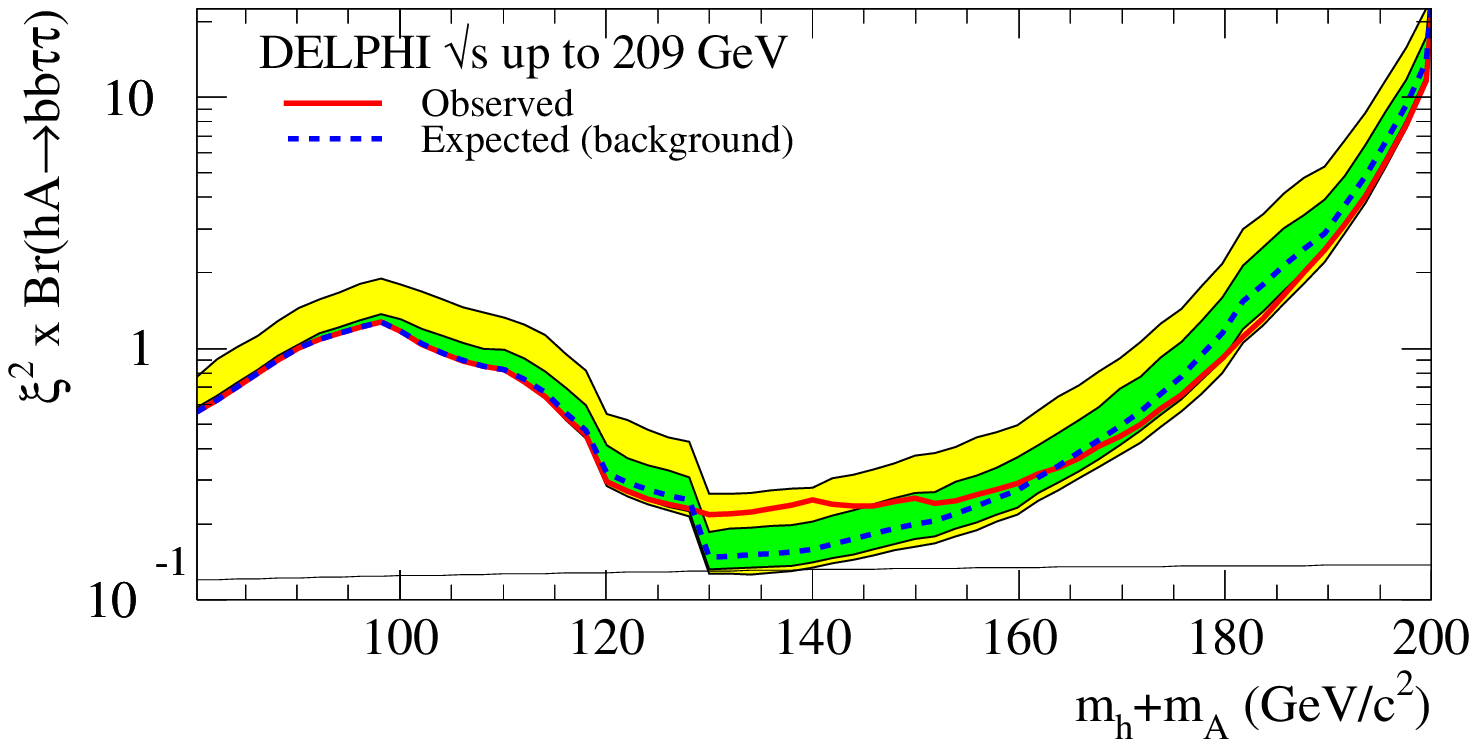,height=9cm}
\caption[]{
95\% CL upper bounds on $\xi^2$, where $\xi$ is the \hA\Zz\ coupling 
factor normalised to its maximal value in the {\sc MSSM}, as functions 
of \mh+\MA.
Results are presented in
the \bbbar \bbbar\ (top) and \bbbar \toto\ (bottom) channels,
for nearly mass-degenerate h and A bosons.  
The limits observed in data (full curve) are shown together with the expected 
median limits in background process experiments (dashed curve).
The bands correspond to the 
68.3\% and 95.0\% confidence intervals from background-only experiments.
The large band asymmetry at low mass in the \bbbar \toto\  
channel reflects the low level of background of the analyses performed 
in this mass range.
The nearly flat thin lines are the values expected in the {\sc MSSM}
\mbox{$ m_{\mathrm h}^{\rm max}$} scenario at \tbeta~=~20.6.
}
\label{fig:ghaz}
\end{center}
\end{figure}

 As for the \hZ\ process, the results of the searches for pair produced
{\sc MSSM} Higgs bosons can be reinterpreted in a more general approach
to set a 95\% CL upper bound on the pair production cross-section.
This was done separately in the \bbbar \bbbar\ and \bbbar \toto\
final states, using in each case the analysis dedicated to the channel
under study.
To achieve the best sensitivity, the results described in this paper 
are combined consistently with those obtained at lower energies at 
LEP2~\cite{ref:pap99,ref:pap98,ref:pap97,ref:pap96,ref:pap95}.
Only the case of nearly mass-degenerate h and A Higgs bosons
is considered since the \hA\ analyses have their highest
sensitivity in that case. 
For each mass hypothesis, the production cross-section 
is decreased with respect to its value in the {\sc MSSM}, 
using the  \mbox{$ m_{\mathrm h}^{\rm max}$} scenario 
 at \tbeta=20 as a reference, 
until a pseudo-confidence level \CLs\ of 5\% is obtained.
Analogously to the SM, the  coupling $\xi$, which
is the   \hA\Zz\ coupling normalised to its maximal value 
in the {\sc MSSM}, is used. $\xi$ is related to the underlying {\sc MSSM} 
parameters $\beta$ and $\alpha$, the Higgs doublet mixing angle,
through the relation: $\xi$~=~cos$(\alpha - \beta)$.
The results are then expressed 
in a way independent from the reference scenario,
as 95\% CL upper bounds on the product of $\xi^2$ and the \hA\ 
branching fractions into the relevant final states. These bounds are
given in Fig.~\ref{fig:ghaz} for sums of masses of the h and A 
bosons from 80 to 200~\GeVcc.

\section{Conclusions}
\label{sec:conclusions}

The 224~\pbinv\ of data taken by DELPHI at 200 - 209~\GeV,
combined with our lower energy data, sets the lower limit  
at 95\% CL on the mass of the Standard Model Higgs boson at:

\[ \MH  > 114.1~\GeVcc  .\]

The following limits are derived in the framework of the {\sc MSSM}
\mbox{$ m_{\mathrm h}^{\rm max}$} and no-mixing scenarios:

\[ \mh > 89.7~\GeVcc \hspace{1cm} \MA > 90.4~\GeVcc  .\]

\noindent
for all values of \tbeta\ above 0.4 and assuming \MA $>$ 12~\GeVcc.

\subsection*{Acknowledgements}
\vskip 3 mm
 We are greatly indebted to our technical 
collaborators, to the members of the CERN-SL Division for the excellent 
performance of the LEP collider, and to the funding agencies for their
support in building and operating the DELPHI detector.\\
We acknowledge in particular the support of \\
Austrian Federal Ministry of Education, Science and Culture,
GZ 616.364/2-III/2a/98, \\
FNRS--FWO, Flanders Institute to encourage scientific and technological 
research in the industry (IWT), Belgium,  \\
FINEP, CNPq, CAPES, FUJB and FAPERJ, Brazil, \\
Czech Ministry of Industry and Trade, GA CR 202/99/1362,\\
Commission of the European Communities (DG XII), \\
Direction des Sciences de la Mati$\grave{\mbox{\rm e}}$re, CEA, France, \\
Bundesministerium f$\ddot{\mbox{\rm u}}$r Bildung, Wissenschaft, Forschung 
und Technologie, Germany,\\
General Secretariat for Research and Technology, Greece, \\
National Science Foundation (NWO) and Foundation for Research on Matter (FOM),
The Netherlands, \\
Norwegian Research Council,  \\
State Committee for Scientific Research, Poland, SPUB-M/CERN/PO3/DZ296/2000,
SPUB-M/CERN/PO3/DZ297/2000, 2P03B 104 19 and 2P03B 69 23(2002-2004)\\
JNICT--Junta Nacional de Investiga\c{c}\~{a}o Cient\'{\i}fica 
e Tecnol$\acute{\mbox{\rm o}}$gica, Portugal, \\
Vedecka grantova agentura MS SR, Slovakia, Nr. 95/5195/134, \\
Ministry of Science and Technology of the Republic of Slovenia, \\
CICYT, Spain, AEN99-0950 and AEN99-0761,  \\
The Swedish Natural Science Research Council,      \\
Particle Physics and Astronomy Research Council, UK, \\
Department of Energy, USA, DE-FG02-01ER41155. \\


\section*{Appendix}
\vskip 3 mm

We give in detail the efficiencies of the signal selection here.

\begin{table} [htbp]
{\small
\begin{center}
\begin{tabular}{cccccccc}  \hline
\MH\       & \hee       & \hmm          & H\toto        & \toto Z       &\multicolumn{2}{c}{\hnn}& \hqq \\
(\GeVcc)    & channel        & channel       & channel       & channel       & Low mass      & High mass        & channel \\ \hline
\multicolumn{7}{c}{First operational period \rs~=~206.5~\GeV} \\ \hline 
  12.0      & 20.6 $\pm$0.6  & 42.1 $\pm$0.7 &               &               &21.2 $\pm$0.6&                  &    \\
  18.0      & 30.4 $\pm$0.7  & 50.5 $\pm$0.7 &               &               &35.1 $\pm$0.7&                  &    \\
  24.0      & 37.6 $\pm$0.7  & 54.7 $\pm$0.7 &               &               &38.6 $\pm$0.7&                  &    \\
  30.0      & 42.2 $\pm$0.7  & 56.9 $\pm$0.7 &               &               &39.1 $\pm$0.7&16.2 $\pm$0.5     &  9.6$\pm$ 0.4  \\
  40.0      & 48.6 $\pm$0.7  & 62.0 $\pm$0.7 &               &               &38.8 $\pm$0.7&26.0 $\pm$0.6     & 18.6$\pm$ 0.6  \\
  50.0      & 51.5 $\pm$0.7  & 64.2 $\pm$0.7 & 13.  $\pm$0.5 &  5.  $\pm$0.3 &42.2 $\pm$0.7&21.5 $\pm$0.6     & 25.5$\pm$ 0.7  \\
  60.0      & 54.8 $\pm$0.7  & 64.9 $\pm$0.7 & 18.  $\pm$0.6 & 12.  $\pm$0.5 &42.3 $\pm$0.7&12.6 $\pm$0.5     & 28.8$\pm$ 0.7  \\
  70.0      & 57.7 $\pm$0.7  & 67.5 $\pm$0.7 & 20.  $\pm$0.6 & 22.  $\pm$0.7 &47.5 $\pm$0.7&12.8 $\pm$0.5     & 28.0$\pm$ 0.7  \\
  80.0      & 57.8 $\pm$0.7  & 67.9 $\pm$0.7 & 20.  $\pm$0.6 & 23.  $\pm$0.7 &54.9 $\pm$0.7&28.0 $\pm$0.6     & 28.3$\pm$ 0.7  \\
  85.0      & 60.3 $\pm$0.7  & 68.7 $\pm$0.7 & 20.  $\pm$0.6 & 24.  $\pm$0.7 &59.8 $\pm$0.7&40.9 $\pm$0.7     & 27.3$\pm$ 0.7  \\
  90.0      & 59.3 $\pm$0.7  & 69.4 $\pm$0.7 & 19.  $\pm$0.6 & 24.  $\pm$0.7 &63.5 $\pm$0.7&51.2 $\pm$0.7     & 27.5$\pm$ 0.7  \\
  95.0      & 60.4 $\pm$0.7  & 69.4 $\pm$0.7 & 19.  $\pm$0.6 & 23.  $\pm$0.7 &65.6 $\pm$0.7&59.5 $\pm$0.7     & 36.1$\pm$ 0.8  \\
 100.0      & 59.0 $\pm$0.7  & 69.8 $\pm$0.6 & 19.  $\pm$0.6 & 21.  $\pm$0.6 &65.2 $\pm$0.7&63.2 $\pm$0.7     & 47.0$\pm$ 1.0  \\
 105.0      & 59.8 $\pm$0.7  & 69.1 $\pm$0.7 & 18.  $\pm$0.6 & 21.  $\pm$0.6 &65.8 $\pm$0.7&67.4 $\pm$0.7     & 54.5$\pm$ 1.1  \\
 110.0      & 60.4 $\pm$0.7  & 69.1 $\pm$0.7 & 18.  $\pm$0.6 & 21.  $\pm$0.6 &64.2 $\pm$0.7&66.6 $\pm$0.7     & 58.8$\pm$ 1.1  \\
 114.0      & 58.4 $\pm$0.7  & 67.2 $\pm$0.7 & 15.  $\pm$0.5 & 19.  $\pm$0.6 &55.9 $\pm$0.7&58.7 $\pm$0.7     & 58.2$\pm$ 1.1  \\
 115.0      & 59.0 $\pm$0.7  & 67.0 $\pm$0.7 & 14.  $\pm$0.5 & 19.  $\pm$0.6 &55.3 $\pm$0.7&58.9 $\pm$0.7     & 56.5$\pm$ 1.1  \\
 116.0      & 56.7 $\pm$0.7  & 64.5 $\pm$0.7 & 14.  $\pm$0.5 & 19.  $\pm$0.6 &55.0 $\pm$0.7&59.3 $\pm$0.7     & 53.1$\pm$ 1.0  \\
 120.0      & 52.4 $\pm$0.7  & 56.3 $\pm$0.7 & 12.  $\pm$0.5 & 18.  $\pm$0.6 &52.3 $\pm$0.7&57.7 $\pm$0.7     & 43.6$\pm$ 0.9  \\
\hline
\multicolumn{7}{c}{Second operational period \rs~=~206.5~\GeV} \\ \hline 
  12.0      & 18.7 $\pm$0.6  & 44.7 $\pm$0.7 &               &               &19.6 $\pm$0.6&                  &    \\
  18.0      & 29.5 $\pm$0.6  & 52.7 $\pm$0.7 &               &               &28.9 $\pm$0.7&                  &    \\
  24.0      & 36.5 $\pm$0.7  & 56.6 $\pm$0.7 &               &               &33.6 $\pm$0.7&                  &    \\
  30.0      & 39.0 $\pm$0.7  & 59.4 $\pm$0.7 &               &               &35.0 $\pm$0.7&14.7 $\pm$0.5     &  8.9$\pm$ 0.4  \\
  40.0      & 43.2 $\pm$0.7  & 63.1 $\pm$0.7 &               &               &38.0 $\pm$0.7&25.2 $\pm$0.6     & 18.2$\pm$ 0.6  \\
  50.0      & 47.7 $\pm$0.7  & 66.1 $\pm$0.7 & 12.  $\pm$0.5 &               &40.3 $\pm$0.7&20.0 $\pm$0.6     & 23.6$\pm$ 0.7  \\
  60.0      & 50.2 $\pm$0.7  & 66.8 $\pm$0.7 & 18.  $\pm$0.6 & 11.  $\pm$0.5 &40.7 $\pm$0.7&11.8 $\pm$0.5     & 26.9$\pm$ 0.7  \\
  70.0      & 52.5 $\pm$0.7  & 69.7 $\pm$0.7 & 20.  $\pm$0.6 & 22.  $\pm$0.7 &44.8 $\pm$0.7&12.3 $\pm$0.5     & 25.7$\pm$ 0.7  \\
  80.0      & 54.8 $\pm$0.7  & 70.3 $\pm$0.7 & 20.  $\pm$0.6 & 23.  $\pm$0.7 &51.1 $\pm$0.7&26.0 $\pm$0.6     & 25.8$\pm$ 0.7  \\
  85.0      & 55.8 $\pm$0.7  & 70.3 $\pm$0.7 & 19.  $\pm$0.6 & 23.  $\pm$0.7 &54.5 $\pm$0.7&36.2 $\pm$0.7     & 26.4$\pm$ 0.7  \\
  90.0      & 53.8 $\pm$0.7  & 70.1 $\pm$1.1 & 20.  $\pm$0.6 & 23.  $\pm$0.7 &60.4 $\pm$0.7&47.6 $\pm$0.7     & 26.5$\pm$ 0.7  \\
  95.0      & 55.7 $\pm$0.7  & 70.7 $\pm$1.1 & 19.  $\pm$0.6 & 23.  $\pm$0.7 &63.2 $\pm$0.7&57.1 $\pm$0.7     & 32.5$\pm$ 0.8  \\
 100.0      & 55.0 $\pm$0.7  & 70.4 $\pm$1.1 & 19.  $\pm$0.6 & 21.  $\pm$0.6 &65.1 $\pm$0.7&62.6 $\pm$0.7     & 44.4$\pm$ 0.9  \\
 105.0      & 55.3 $\pm$0.7  & 70.2 $\pm$1.1 & 17.  $\pm$0.6 & 22.  $\pm$0.7 &64.1 $\pm$0.7&64.8 $\pm$0.7     & 51.9$\pm$ 1.0  \\
 110.0      & 56.2 $\pm$0.7  & 68.9 $\pm$1.1 & 17.  $\pm$0.6 & 19.  $\pm$0.6 &60.6 $\pm$0.7&63.8 $\pm$0.7     & 56.9$\pm$ 1.1  \\
 114.0      & 55.9 $\pm$0.7  & 68.3 $\pm$1.1 & 15.  $\pm$0.5 & 19.  $\pm$0.6 &54.3 $\pm$0.7&58.0 $\pm$0.7     & 55.2$\pm$ 1.1  \\
 115.0      & 55.0 $\pm$0.7  & 67.1 $\pm$1.1 & 15.  $\pm$0.5 & 19.  $\pm$0.6 &53.6 $\pm$0.7&58.1 $\pm$0.7     & 54.9$\pm$ 1.0  \\
 116.0      & 54.1 $\pm$0.7  & 64.8 $\pm$1.1 & 13.  $\pm$0.5 & 18.  $\pm$0.6 &53.1 $\pm$0.7&57.9 $\pm$0.7     & 53.1$\pm$ 1.0  \\
 120.0      & 48.2 $\pm$0.7  & 54.8 $\pm$1.1 & 12.  $\pm$0.5 & 19.  $\pm$0.6 &48.9 $\pm$0.7&56.1 $\pm$0.7     & 42.6$\pm$ 0.9  \\
\hline
\end{tabular}
\caption[]{\ZH\ channels: 
 efficiencies (in \%) of the selection in the two operational periods, as a function of the mass of 
the Higgs boson. The quoted errors are statistical only. Only efficiencies higher than 5\% are shown.
}
\label{ta:hzeff}
\end{center}
}
\end{table}

\newpage

\begin{table} [htbp]
\begin{center}
\begin{tabular}{cc|c|c|c}  \hline
 \MA  & \mh  & \rs~=~199.6~\GeV\ & \rs~=~206.5~\GeV\ & \rs~=~206.5~\GeV\ \\
(\GeVcc)  &(\GeVcc)  &  &  1st period &  2nd period \\ \hline
  40.0 &  40.0   & 17.3 $\pm$ 0.5 & 11.4 $\pm$ 0.5  & 11.4 $\pm$ 0.5  \\
  50.0 &  50.0   & 62.9 $\pm$ 1.1 & 61.2 $\pm$ 1.1  & 59.7 $\pm$ 0.9  \\
  60.0 &  60.0   & 74.4 $\pm$ 1.2 & 71.7 $\pm$ 1.2  & 70.0 $\pm$ 0.9  \\
  70.0 &  70.0   & 78.6 $\pm$ 1.2 & 77.5 $\pm$ 1.2  & 76.4 $\pm$ 0.9  \\
  80.0 &  80.0   & 85.3 $\pm$ 1.3 & 85.0 $\pm$ 1.3  & 83.3 $\pm$ 1.1  \\
  85.0 &  85.0   & 87.3 $\pm$ 1.3 & 88.9 $\pm$ 1.3  & 86.8 $\pm$ 1.0  \\
  90.0 &  90.0   & 89.0 $\pm$ 1.4 & 89.4 $\pm$ 1.3  & 88.2 $\pm$ 1.2  \\
  95.0 &  95.0   & 88.0 $\pm$ 1.3 & 88.4 $\pm$ 1.3  & 87.4 $\pm$ 0.9  \\
 100.0 & 100.0   &                & 86.8 $\pm$ 1.4  & 84.8 $\pm$ 0.9  \\
 103.0 & 103.0   &                & 82.7 $\pm$ 1.3  & 81.6 $\pm$ 0.9  \\
  12.0 &  70.0   & 24.6 $\pm$ 0.7 & 23.2 $\pm$ 0.7  & 21.8 $\pm$ 0.6  \\
  12.0 & 110.0   & 61.2 $\pm$ 1.1 & 59.6 $\pm$ 1.1  & 53.7 $\pm$ 1.0  \\
  12.0 & 150.0   & 51.4 $\pm$ 1.0 & 56.0 $\pm$ 1.1  & 53.9 $\pm$ 0.8  \\
  12.0 & 170.0   & 37.2 $\pm$ 0.9 & 43.4 $\pm$ 0.9  & 41.4 $\pm$ 0.9  \\
  12.0 & 194.0   &                & 12.9 $\pm$ 0.5  & 12.1 $\pm$ 0.4  \\
  30.0 &  50.0   & 20.3 $\pm$ 0.6 & 15.4 $\pm$ 0.6  & 15.2 $\pm$ 0.4  \\
  30.0 &  90.0   & 66.8 $\pm$ 1.2 & 67.5 $\pm$ 1.2  & 65.7 $\pm$ 1.1  \\
  30.0 & 110.0   & 71.9 $\pm$ 1.2 & 72.0 $\pm$ 1.2  & 68.5 $\pm$ 1.2  \\
  30.0 & 150.0   & 64.6 $\pm$ 1.3 & 69.4 $\pm$ 1.2  & 64.9 $\pm$ 1.1  \\
  30.0 & 176.0   &                & 29.6 $\pm$ 0.8  & 29.9 $\pm$ 0.6  \\
  40.0 &  50.0   & 51.6 $\pm$ 1.0 & 47.5 $\pm$ 0.9  & 46.2 $\pm$ 0.7  \\
  50.0 &  60.0   & 68.0 $\pm$ 1.2 & 66.8 $\pm$ 1.1  & 66.3 $\pm$ 1.0  \\
  50.0 &  90.0   & 79.0 $\pm$ 1.3 & 78.4 $\pm$ 1.3  & 76.3 $\pm$ 1.1  \\
  50.0 & 110.0   & 82.1 $\pm$ 1.3 & 81.2 $\pm$ 1.3  & 79.4 $\pm$ 1.0  \\
  50.0 & 130.0   & 78.2 $\pm$ 1.2 & 79.9 $\pm$ 1.3  & 78.3 $\pm$ 1.2  \\
  50.0 & 156.0   &                & 61.1 $\pm$ 1.1  & 59.4 $\pm$ 0.8  \\
  60.0 &  70.0   & 77.6 $\pm$ 1.3 & 75.1 $\pm$ 1.2  & 74.0 $\pm$ 1.0  \\
  60.0 &  80.0   & 79.5 $\pm$ 1.3 & 78.1 $\pm$ 1.2  & 77.9 $\pm$ 1.0  \\
  60.0 &  90.0   & 82.7 $\pm$ 1.3 & 82.1 $\pm$ 1.3  & 79.6 $\pm$ 1.0  \\
  60.0 & 100.0   & 81.9 $\pm$ 1.3 & 73.7 $\pm$ 1.2  & 80.5 $\pm$ 1.2  \\
  70.0 &  80.0   & 82.6 $\pm$ 1.4 & 80.9 $\pm$ 1.3  & 79.3 $\pm$ 1.1  \\
  70.0 &  90.0   & 86.1 $\pm$ 1.3 & 83.8 $\pm$ 1.3  & 82.0 $\pm$ 1.1  \\
  70.0 & 110.0   & 85.7 $\pm$ 1.3 & 85.3 $\pm$ 1.3  & 86.0 $\pm$ 1.0  \\
  70.0 & 130.0   &                & 83.3 $\pm$ 1.3  & 81.4 $\pm$ 1.0  \\
  70.0 & 136.0   &                & 77.5 $\pm$ 1.2  & 78.0 $\pm$ 0.9  \\
  80.0 &  85.0   & 86.5 $\pm$ 1.3 & 85.7 $\pm$ 1.4  & 85.4 $\pm$ 1.0  \\
  80.0 &  90.0   & 88.6 $\pm$ 1.3 & 89.4 $\pm$ 1.3  & 85.8 $\pm$ 1.0  \\
  80.0 & 100.0   & 88.8 $\pm$ 1.4 & 90.1 $\pm$ 1.3  & 86.8 $\pm$ 1.1  \\
  85.0 &  90.0   & 89.5 $\pm$ 1.3 & 88.8 $\pm$ 1.3  & 87.1 $\pm$ 1.0  \\
  85.0 &  95.0   & 89.0 $\pm$ 1.3 & 89.2 $\pm$ 1.3  & 88.5 $\pm$ 1.3  \\
  90.0 &  95.0   & 89.0 $\pm$ 1.3 & 89.0 $\pm$ 1.3  & 87.4 $\pm$ 1.0  \\
  90.0 & 100.0   & 87.8 $\pm$ 1.3 & 89.6 $\pm$ 1.3  & 87.5 $\pm$ 1.0  \\
  90.0 & 110.0   &                & 85.7 $\pm$ 1.3  & 84.7 $\pm$ 1.0  \\
  90.0 & 116.0   &                & 82.1 $\pm$ 1.3  & 80.8 $\pm$ 1.0  \\

\hline\end{tabular}
\caption[]{\hA\ four-jet channel: 
  efficiencies of the selection (in \%) 
  at \rs~=~199.6~\GeV\  and \rs~=~206.5~\GeV\ as a function of the masses of the A and h bosons,
  from simulated samples corresponding
  to various mass differences between the two bosons. 
  The quoted errors are statistical only.}
\label{ta:scaneff}
\end{center}
\end{table}

\begin{table} [htbp]
\begin{center}
\begin{tabular}{cccc}  \hline \\[-.4cm]
&& \Abb & \Acc \\ \hline
 \MA & \mh & Efficiency & Efficiency \\
(\GeVcc)   & (\GeVcc)  & (\%) & (\%) \\ \hline
\hline
\multicolumn{4}{c}{First Period} \\
 \hline
  12.0 & 30.0   & 21.8 $\pm$ 0.4 &  7.3 $\pm$ 0.3\\
  12.0 & 50.0   & 49.3 $\pm$ 0.5 & 20.0 $\pm$ 0.4\\
  12.0 & 70.0   & 54.7 $\pm$ 0.5 & 21.4 $\pm$ 0.4\\
  12.0 & 90.0   & 76.3 $\pm$ 0.4 & 33.4 $\pm$ 0.4\\
  12.0 & 105.0  & 79.7 $\pm$ 0.4 & 46.2 $\pm$ 0.5\\
  20.0 & 50.0   & 45.5 $\pm$ 0.5 & 18.0 $\pm$ 0.5\\
  20.0 & 70.0   & 57.4 $\pm$ 0.5 & 23.4 $\pm$ 0.5\\
  20.0 & 90.0   & 72.3 $\pm$ 0.5 & 32.8 $\pm$ 0.5\\
  20.0 & 105.0  & 81.7 $\pm$ 0.4 & 49.1 $\pm$ 0.5\\
  30.0 & 70.0   & 60.8 $\pm$ 0.5 & 26.5 $\pm$ 0.5\\
  30.0 & 90.0   & 72.9 $\pm$ 0.4 & 32.0 $\pm$ 0.5 \\
  30.0 & 105.0  & 79.6 $\pm$ 0.4 & 45.3 $\pm$ 0.5\\ 
  40.0 & 90.0   & 74.3 $\pm$ 0.4 & 34.4 $\pm$ 0.5\\
  40.0 & 105.0  & 79.8 $\pm$ 0.4 & 39.8 $\pm$ 0.5\\
  50.0 & 105.0  & 80.7 $\pm$ 0.4 & 42.9 $\pm$ 0.5 \\ 
\hline
\multicolumn{4}{c}{Second Period} \\
 \hline
 \hline
  12.0 & 30.0   & 20.2 $\pm$ 0.4 &  6.7 $\pm$ 0.3\\
  12.0 & 50.0   & 48.6 $\pm$ 0.5 & 19.0 $\pm$ 0.4\\
  12.0 & 70.0   & 53.4 $\pm$ 0.5 & 20.9 $\pm$ 0.4\\
  12.0 & 90.0   & 75.3 $\pm$ 0.4 & 31.4 $\pm$ 0.5\\
  12.0 & 105.0  & 78.9 $\pm$ 0.4 & 44.6 $\pm$ 0.5\\
  20.0 & 50.0   & 43.8 $\pm$ 0.5 & 17.0 $\pm$ 0.4\\
  20.0 & 70.0   & 55.7 $\pm$ 0.5 & 22.8 $\pm$ 0.5\\
  20.0 & 90.0   & 70.4 $\pm$ 0.5 & 31.8 $\pm$ 0.5\\
  20.0 & 105.0  & 80.6 $\pm$ 0.4 & 48.0 $\pm$ 0.5\\
  30.0 & 70.0   & 53.1 $\pm$ 0.5 & 25.4 $\pm$ 0.4\\
  30.0 & 90.0   & 71.3 $\pm$ 0.5 & 31.1 $\pm$ 0.5 \\
  30.0 & 105.0  & 78.3 $\pm$ 0.4 & 44.9 $\pm$ 0.5\\ 
  40.0 & 90.0   & 73.2 $\pm$ 0.4 & 33.6 $\pm$ 0.5\\
  40.0 & 105.0  & 78.3 $\pm$ 0.4 & 38.4 $\pm$ 0.5\\
  50.0 & 105.0  & 78.8 $\pm$ 0.5 & 41.5 $\pm$ 0.5 \\ 

\hline\end{tabular}
\caption[]{(\hAA )(\Zqq ) channels with \Abb\ or \Acc:
  efficiencies of the selection (in~\%)
  at \rs~=~206.5~\GeV\ as a function of the masses of the A and h bosons. 
  The quoted errors are statistical only.}
\label{ta:aaqqeff}
\end{center}
\end{table}

\begin{table} [htbp]
\begin{center}
\begin{tabular}{c|c|cc|cc}  \hline
 & \multicolumn{1}{c|}{\rs~=~199.6~\GeV} &
 \multicolumn{2}{c|}{\rs~=~206.5~\GeV }  &
 \multicolumn{2}{c}{\rs~=~206.5~\GeV } \\
 & \multicolumn{1}{c|}{} &
 \multicolumn{2}{c|}{1st period }  &
 \multicolumn{2}{c}{2nd period }
 \\  \hline
 \MA   & Four-jet   & Four-jet         & Tau           &  Four-jet     & Tau\\
(\GeVcc)& channel  & channel           & channel       &       channel & channel \\ \hline
\multicolumn{6}{c}{\tbeta\ = 50} \\ \hline 
 40.0  &  13.6 $\pm$ 0.5  &  9.8 $\pm$ 0.5 &               & 10.2 $\pm$ 0.5 &               \\  
 50.0  &  54.8 $\pm$ 1.0  & 52.6 $\pm$ 1.1 &               & 52.2 $\pm$ 1.0 &               \\  
 60.0  &  70.0 $\pm$ 1.2  & 66.9 $\pm$ 1.2 &  5.0$\pm$ 0.3 & 66.8 $\pm$ 1.2 &  4.6$\pm$ 0.3 \\  
 70.0  &  77.9 $\pm$ 1.2  & 76.3 $\pm$ 1.2 & 12.5$\pm$ 0.5 & 74.3 $\pm$ 1.3 & 13.5$\pm$ 0.5 \\  
 80.0  &  82.9 $\pm$ 1.7  & 82.3 $\pm$ 1.3 & 21.8$\pm$ 0.7 & 80.0 $\pm$ 1.4 & 21.2$\pm$ 0.6 \\  
 85.0  &  84.3 $\pm$ 1.4  & 84.2 $\pm$ 1.3 & 22.4$\pm$ 0.7 & 83.0 $\pm$ 1.3 & 19.7$\pm$ 0.6 \\  
 90.0  &  85.9 $\pm$ 1.4  & 85.2 $\pm$ 1.3 & 21.4$\pm$ 0.6 & 84.1 $\pm$ 1.4 & 21.1$\pm$ 0.6 \\  
 95.0  &  84.3 $\pm$ 1.4  & 86.0 $\pm$ 1.3 & 21.0$\pm$ 0.6 & 84.5 $\pm$ 1.4 & 20.4$\pm$ 0.6 \\  
100.0  &                  & 83.8 $\pm$ 1.4 & 19.0$\pm$ 0.6 & 82.1 $\pm$ 1.4 & 17.5$\pm$ 0.6 \\  
\hline
\end{tabular}
\caption[]{\hA\ channels: 
  efficiencies of the selection (in \%) 
  at \rs~=~199.6~\GeV\ and \rs~=~206.5~\GeV\ as a function of the mass of the
A boson for    \tbeta\ = 50. The efficiencies are defined relative to the
$ {\mathrm b \bar{b}} {\mathrm b \bar{b}}$ or
$\tau^+\tau^- {\mathrm b \bar{b}}$ final state.
 The 1999 data has been reanalysed in the four-jet channel  only.
  The quoted errors are statistical only.}
\label{ta:haeff}
\end{center}
\end{table}

\begin{table} [htbp]
\begin{center}
\begin{tabular}{c|c|c|c}  \hline
 \mh  & \rs~=~199.6~\GeV & \rs~=~206.5~\GeV   & \rs~=~206.5~\GeV  \\
(\GeVcc) &  & 1st period   & 2nd period \\  \hline               
 70.0  &  60.7 $\pm$ 0.7  & 60.2 $\pm$ 0.7 & 56.7 $\pm$ 0.7  \\  
 80.0  &  66.1 $\pm$ 0.7  & 64.1 $\pm$ 0.7 & 60.9 $\pm$ 0.7 \\  
 85.0  &  68.6 $\pm$ 0.7  & 65.5 $\pm$ 0.7 & 64.4 $\pm$ 0.7  \\  
 90.0  &  66.8 $\pm$ 0.7  & 66.6 $\pm$ 0.7 & 66.4 $\pm$ 0.7  \\  
 95.0  &  67.3 $\pm$ 0.7  & 67.1 $\pm$ 0.7 & 65.5 $\pm$ 0.7  \\  
100.0  &  64.4 $\pm$ 0.7  & 67.9 $\pm$ 0.7 & 65.8 $\pm$ 0.7  \\  
\hline
\end{tabular}
\caption[]{\hZ\ four-jet channel: 
  efficiencies of the \hA\ four-b selection (in \%) 
  at \rs~=~199.6~\GeV\ and \rs~=~206.5~\GeV\ as a function of 
  the mass of the h boson. 
  The efficiencies are defined as for the \hZ\ four-jet selection, 
  relative to the \lhqq\ final-state with all {\sc SM} decay modes
  allowed but that in $\tau\tau$ pairs. 
  The quoted errors are statistical only.}
\label{ta:ad4beff}
\end{center}
\end{table}

\begin{table} [hbtp]
\begin{center}
\begin{tabular}{cc|c|c|c}  \hline
 \MA  & \mh  & \rs~=~199.6~\GeV\ & \rs~=~206.5~\GeV\ & \rs~=~206.5~\GeV\ \\
(\GeVcc)  &(\GeVcc)  &  &  1st period &  2nd period \\ \hline
  70.0 &  70.0   & 52.4 $\pm$ 0.7 & 52.4 $\pm$ 0.7  & 51.3 $\pm$ 0.5  \\
  70.0 &  80.0   & 55.8 $\pm$ 0.8 & 54.4 $\pm$ 0.7  & 53.0 $\pm$ 0.6  \\
  70.0 &  85.0   & 56.5 $\pm$ 0.7 & 55.4 $\pm$ 0.7  & 52.9 $\pm$ 0.7  \\
  70.0 &  90.0   & 58.4 $\pm$ 0.7 & 56.5 $\pm$ 0.7  & 55.4 $\pm$ 0.6  \\
  70.0 & 100.0   & 65.3 $\pm$ 0.7 & 63.4 $\pm$ 0.7  & 62.2 $\pm$ 0.6  \\
  70.0 & 110.0   & 71.3 $\pm$ 0.6 & 69.6 $\pm$ 0.7  & 69.6 $\pm$ 0.5  \\
  70.0 & 130.0   &                & 69.1 $\pm$ 0.7  & 68.2 $\pm$ 0.5  \\
  70.0 & 136.0   &                & 57.0 $\pm$ 0.7  & 57.0 $\pm$ 0.5  \\
  80.0 &  80.0   & 56.7 $\pm$ 0.7 & 57.0 $\pm$ 0.7  & 55.0 $\pm$ 0.6  \\
  80.0 &  85.0   & 61.0 $\pm$ 0.7 & 60.1 $\pm$ 0.7  & 58.2 $\pm$ 0.5  \\
  80.0 &  90.0   & 67.4 $\pm$ 0.7 & 67.0 $\pm$ 0.7  & 63.8 $\pm$ 0.5  \\
  80.0 & 100.0   & 78.7 $\pm$ 0.6 & 79.6 $\pm$ 0.6  & 73.7 $\pm$ 0.5  \\
  85.0 &  85.0   & 67.0 $\pm$ 0.7 & 67.4 $\pm$ 0.7  & 64.2 $\pm$ 0.5  \\
  85.0 &  90.0   & 75.2 $\pm$ 0.6 & 73.0 $\pm$ 0.7  & 71.3 $\pm$ 0.5  \\
  85.0 &  95.0   & 79.2 $\pm$ 0.6 & 78.1 $\pm$ 0.6  & 75.9 $\pm$ 0.6  \\
  90.0 &  90.0   & 80.8 $\pm$ 0.6 & 78.0 $\pm$ 0.6  & 76.3 $\pm$ 0.5  \\
  90.0 &  95.0   & 81.8 $\pm$ 0.5 & 81.1 $\pm$ 0.6  & 79.1 $\pm$ 0.5  \\
  90.0 & 100.0   & 79.6 $\pm$ 0.6 & 80.4 $\pm$ 0.6  & 80.8 $\pm$ 0.4  \\
  90.0 & 110.0   &                & 78.2 $\pm$ 0.6  & 77.7 $\pm$ 0.4  \\
  90.0 & 116.0   &                & 65.5 $\pm$ 0.7  & 64.7 $\pm$ 0.5  \\
  95.0 &  95.0   & 81.3 $\pm$ 0.6 & 82.1 $\pm$ 0.5  & 80.7 $\pm$ 0.4  \\
 100.0 & 100.0   &                & 79.2 $\pm$ 0.6  & 78.2 $\pm$ 0.4  \\
 103.0 & 103.0   &                & 67.2 $\pm$ 0.7  & 66.8 $\pm$ 0.5  \\

\hline\end{tabular}
\caption[]{\hA\ four-b channel : 
  efficiencies of the low mass \hZ\ four-jet selection (in \%) 
  at \rs~=~199.6~\GeV\ and \rs~=~206.5~\GeV\ as a function of
  the masses of the A and h bosons.
  The efficiencies are defined as for the \hA\ four-jet selection, that
  is relative to the  $ {\mathrm b \bar{b}} {\mathrm b \bar{b}}$
  final state.
  The quoted errors are statistical only.}
\label{ta:adhqeff}
\end{center}
\end{table}



\clearpage

\newpage

\clearpage

\end{document}